\documentclass[english]{article}
\usepackage[latin9]{inputenc}
\usepackage{geometry}
\usepackage{color}
\usepackage{float}
\usepackage{amsmath}
\usepackage{amssymb}
\usepackage{graphicx}
\usepackage{setspace}
\usepackage[authoryear]{natbib}

\makeatletter

\providecommand{\tabularnewline}{\\}

\@ifundefined{showcaptionsetup}{}{%
 \PassOptionsToPackage{caption=false}{subfig}}
\usepackage{subfig}
\makeatother

\usepackage{babel}
\usepackage[bottom, symbol]{footmisc}
\begin{document}

\title{\textbf{Various methods for queue length and traffic volume estimation
using probe vehicle trajectories}\footnote{\textcopyright2019. This manuscript is made available under the CC-BY-NC-ND 4.0 license (http://creativecommons.org/licenses/by-nc-nd/4.0/).}}

\author{Yan Zhao\textsuperscript{a}, Jianfeng Zheng\textsuperscript{b,}\footnote{Corresponding author, email: zhengjf@umich.edu.},
Wai Wong\textsuperscript{c}, Xingmin Wang\textsuperscript{c}, Yuan
Meng\textsuperscript{b}, Henry X. Liu\textsuperscript{b,c,d}\date{}}

\maketitle
\begin{singlespace}
\textsuperscript{a}\textit{Department of Mechanical Engineering,
University of Michigan, Ann Arbor, MI, USA}

\textsuperscript{b}\textit{Didi Chuxing Inc., Beijing, China}

\textsuperscript{c}\textit{Department of Civil and Environmental
Engineering, University of Michigan, Ann Arbor, MI, USA}

\textsuperscript{d}\textit{University of Michigan Transportation
Research Institute, University of Michigan, Ann Arbor, MI, USA}
\end{singlespace}

\section*{Abstract}

The rapid development of connected vehicle technology and the emergence
of ride-hailing services have enabled the collection of a tremendous
amount of probe vehicle trajectory data. Due to the large scale, the
trajectory data have become a potential substitute for the widely
used fixed-location sensors in terms of the performance measures of
transportation networks. Specifically, for traffic volume and queue
length estimation, most of the trajectory data based methods in the
existing literature either require high market penetration of the
probe vehicles to identify the shockwave or require the prior information
about the queue length distribution and the penetration rate, which
may not be feasible in the real world. To overcome the limitations
of the existing methods, this paper proposes a series of novel methods
based on probability theory. By exploiting the stopping positions
of the probe vehicles in the queues, the proposed methods try to establish
and solve a single-variable equation for the penetration rate of the
probe vehicles. Once the penetration rate is obtained, it can be used
to project the total queue length and the total traffic volume. The
validation results using both simulation data and real-world data
show that the methods would be accurate enough for assistance in performance
measures and traffic signal control at intersections, even when the
penetration rate of the probe vehicles is very low. \\
\\
\textbf{\textit{Keywords}}\textit{:} Probe vehicle, Queue length estimation,
Penetration rate, Traffic volume estimation

\pagebreak{}

\section{Introduction and Motivation}

Traffic volumes and queue lengths are important performance measures
for signalized intersections. Conventional approaches for traffic
volume measurement and queue length estimation are primarily based
on fixed-location sensors, such as loop detectors \citep{liu2009real,lee2015real,an2018real}.
However, the installation and maintenance of the fixed-location sensors
are very costly, calling for urgent needs of new alternatives of data
sources. This gap can be now fulfilled, thanks to the rapid development
of connected vehicle technology and the emergence of ride-hailing
services. The global positioning system (GPS) devices on the connected
vehicles or the smartphones in the ride-hailing vehicles could record
the trajectories of these probe vehicles, providing rich information
about the traffic conditions in transportation networks. 

Based on the probe vehicle trajectory data, a wide range of methods
have been proposed for estimating the queue lengths and traffic volumes
at the signalized intersections \citep{guo2019urban}. A stream of
literature solves the problem from the perspective of probability
theory and statistics. \citet{comert2009queue} showed that given
the penetration rate of the probe vehicles and the distribution of
queue lengths, the positions of the last probe vehicles in the queues
alone would be sufficient for cycle-by-cycle queue length estimation.
\citet{comert2009queue} also analyzed the relationship between the
probe vehicle market penetration ratio and estimation accuracy. \citet{comert2011analytical}
extended their work to both spatial and temporal dimensions by considering
the time when the probe vehicles joined the queues. In 2013, Comert
studied the effect of the data from stop line detection \citep{comert2013effect}
and proposed another simple analytical model \citep{comert2013simple}.
\citet{li2013estimating} formulated the dynamics of the queue length
as a state transition process and employed a Kalman filter to estimate
the queue length cycle by cycle. With the assumption of Poisson distribution,
\citet{comert2016queue} summarized a series of methods of queue length
estimation and penetration rate estimation and evaluated the estimators
systematically. As for traffic volume estimation, \citet{zheng2017estimating}
applied maximum likelihood estimation, assuming the vehicle arrivals
at the intersections follow a time-varying Poisson process. The model
was validated using the trajectory data collected from connected vehicles
and taxis. \citet{zhan2017citywide} studied citywide traffic volume
estimation using large-scale trajectory data, by combining some machine
learning techniques and the traditional traffic flow theory. \citet{wang2019traffic}
constructed a three-layer Bayesian network to capture the relationship
between vehicle arrival processes and the timing information in probe
vehicle trajectory data. The average arrival rate was inferred from
the Bayesian network by applying the Expectation-Maximization algorithm.
There is also a stream of literature that applies the shockwave theory
to probe vehicle data \citep{ban2011real,cetin2012estimating,hao2015kinematic,hao2015long,ramezani2015queue,li2017real,rompis2018probe},
or combines probe vehicle data and loop detector data \citep{badillo2012queue,cai2014shock,wang2017shockwave,shahrbabaki2018data},
to estimate or predict the queue lengths. Since these studies are
not closely related to this paper methodologically, they will not
be introduced in detail.

Most of the existing literature introduced above on queue length estimation
focuses on cycle-by-cycle estimation and requires the prior information
about the penetration rate of the probe vehicles and the distribution
of queue lengths. However, the prior information is usually not available.
Although a recent study by \citet{wong2018volume} proposed a novel
method that provides an unbiased estimator for the probe vehicle penetration
rate solely based on probe vehicle trajectory data, the method cannot
handle the cases when some of the queues are empty. As for traffic
volume estimation, the model developed by \citet{zheng2017estimating}
assumes the vehicle arrivals in each cycle follow a time-varying Poisson
process, which might not be reasonable in over-saturation cases when
the arrival process, the queueing process, and the departure process
are all different. Although the method proposed by \citet{zhan2017citywide}
can be applied in large scale, it requires the ground-truth traffic
volume data on some road segments to build a connection between their
high-level features and the actual volume categories, which implies
that the method depends on not only the trajectory data but also other
sources of data. \citet{zhao2019volume} proposed a simplified method
of finding the penetration rate of the probe vehicles based on Bayes'
theorem. Extending the method in \citet{zhao2019volume}, this paper
aims to propose a general framework and a series of methods that can
estimate queue length and traffic volume both accurately and efficiently.

Estimating the states of the whole population from a small portion
of it \citep{wong2015systematic,wong2016biased}, in nature, has to
build a connection between the small portion and the whole population
by their common features. When the traffic is flowing, it is difficult
to infer how many regular vehicles are around the probe vehicles.
Consequently, it is almost impossible to estimate the penetration
rate of the probe vehicles in the traffic. However, when the vehicles
are stopping at the intersections, because the empirical value of
the space headway is usually around 7.5 m/veh, the number of vehicles
in front of the last probe vehicle can be roughly inferred. Although
the number of vehicles behind the last probe vehicle is still unknown,
the incomplete information could still provide an opportunity to estimate
the penetration rate of the probe vehicles. According to the penetration
rate, the total queue length and the total traffic volume can be projected
by scaling up the number of probe vehicles in the queues and in the
traffic, respectively \citep{wong2016evaluation,wong2018bootstrap}.
The proposed methods in this paper take the stopping positions at
the intersections as the common characteristics between the probe
vehicles and the regular vehicles. Since the proposed methods in this
paper have few external dependencies, they could overcome the limitations
of the existing methods and be applied to a broader range of scenarios.
The methods have been validated by both simulation and large-scale
real-world data, showing good accuracy.

The rest of this paper is organized as follows. In Section \ref{sec: Problem-Statement},
a detailed description of the problem will be given. Depending upon
the existence of the probe vehicles, the queues over different cycles
will be categorized into two classes: the observable queues (with
probe vehicles) and the hidden queues (without probe vehicles). It
will also be shown that the total traffic volume and the total queue
length can be easily obtained once the probe vehicle penetration rate
is known. Section \ref{sec: Q_obs} will present four different estimators
of the total length of the observable queues. Section \ref{sec: Q_hid}
will present two different estimators of the total length of the hidden
queues. In Section \ref{sec: pr}, two methods for estimating the
penetration rate of the probe vehicles will be proposed, which combine
the various estimators presented in Section \ref{sec: Q_obs} and
Section \ref{sec: Q_hid}. The proposed methods are validated and
evaluated in Section \ref{sec: validation}. Finally, there will be
some concluding remarks in Section \ref{sec: discussion}.

\section{Problem Statement\label{sec: Problem-Statement}}

When the vehicles are stopping at the intersections due to the traffic
lights, some vehicles in the queue might be the probe vehicles of
which the trajectories could be recorded by the onboard GPS devices.
For a specific movement and a specific time slot, the vehicle arrival
process is assumed to be stationary; the probe vehicles are assumed
to be homogeneously mixed with other vehicles. Let $p$ denote the
penetration rate of the probe vehicles, that is, when arbitrarily
selecting a vehicle from the queue, its probability of being a probe
vehicle is $p$, where $p\in(0,1)$. 

Suppose the trajectory data of the probe vehicles are collected for
$C$ cycles. In each cycle, the positions of the probe vehicles in
the queue can be easily extracted from the trajectory data. The average
space headway when vehicles are stopping at the intersections is assumed
to be known empirically, which is a common assumption in the relevant
literature. Then, with the knowledge of the position of the stop bar,
the number of vehicles in front of the last probe vehicle can also
be inferred, although the number of vehicles behind the last probe
vehicle is still unknown. Denote the queue length in the $i$th cycle
by a random variable $Q_{i},\forall i\in\{1,2,\dots,C\}$. Denote
the number of probe vehicles in the $i$th cycle by a random variable
$N_{i}$. Denote the observed partial queue in the $i$th cycle by
a tuple $q_{i}$ consisting of ``0''s and ``1''s which represent
regular vehicles and probe vehicles, respectively. Denote the length
of the observed partial queue by $\left|q_{i}\right|$. Apparently,
$Q_{i}\ge\left|q_{i}\right|\ge N_{i},\forall i\in\{1,2,\dots,C\}$. 

\begin{figure}[h]
\begin{centering}
\includegraphics[width=0.8\textwidth]{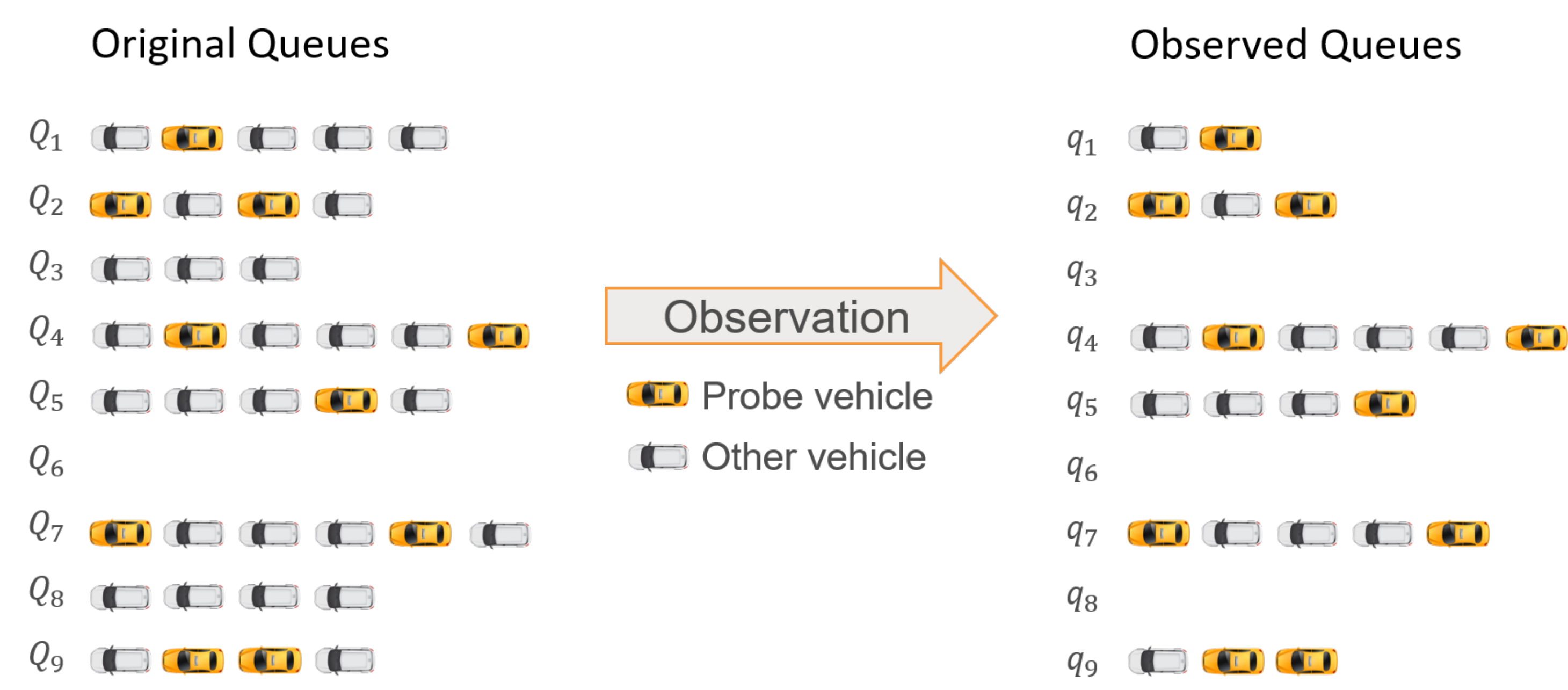}
\par\end{centering}
\caption{\label{fig: Observation-process}Observation process}
\end{figure}

Figure \ref{fig: Observation-process} illustrates what can be easily
inferred from the trajectory data. $Q_{1}$, $Q_{2}$, $Q_{4}$, $Q_{5}$,
$Q_{7}$, and $Q_{9}$ are (partially) observable because of the probe
vehicles in the queues. $Q_{3}$, $Q_{6}$, and $Q_{8}$ are hidden
because there are no probe vehicles. Denote the total length of the
observable queues and the total length of the hidden queues by $Q^{obs}$
and $Q^{hid}$, respectively. In Figure \ref{fig: Observation-process},
$Q^{obs}=Q_{1}+Q_{2}+Q_{4}+Q_{5}+Q_{7}+Q_{9}=30$ and $Q^{hid}=Q_{3}+Q_{6}+Q_{8}=7$.
In the $i$th cycle, if the queue is observable, then denote the positions
of the first and the last probe vehicles by $S_{i}$ and $T_{i}$,
respectively. 

Define a binary random variable $X_{i}^{l}$ to indicate if the queue
length in the $i$th cycle is $l$, that is, 
\begin{equation}
X_{i}^{l}=\begin{cases}
1, & Q_{i}=l\\
0, & Q_{i}\ne l
\end{cases},\label{eq: X_def}
\end{equation}
where $l\in\left\{ 0,1,\dots,L_{max}\right\} $ and $L_{max}$ is
an upper bound of the queue length. Denote the number of queues of
length $l$ in all the cycles by $C_{l}$. Obviously, $C=\sum_{l=0}^{L_{max}}C_{l}$
and $C_{l}=\sum_{i=1}^{C}X_{i}^{l}$. 

To estimate the traffic volume and the total queue length for a specific
movement and a specific time slot, the key step is to find the penetration
rate $p$. Denote the total number of probe vehicles in the queues
by $Q^{probe}$ and denote the traffic volume of probe vehicles by
$V^{probe}$. Since $Q^{probe}$ and $V^{probe}$ can be easily obtained
from the trajectory data by counting the number of probe vehicles
in the queues and in the traffic flows, once $p$ is known, equation
(\ref{eq: est_Qall}) and equation (\ref{eq: est_Vall}) can give
an estimate of the total queue length and the total traffic volume,
respectively \citep{wong2018bootstrap,wong2018unbiased,zhao2019volume}. 

\begin{equation}
\hat{Q}^{all}=\frac{Q^{probe}}{p}.\label{eq: est_Qall}
\end{equation}

\begin{equation}
\hat{V}^{all}=\frac{V^{probe}}{p}.\label{eq: est_Vall}
\end{equation}

Table \ref{tab:Notations} summarizes the notations defined above. 

\begin{table}[h]
\caption{\label{tab:Notations}Notations}

\centering{}%
\begin{tabular}{ll}
\hline 
Notation & Description\tabularnewline
\hline 
$C$ & The total number of cycles\tabularnewline

$Q_{i}$ & The queue length in the $i$th cycle\tabularnewline

$q_{i}$ & The observed partial queue in the $i$th cycle\tabularnewline

$N_{i}$ & The number of probe vehicles in the $i$th cycle\tabularnewline

$S_{i}$ & The position of the first probe vehicle in the $i$th cycle\tabularnewline

$T_{i}$ & The position of the last probe vehicle in the $i$th cycle\tabularnewline

$X_{i}^{l}$ & A binary variable to indicate if the queue length in the $i$th cycle
is $l$\tabularnewline

$C_{l}$ & The total number of queues of length $l$\tabularnewline

$L_{max}$ & An upper bound of the queue length\tabularnewline

$Q^{obs}$ & The total length of all the (partially) observable queues\tabularnewline

$Q^{hid}$ & The total length of the hidden queues \tabularnewline

$Q^{probe}$ & The total number of probe vehicles in all the queues\tabularnewline

$V^{probe}$ & The traffic volume of probe vehicles\tabularnewline
\hline 
\end{tabular}
\end{table}

\section{Estimation of $Q^{obs}$\label{sec: Q_obs}}

$Q^{obs}$ can be estimated through two approaches. Estimator 1, 2,
and 3 are based on the fact that the probe vehicles are expected to
segregate the regular vehicles equally. These estimators only require
the number of stopping probe vehicles in each cycle and the stopping
positions of the first and the last probe vehicles in the queues,
all of which can be easily extracted from the trajectory data. Therefore,
the estimators are constant values. By contrast, estimator 4 is based
on Bayes' theorem, which relies on the penetration rate $p$. Thus,
estimator 4 is a function of $p$. 

\subsection{Estimator 1 using the first probe vehicles in the queues}
\begin{description}
\item [{Theorem}] 1:
\end{description}
For any integer $n_{i}\ge1$, given that $N_{i}=n_{i}$ in the $i$th
cycle, 
\begin{equation}
\mathbb{E}(Q_{i}\mid N_{i}=n_{i})=\mathbb{E}(S_{i}\mid N_{i}=n_{i})(n_{i}+1)-1.
\end{equation}

The proof is in Appendix A.

Theorem 1 states that given the number of probe vehicles in an observable
queue, the expected queue length can be obtained from the expected
stopping position of the first probe vehicle. Based on Theorem 1,
given the number of probe vehicles in each cycle, the expected total
length of the observable queues can be expressed as 
\begin{align}
\sum_{i:n_{i}\ne0}\mathbb{E}(Q_{i}\mid N_{i}=n_{i}) & =\sum_{i:n_{i}\ne0}\left(\mathbb{E}\left(S_{i}\mid N_{i}=n_{i}\right)\left(n_{i}+1\right)-1\right).\\
 & =\sum_{i:n_{i}\ne0}\mathbb{E}\left(S_{i}\mid N_{i}=n_{i}\right)\left(n_{i}+1\right)-\sum_{i:n_{i}\ne0}1\\
 & =\sum_{j=1}^{L_{max}}\sum_{i:n_{i}=j}\mathbb{E}\left(S_{i}\mid N_{i}=j\right)\left(j+1\right)-\sum_{i:n_{i}\ne0}1\\
 & =\sum_{j=1}^{L_{max}}\left(j+1\right)\sum_{i:n_{i}=j}\mathbb{E}\left(S_{i}\mid N_{i}=j\right)-\sum_{i:n_{i}\ne0}1.
\end{align}
Therefore, given the position of the first stopping probe vehicle
$S_{i}=s_{i}$ in the $i$th cycle, $\forall i\in\{1,2,,\dots,C\}$,
by substituting the sample mean $\frac{\sum_{i:n_{i}=j}s_{i}}{\sum_{i:n_{i}=j}1}$
for the expected value $\mathbb{E}\left(S_{i}\mid N_{i}=j\right),\forall j\ge1$,
$Q^{obs}$ can be estimated by 
\begin{align}
\hat{Q}_{1}^{obs} & =\sum_{j=1}^{L_{max}}\left(j+1\right)\sum_{i:n_{i}=j}s_{i}-\sum_{i:n_{i}\ne0}1\\
 & =\sum_{j=1}^{L_{max}}\sum_{i:n_{i}=j}s_{i}\left(j+1\right)-\sum_{i:n_{i}\ne0}1\\
 & =\sum_{i:n_{i}\ne0}s_{i}\left(n_{i}+1\right)-\sum_{i:n_{i}\ne0}1\\
 & =\sum_{i:n_{i}\ne0}\left(s_{i}\left(n_{i}+1\right)-1\right).
\end{align}

\subsection{Estimator 2 using the last probe vehicles in the queues}
\begin{description}
\item [{Theorem}] 2:
\end{description}
For any integer $n_{i}\ge1$, given that $N_{i}=n_{i}$ in the $i$th
cycle,

\begin{equation}
\mathbb{E}(Q_{i}\mid N_{i}=n_{i})=\mathbb{E}(T_{i}\mid N_{i}=n_{i})\frac{n_{i}+1}{n_{i}}-1.
\end{equation}

The proof is in Appendix A.

Theorem 2 states that given the number of probe vehicles in an observable
queue, the expected queue length can be obtained from the expected
stopping position of the last probe vehicle. Based on Theorem 2, given
the number of probe vehicles in each cycle, the expected total length
of observable queues can be expressed as 
\begin{equation}
\sum_{i:n_{i}\ne0}\mathbb{E}(Q_{i}\mid N_{i}=n_{i})=\sum_{i:n_{i}\ne0}\left(\mathbb{E}(T_{i}\mid N_{i}=n_{i})\frac{n_{i}+1}{n_{i}}-1\right).
\end{equation}
Following the similar derivations with estimator 1, given the position
of the last stopping probe vehicle $T_{i}=t_{i}$ in the $i$th cycle,
$\forall i\in\{1,2,\dots,C\}$, by substituting the sample mean $\frac{\sum_{i:n_{i}=j}t_{i}}{\sum_{i:n_{i}=j}1}$
for the expected value $\mathbb{E}\left(T_{i}\mid N_{i}=j\right),\forall j\ge1$,
$Q^{obs}$ can be estimated by 
\begin{equation}
\hat{Q}_{2}^{obs}=\sum_{i:n_{i}\ne0}\left(t_{i}\frac{n_{i}+1}{n_{i}}-1\right).
\end{equation}

\subsection{Estimator 3 using the first and the last probe vehicles in the queues}
\begin{description}
\item [{Theorem}] 3:
\end{description}
For any integer $n_{i}\ge1$, given that $N_{i}=n_{i}$ in the $i$th
cycle,

\begin{equation}
\mathbb{E}(Q_{i}\mid N_{i}=n_{i})=\mathbb{E}(S_{i}\mid N_{i}=n_{i})+\mathbb{E}(T_{i}\mid N_{i}=n_{i})-1,
\end{equation}

\begin{equation}
\mathbb{E}(Q_{i}\mid N_{i}\ge1)=\mathbb{E}(S_{i}\mid N_{i}\ge1)+\mathbb{E}(T_{i}\mid N_{i}\ge1)-1.
\end{equation}

The proof is in Appendix A. 

Theorem 3 states that given the number of probe vehicles in an observable
queue, the expected queue length can be obtained from the expected
stopping positions of the first and the last probe vehicles. Based
on Theorem 3, given the number of probe vehicles in each cycle, the
expected total length of the observable queues can be expressed as 

\begin{equation}
\sum_{i:n_{i}\ne0}\mathbb{E}(Q_{i}\mid N_{i}=n_{i})=\sum_{i:n_{i}\ne0}\left(\mathbb{E}(S_{i}\mid N_{i}=n_{i})+\mathbb{E}(T_{i}\mid N_{i}=n_{i})-1\right).
\end{equation}
Therefore, by substituting the sample means $\frac{\sum_{i:n_{i}=j}s_{i}}{\sum_{i:n_{i}=j}1}$
and $\frac{\sum_{i:n_{i}=j}t_{i}}{\sum_{i:n_{i}=j}1}$ for the expected
values $\mathbb{E}(S_{i}\mid N_{i}=n_{i})$ and $\mathbb{E}(T_{i}\mid N_{i}=n_{i}),\forall j\ge1$,
respectively, $Q^{obs}$ can be estimated by 
\begin{equation}
\hat{Q}_{3}^{obs}=\sum_{i:n_{i}\ne0}\left(s_{i}+t_{i}-1\right).
\end{equation}

\begin{figure}[h]
\begin{centering}
\includegraphics[scale=0.5]{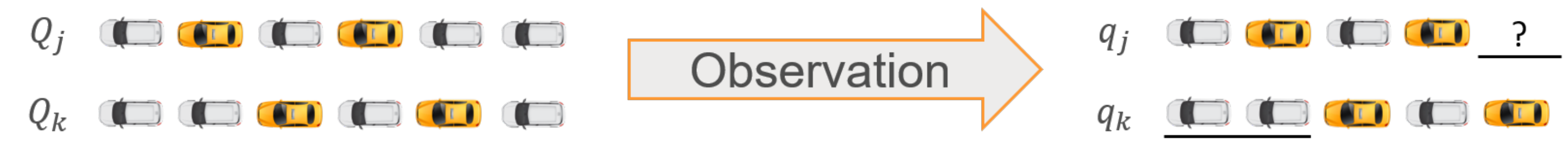}
\par\end{centering}
\caption{\label{fig: Q_obs_3}The missing information compensated by another
queue}
\end{figure}

The mechanism behind $\hat{Q}_{3}^{obs}$ is intuitive. Take Figure
\ref{fig: Q_obs_3} for example. The queue in the $k$th cycle is
the reverse of the queue in the $j$th cycle, which implies that the
number of vehicles behind the last probe vehicle in the $j$th cycle
is equal to the number of vehicles in front of the first probe vehicle
in the $k$th cycle. Because of the symmetry, these two queues have
the same probability of occurring. Therefore, even though the number
of vehicles behind the last probe vehicle in a cycle is unknown, as
long as the sample size is sufficient, the missing number could be
compensated by the number of vehicles in front of the first probe
vehicle in another cycle. Essentially, $\hat{Q}_{3}^{obs}$ is obtained
by summing up the position of the last probe vehicle $t_{i}$ and
the number of vehicles in front of the first probe vehicle $s_{i}-1$,
which could be regarded as a compensation of the missing vehicles
in the rear.

\subsection{Estimator 4 based on Bayes' theorem}

\begin{figure}[h]
\begin{centering}
\includegraphics[width=1\textwidth]{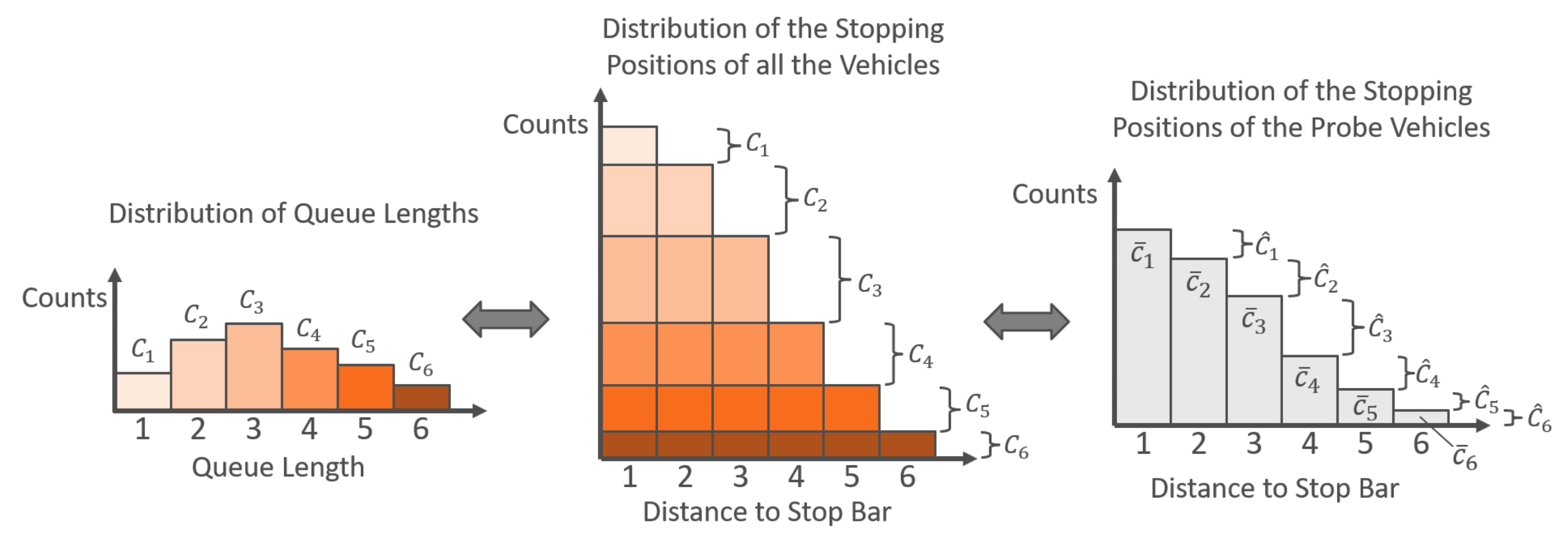}
\par\end{centering}
\caption{\label{fig: get_PQ}The relationship between the distributions of
queue lengths and stopping positions}
\end{figure}

Given all the observed partial queues, as derived in \citet{zhao2019volume},
the conditional expectation of the total length of the observable
queues can be expressed as
\begin{align}
\sum_{i:n_{i}\ne0}\mathbb{E}(Q_{i}\mid q_{i}) & =\sum_{i:n_{i}\ne0}\sum_{l=1}^{L_{max}}\frac{P(Q_{i}=l)P(q_{i}\mid Q_{i}=l)}{\sum_{j=0}^{L_{max}}P(Q_{i}=j)P(q_{i}\mid Q_{i}=j)}l\label{eq: expected_obs_queues1}\\
 & =\sum_{i:n_{i}\ne0}\sum_{l=\left|q_{i}\right|}^{L_{max}}\frac{\mathbb{E}(C_{l})p^{n_{i}}\left(1-p\right)^{l-n_{i}}}{\sum_{j=\left|q_{i}\right|}^{L_{max}}\mathbb{E}(C_{j})p^{n_{i}}\left(1-p\right)^{j-n_{i}}}l\\
 & =\sum_{i:n_{i}\ne0}\sum_{l=\left|q_{i}\right|}^{L_{max}}\frac{p\mathbb{E}(C_{l})}{\sum_{j=\left|q_{i}\right|}^{L_{max}}p\mathbb{E}(C_{j})\left(1-p\right)^{j-l}}l.\label{eq: expected_obs_queues}
\end{align}
$C_{l}$, the number of cycles with queues of length $l$, equals
to the difference between the count of stopping vehicles at position
$l+1$ and the count of stopping vehicles at position $l$, as illustrated
by the first two diagrams in Figure \ref{fig: get_PQ}. Since the
probe vehicles are assumed to be homogeneously mixed with other vehicles,
the histogram of the stopping positions of the probe vehicles is a
$p$ scaled-down version of the histogram of the stopping positions
of all the vehicles. Therefore, $\hat{C}_{l}$, the difference between
$\bar{c}_{l}$, the count of stopping probe vehicles at position $l+1$,
and $\bar{c}_{l+1}$, the count of stopping probe vehicles at position
$l$, can be used to approximate $p\mathbb{E}(C_{l})$. When the difference
is negative, a least-squares method can be applied to ensure the nonnegativity
of $\hat{C}_{l}$ \citep{zhao2019volume}.

Once $\hat{C}_{l}$ is obtained, replacing $p\mathbb{E}(C_{l})$ in
equation (\ref{eq: expected_obs_queues}) by its approximation $\hat{C}_{l}$
gives an estimate of $Q^{obs}$ 
\begin{equation}
\hat{Q}_{4}^{obs}(p)=\sum_{i:n_{i}\ne0}\sum_{l=\left|q_{i}\right|}^{L_{max}}\frac{\hat{C}_{l}}{\sum_{j=\left|q_{i}\right|}^{L_{max}}\hat{C}_{j}\left(1-p\right)^{j-l}}l,\label{eq: Q^obs_4}
\end{equation}
which is a function of the penetration rate $p$.

\section{Estimation of $Q^{hid}$\label{sec: Q_hid}}

After estimating $Q^{obs}$, the following question is how to estimate
$Q^{hid}$, as there is no probe vehicle in the corresponding cycles.
Fortunately, the fact that no probe vehicle is in the queues also
contains information. In this section, two estimators of $Q^{hid}$
will be presented. Similar to $\hat{Q}_{4}^{obs}(p)$, estimator 1
of $Q^{hid}$ applies Bayes' theorem to the hidden queues directly.
Estimator 2 utilizes the ratio between the probability of being observable
and the probability of being hidden for each queue, to estimate the
total length of the hidden queues. 

\subsection{Estimator 1 based on Bayes' theorem}

Similar to equation (\ref{eq: expected_obs_queues}), given the fact
that no probe vehicle is observed in the hidden queues, the expected
total length of the hidden queues can be expressed as 
\begin{align}
\sum_{i:n_{i}=0}\mathbb{E}(Q_{i}\mid q_{i}) & =\sum_{i:n_{i}=0}\sum_{l=0}^{L_{max}}\frac{P(Q_{i}=l)P(q_{i}\mid Q_{i}=l)}{\sum_{j=0}^{L_{max}}P(Q_{i}=l)P(q_{i}\mid Q_{i}=j)}l.\\
 & =\sum_{i:n_{i}\ne0}\sum_{l=0}^{L_{max}}\frac{p\mathbb{E}(C_{l})}{\sum_{j=0}^{L_{max}}p\mathbb{E}(C_{j})\left(1-p\right)^{j-l}}l.
\end{align}
Therefore, an estimator of $Q^{hid}$ can be given by
\begin{align}
\hat{Q}_{1}^{hid}(p) & =\sum_{i:n_{i}=0}\sum_{l=0}^{L_{max}}\frac{\hat{C}_{l}}{\sum_{j=0}^{L_{max}}\hat{C}_{j}\left(1-p\right)^{j-l}}l.\label{eq: Q^hid_2}
\end{align}
Please note that different from equation (\ref{eq: Q^obs_4}), the
summation over $l$ in equation (\ref{eq: Q^hid_2}) starts from $0$,
because when $q_{i}$ is an empty tuple, 
\begin{equation}
P(q_{i}\mid Q_{i}=l)=(1-p)^{l}.
\end{equation}
Here shows how to find $\hat{C}_{0}$, an estimate of $p\mathbb{E}(C_{0})$.

In all the queues, the expected counts of queues of length 0 is
\begin{equation}
\mathbb{E}(C_{0})=C-\sum_{l=1}^{L_{max}}\mathbb{E}(C_{l}).
\end{equation}
Therefore, multiplying $p$ on the two sides of the equation gives
\begin{equation}
p\mathbb{E}(C_{0})=pC-\sum_{l=1}^{L_{max}}p\mathbb{E}(C_{l}).
\end{equation}
$\hat{C}_{0}$, an estimate of $p\mathbb{E}(C_{0})$, can be easily
given by
\begin{equation}
\hat{C}_{0}=pC-\sum_{l=1}^{L_{max}}\hat{C}_{l}.\label{eq: C0}
\end{equation}
All the parameters except $p$ on the right-hand side of equation
(\ref{eq: Q^hid_2}) can be calculated, therefore, $\hat{Q}_{1}^{hid}(p)$
is a function of only $p$. If stop line detection (such as loop detectors)
data are available, $\hat{C}_{0}$ can be more easily obtained.

\subsection{Estimator 2 using the probabilities of being observed and being hidden}

Among the observable queues, $\forall l\in\{1,2,\dots,L_{max}\}$,
the expected counts of queues of length $l$ can be expressed as 
\begin{align}
\sum_{i:n_{i}\ne0}\mathbb{E}\left(X_{i}^{l}\mid q_{i}\right) & =\sum_{i:n_{i}\ne0}\left(P(X_{i}^{l}=1\mid q_{i})\cdot1+P(X_{i}^{l}=0\mid q_{i})\cdot0\right)\\
 & =\sum_{i:n_{i}\ne0}\left(P(Q_{i}=l\mid q_{i})\cdot1+P(Q_{i}\ne l\mid q_{i})\cdot0\right)\\
 & =\sum_{i:n_{i}\ne0}P(Q_{i}=l\mid q_{i}).
\end{align}
For a queue of length $l$, the probability of being hidden (without
any probe vehicle) is $(1-p)^{l}$; the probability of being observed
(with at least one probe vehicle) is $1-(1-p)^{l}$. Therefore, the
expected total length of the hidden queues can be estimated by
\begin{align}
\sum_{l=1}^{L_{max}}\left(\frac{(1-p)^{l}}{1-(1-p)^{l}}\sum_{i:n_{i}\ne0}\mathbb{E}\left(X_{i}^{l}\mid q_{i}\right)\right)l & =\sum_{l=1}^{L_{max}}\frac{(1-p)^{l}}{1-(1-p)^{l}}\sum_{i:n_{i}\ne0}P(Q_{i}=l\mid q_{i})l\label{eq: Q^hid_1_mid0}\\
 & =\sum_{i:n_{i}\ne0}\sum_{l=1}^{L_{max}}\frac{(1-p)^{l}}{1-(1-p)^{l}}\frac{P(Q_{i}=l)P(q_{i}\mid Q_{i}=l)}{\sum_{j=0}^{L_{max}}P(Q_{i}=j)P(q_{i}\mid Q_{i}=j)}l\label{eq: Q^hid_1_mid2}\\
 & =\sum_{i:n_{i}\ne0}\sum_{l=\left|q_{i}\right|}^{L_{max}}\frac{(1-p)^{l}}{1-(1-p)^{l}}\frac{p\mathbb{E}(C_{l})}{\sum_{j=\left|q_{i}\right|}^{L_{max}}p\mathbb{E}(C_{j})\left(1-p\right)^{j-l}}l.
\end{align}
Then, an estimator of $Q^{hid}$, the total length of the hidden queues,
can be defined as 
\begin{align}
\hat{Q}_{2}^{hid}(p) & =\sum_{i:n_{i}\ne0}\sum_{l=\left|q_{i}\right|}^{L_{max}}\frac{\left(1-p\right)^{l}}{1-\left(1-p\right)^{l}}\frac{\hat{C}_{l}}{\sum_{j=\left|q_{i}\right|}^{L_{max}}\hat{C}_{j}\left(1-p\right)^{j-l}}l.\label{eq: Q^hid_1}
\end{align}

\section{Estimation of Penetration Rate\label{sec: pr}}

In this section, two different methods for penetration rate estimation
will be presented. The methodology is to establish an equation with
only a single unknown variable $p$ using the estimators developed
in the previous sections. Then, an estimate of $p$ can be obtained
by solving the equation. Method 1 is based upon the equivalence between
the different estimators. Method 2 exploits the fact that the portion
of probe vehicles in the queues is approximately equal to the penetration
rate.

\subsection{Method 1}

When estimating $Q^{obs}$, estimator 1, 2, and 3 can generate constant
results, whereas estimator 4 is a function of $p$. Since the four
estimators are of the same variable $Q^{obs}$, it is intuitive to
establish the following single-variable equation
\begin{equation}
\hat{Q}_{i}^{obs}=\hat{Q}_{4}^{obs}(p),\forall i=1,2,3.\label{eq: est_p_1}
\end{equation}
Solving the equation will yield an estimate of the penetration rate
$p$. Similarly, when estimating $Q_{hid}$, both estimator 1 and
estimator 2 are functions of $p$. Therefore, another single-variable
equation can be given by

\begin{equation}
\hat{Q}_{1}^{hid}(p)=\hat{Q}_{2}^{hid}(p).\label{eq: est_p_2-2}
\end{equation}

A more general formulation of this method can be expressed as follows. 

\begin{equation}
\hat{Q}_{i}^{obs}(p)+\hat{Q}_{j}^{hid}(p)=\hat{Q}_{m}^{obs}(p)+\hat{Q}_{n}^{hid}(p).\label{eq: est_p_3-1}
\end{equation}
As long as it is an equation with a single unknown variable $p$,
solving it will give an estimate of the penetration rate. Both the
left-hand side and the right-hand side of equation (\ref{eq: est_p_3-1})
can be regarded as estimators of the total queue length.

\subsection{Method 2}

Another way to establish a single-variable equation for $p$ is shown
by equation (\ref{eq: est_p_4}).

\begin{equation}
\frac{Q^{probe}}{\hat{Q}_{i}^{obs}(p)+\hat{Q}_{j}^{hid}(p)}=p,\forall i=1,2,3,4,\forall j=1,2,\label{eq: est_p_4}
\end{equation}
The left-hand side of equation (\ref{eq: est_p_4}) could be interpreted
as an estimate of the portion of probe vehicles in the queues. The
right-hand side is the penetration rate which should be approximately
equal to the left-hand side. Similarly, solving this equation yields
an estimate of $p$. 

In practice, it is usually hard to find $p$ by solving equation (\ref{eq: est_p_1}),
(\ref{eq: est_p_2-2}), (\ref{eq: est_p_3-1}), or (\ref{eq: est_p_4})
directly. Instead, an iterative algorithm should be applied. One may
search $p$ from an upper bound to 0 with a small step size until
the difference between the left-hand side and the right-hand side
reaches certain stopping criteria. The upper bound can be taken as
$\frac{Q^{probe}}{\sum_{i}\left|q_{i}\right|}$ since it is an overestimate
of the penetration rate $p$.

Once $p$ is estimated, equation (\ref{eq: est_Qall}) and equation
(\ref{eq: est_Vall}) can be used to estimate the total queue length
and the total traffic volume, respectively.

\section{Validation and Evaluation\label{sec: validation}}

\subsection{Simulation}

The focus of this test is on the estimation of penetration rate and
queue length. Unlike the existing methods \citep{comert2009queue,comert2016queue,zheng2017estimating},
the proposed methods in this paper do not require the prior information
about the penetration rate and the queue length distribution. For
demonstration purposes, the testing dataset is generated by a simulation
of Poisson processes, although any other stochastic process can also
be applied. The penetration rate of the probe vehicles is enumerated
from 0.01 to 0.99 with a step size of 0.01 in each test, in order
to test the robustness of the proposed methods.

\subsubsection{The comparison of different methods}

\begin{figure}[H]
\begin{centering}
\subfloat[]{\begin{centering}
\includegraphics[width=0.4\textwidth]{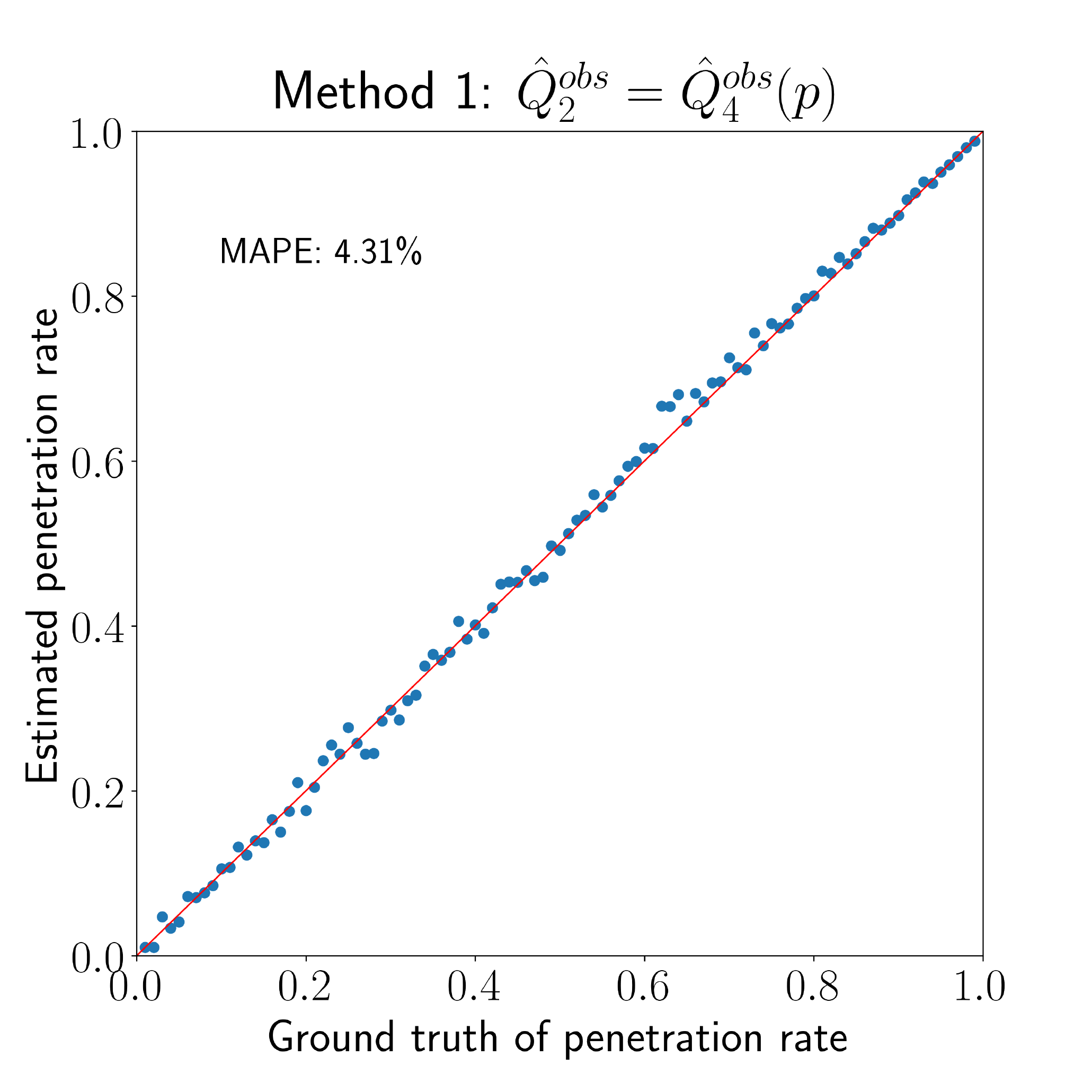}
\par\end{centering}
}\subfloat[]{\begin{centering}
\includegraphics[width=0.4\textwidth]{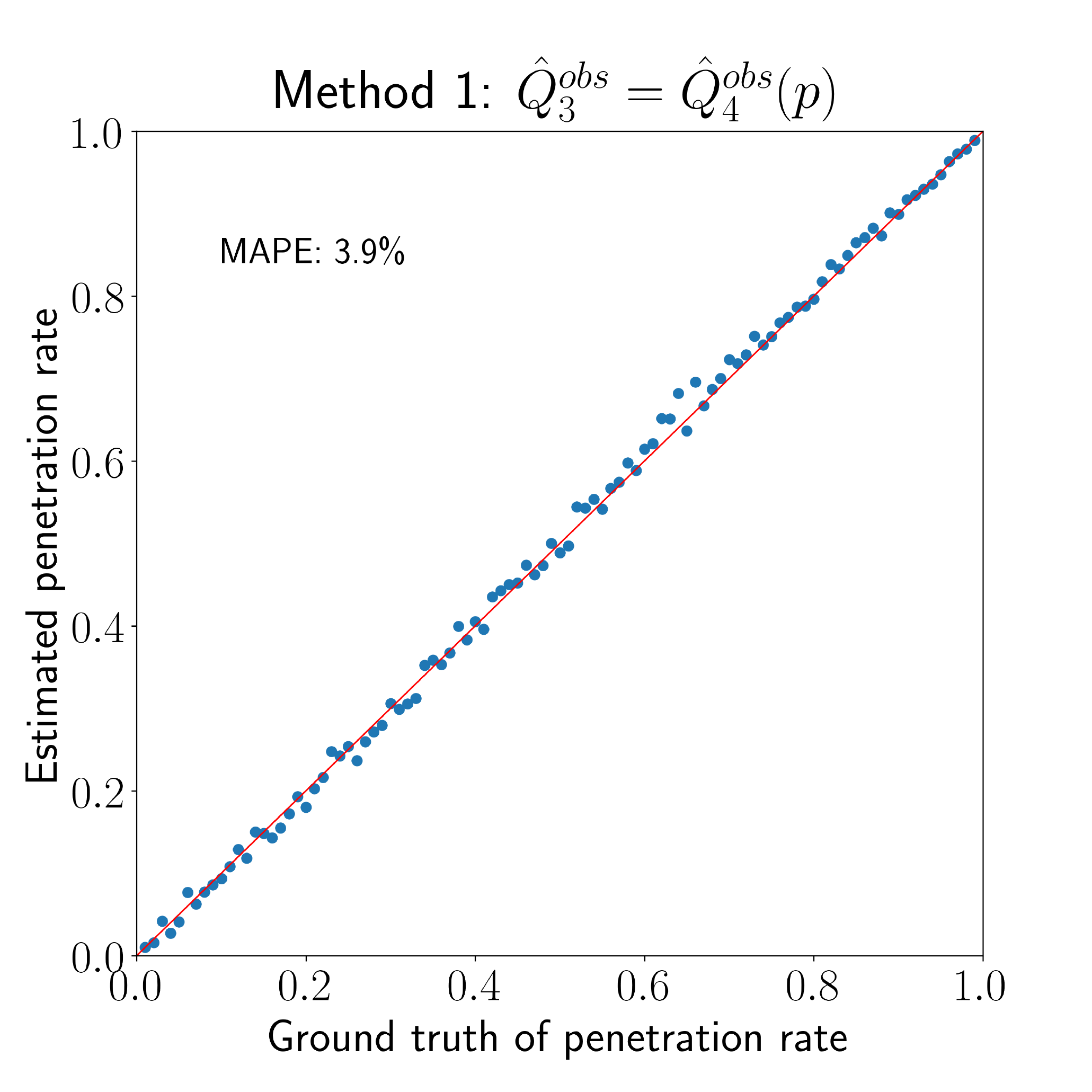}
\par\end{centering}
}
\par\end{centering}
\begin{centering}
\subfloat[]{\begin{centering}
\includegraphics[width=0.4\textwidth]{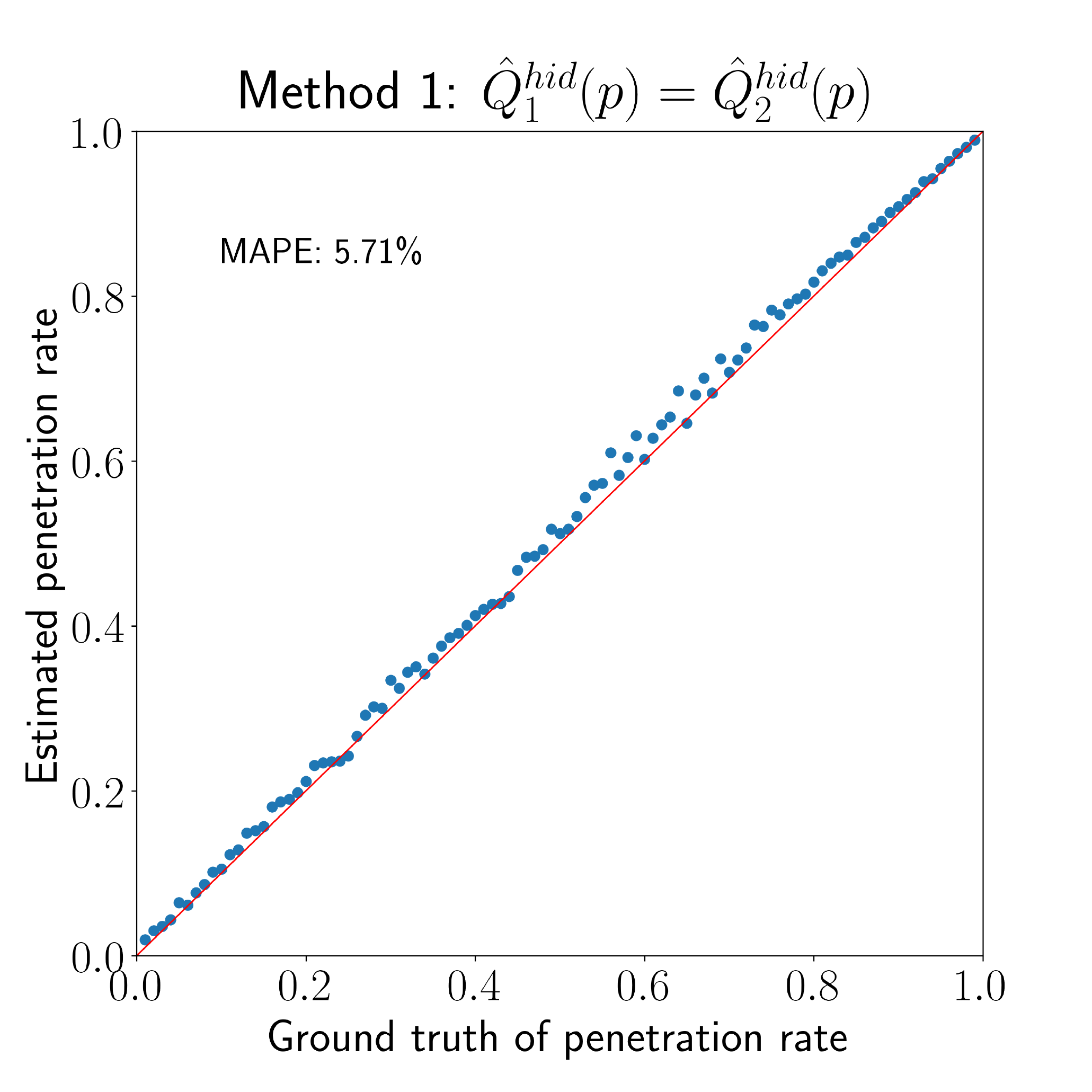}
\par\end{centering}
}\subfloat[]{\begin{centering}
\includegraphics[width=0.4\textwidth]{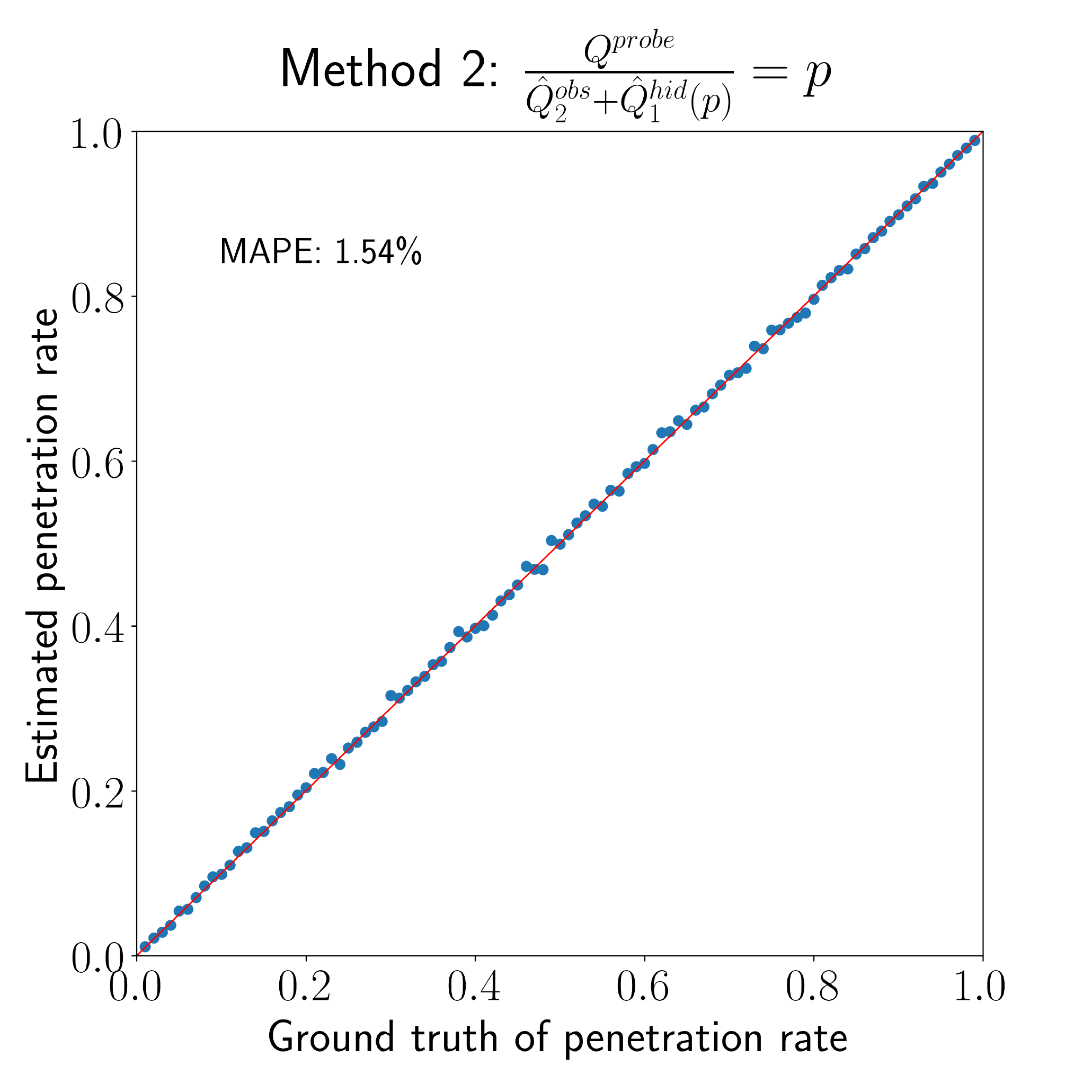}
\par\end{centering}
}
\par\end{centering}
\begin{centering}
\subfloat[]{\begin{centering}
\includegraphics[width=0.4\textwidth]{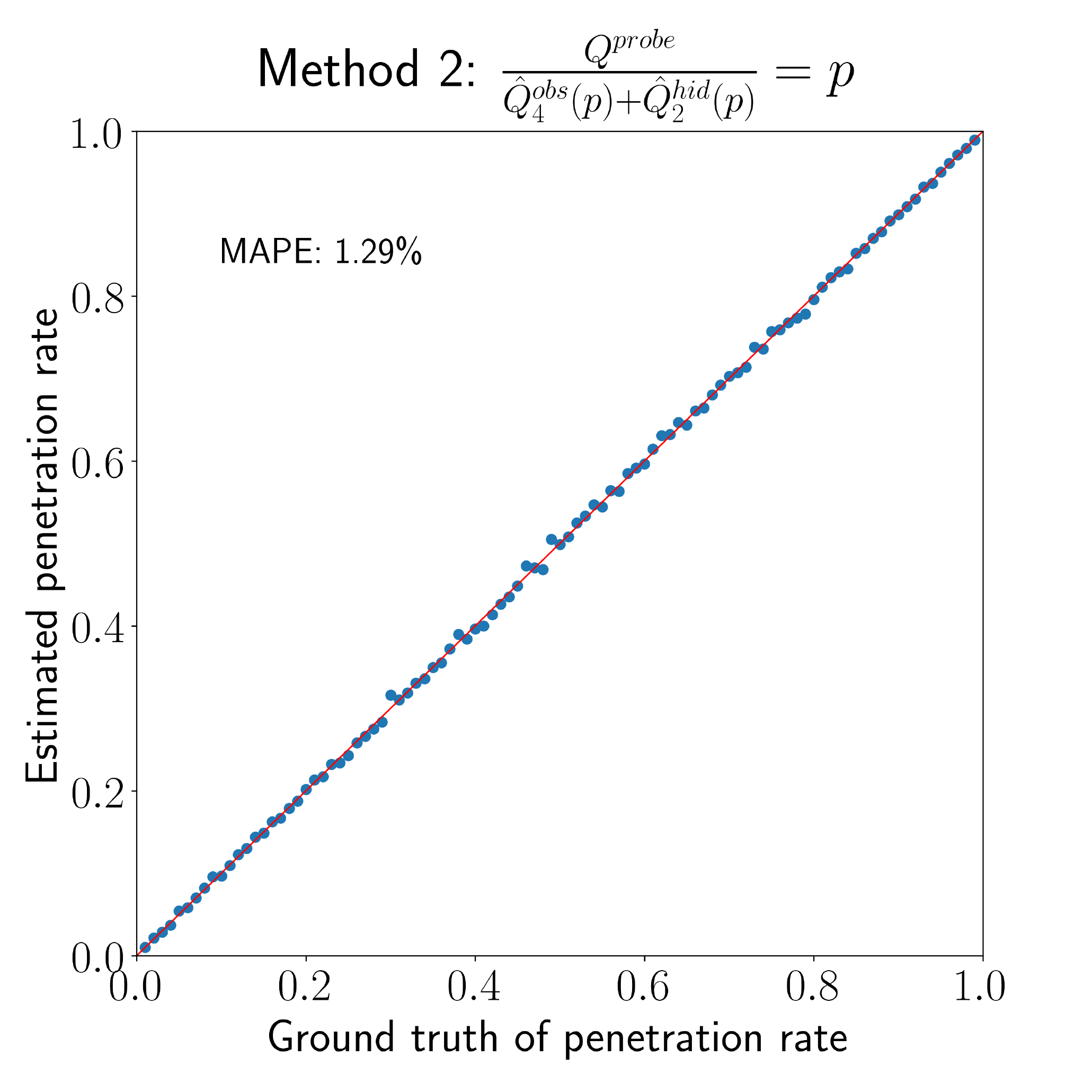}
\par\end{centering}
}\subfloat[]{\begin{centering}
\includegraphics[width=0.4\textwidth]{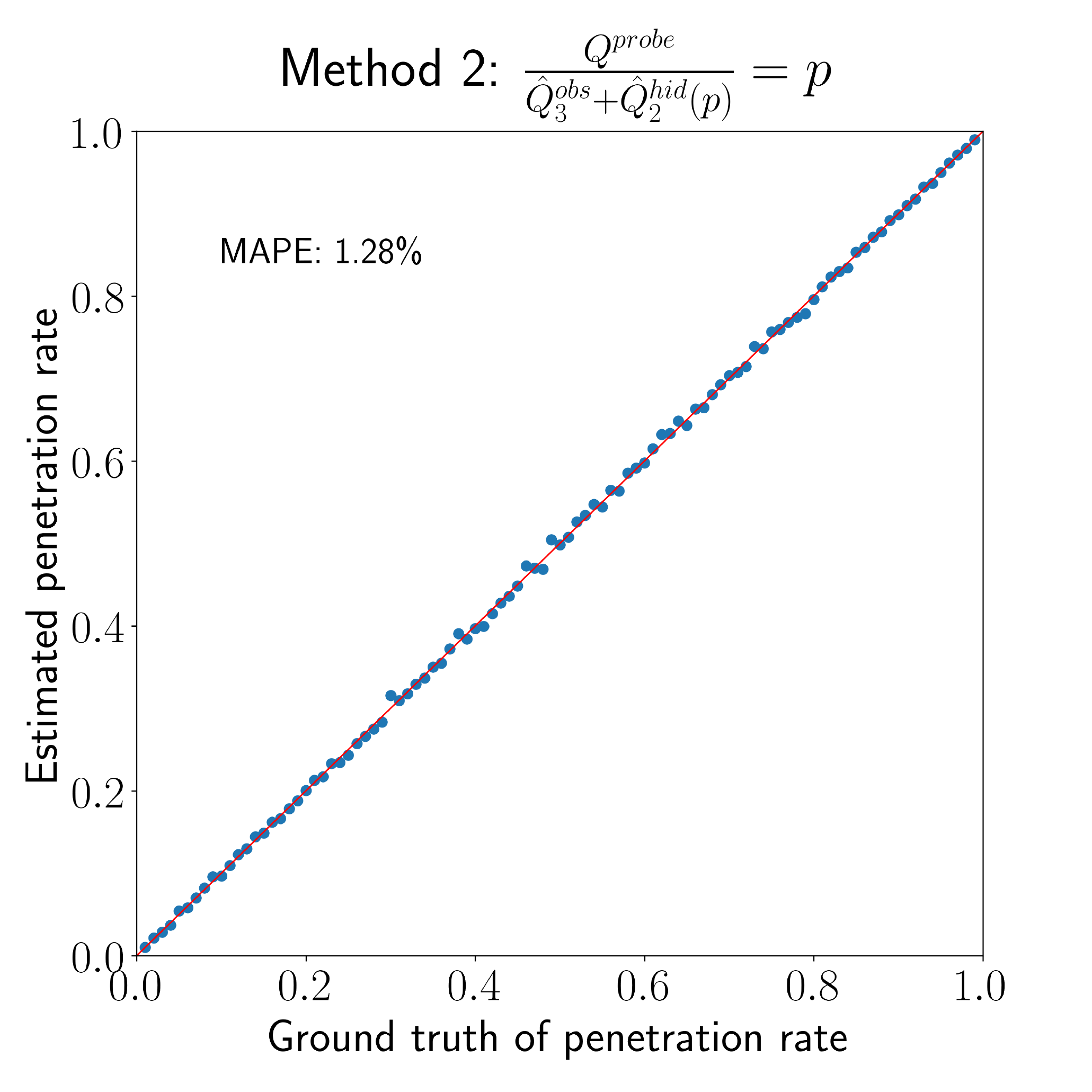}
\par\end{centering}
}
\par\end{centering}
\caption{\label{fig: pr_results}The results of penetration rate estimation
using different methods }
\end{figure}

\begin{figure}[H]
\begin{centering}
\subfloat[]{\begin{centering}
\includegraphics[width=0.4\textwidth]{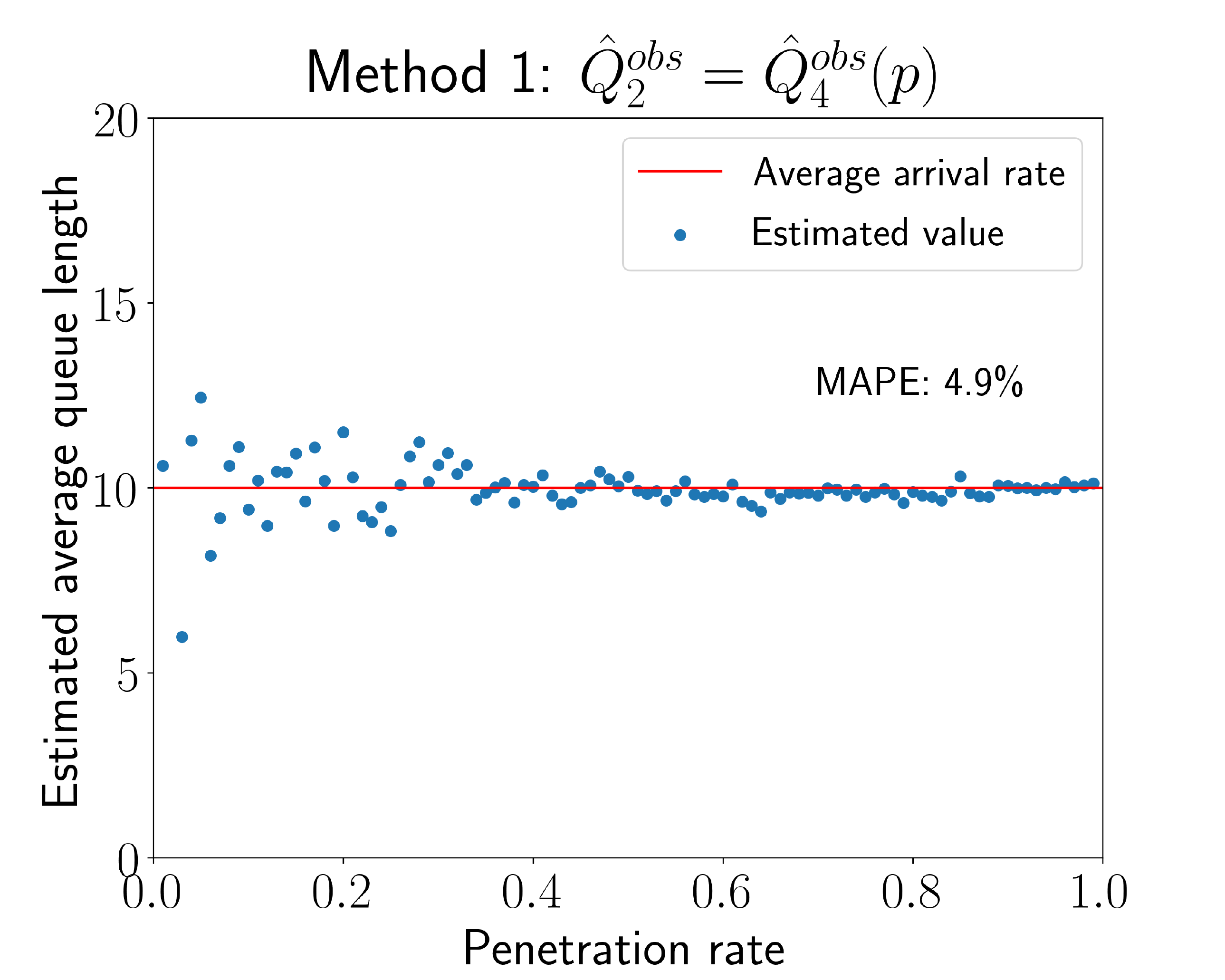}
\par\end{centering}
}\subfloat[]{\begin{centering}
\includegraphics[width=0.4\textwidth]{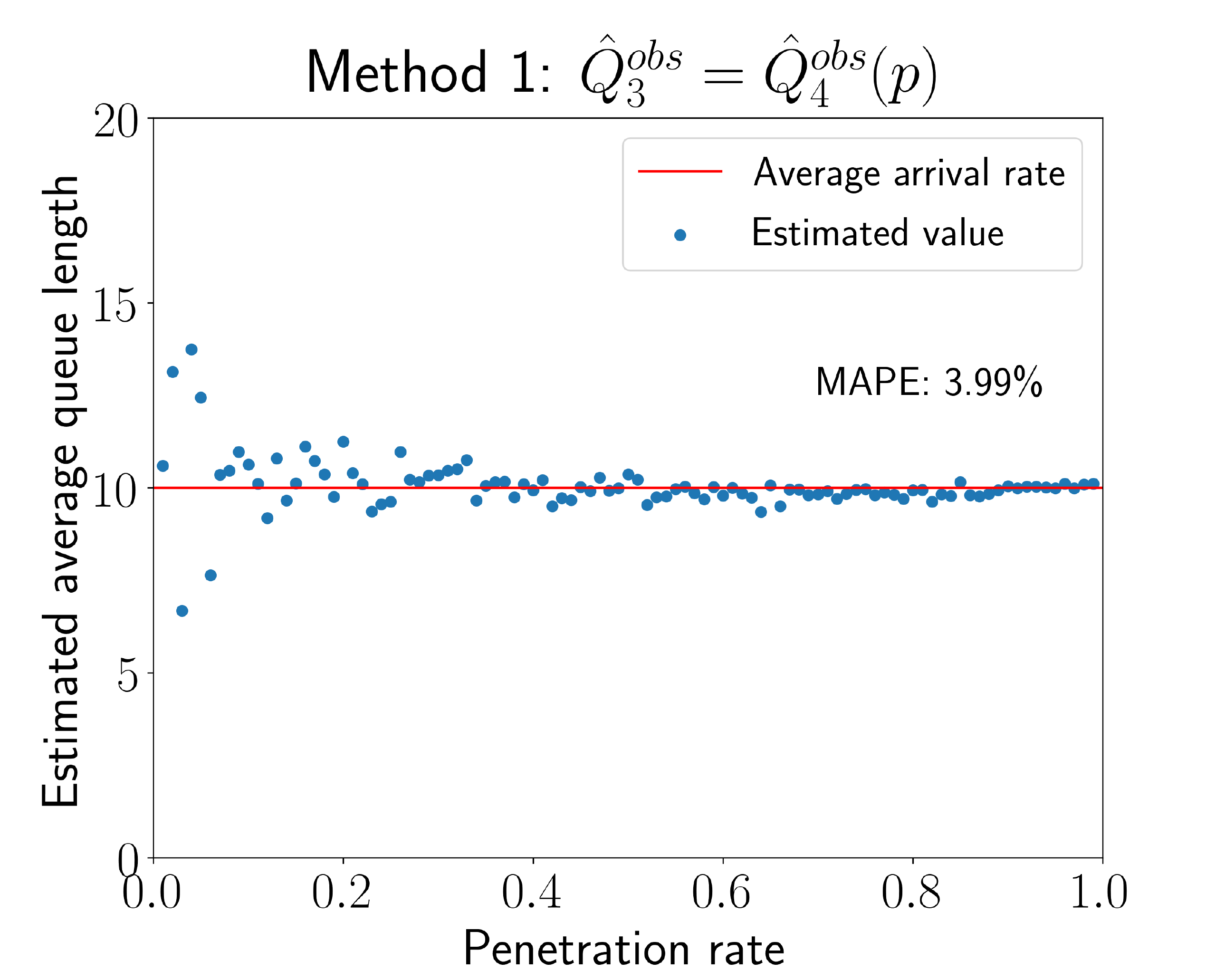}
\par\end{centering}
}
\par\end{centering}
\begin{centering}
\subfloat[]{\begin{centering}
\includegraphics[width=0.4\textwidth]{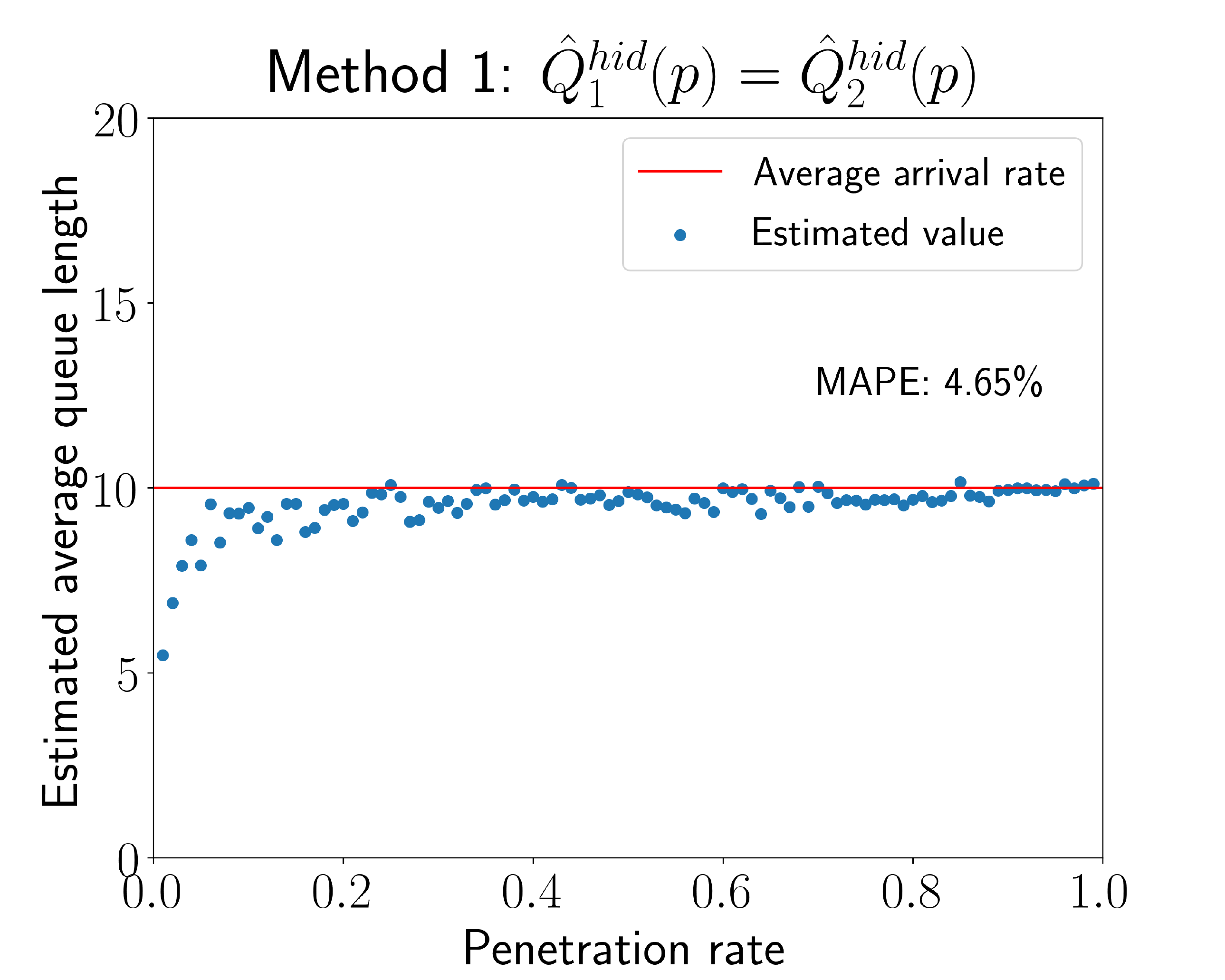}
\par\end{centering}
}\subfloat[]{\begin{centering}
\includegraphics[width=0.4\textwidth]{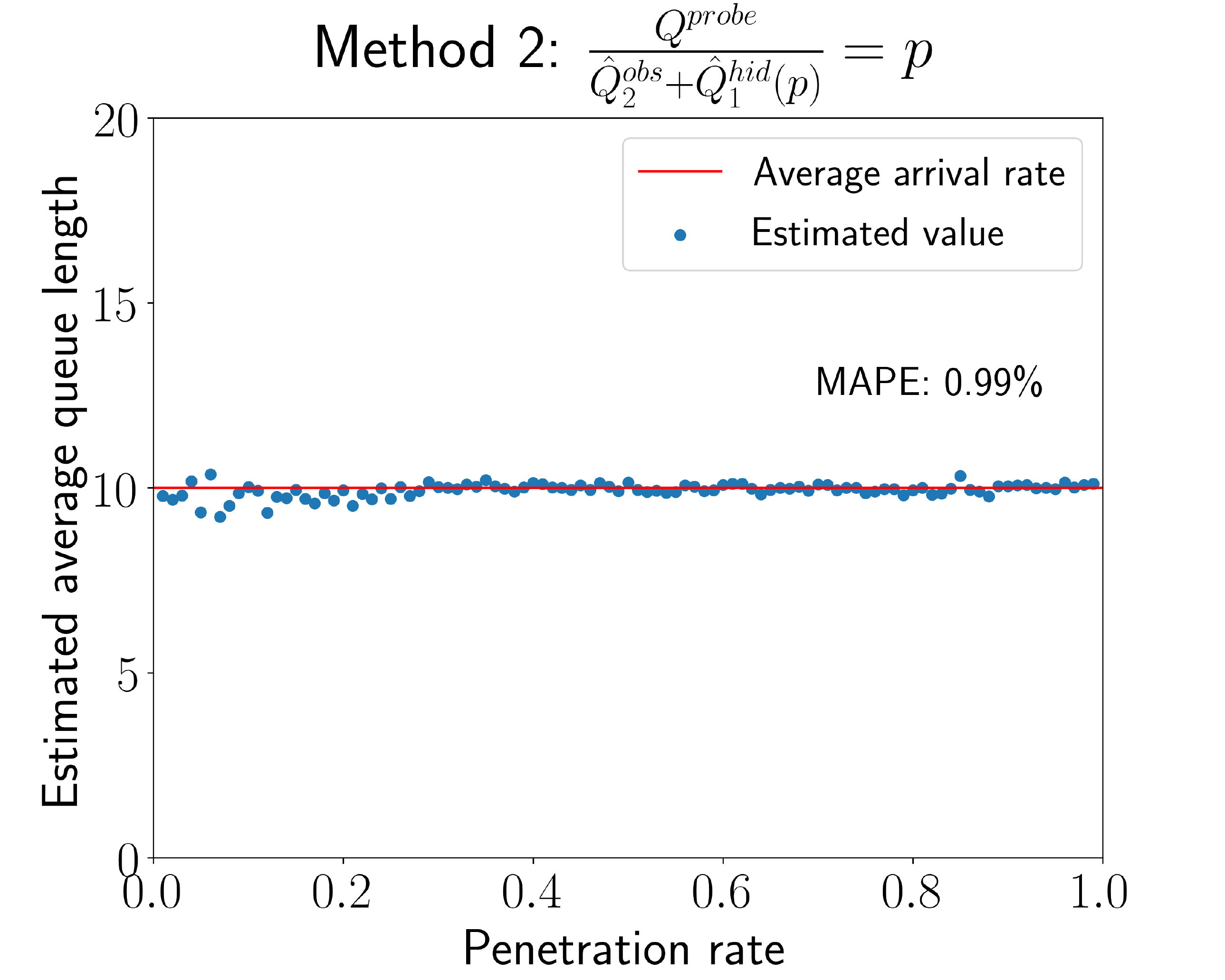}
\par\end{centering}
}
\par\end{centering}
\begin{centering}
\subfloat[]{\begin{centering}
\includegraphics[width=0.4\textwidth]{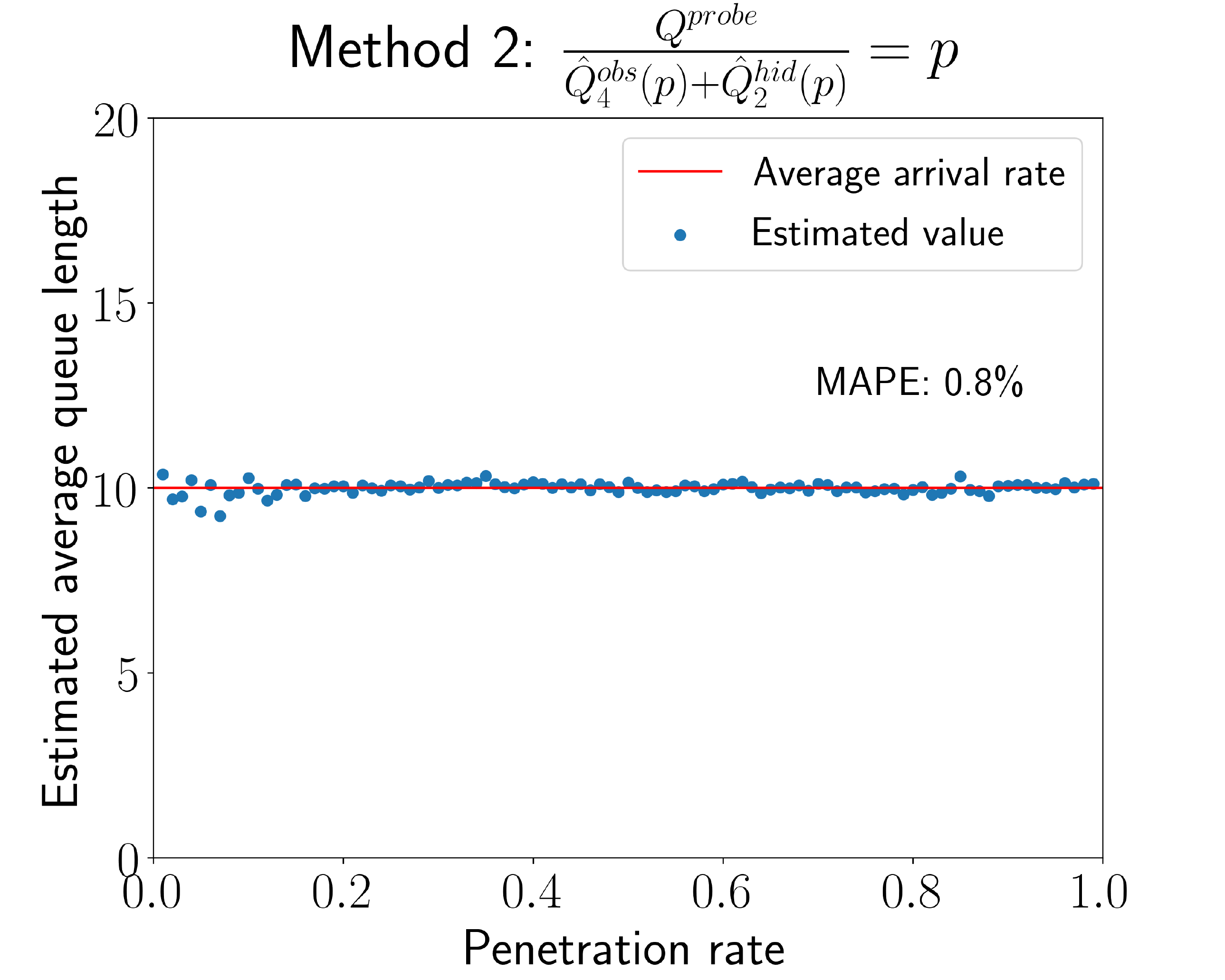}
\par\end{centering}
}\subfloat[]{\begin{centering}
\includegraphics[width=0.4\textwidth]{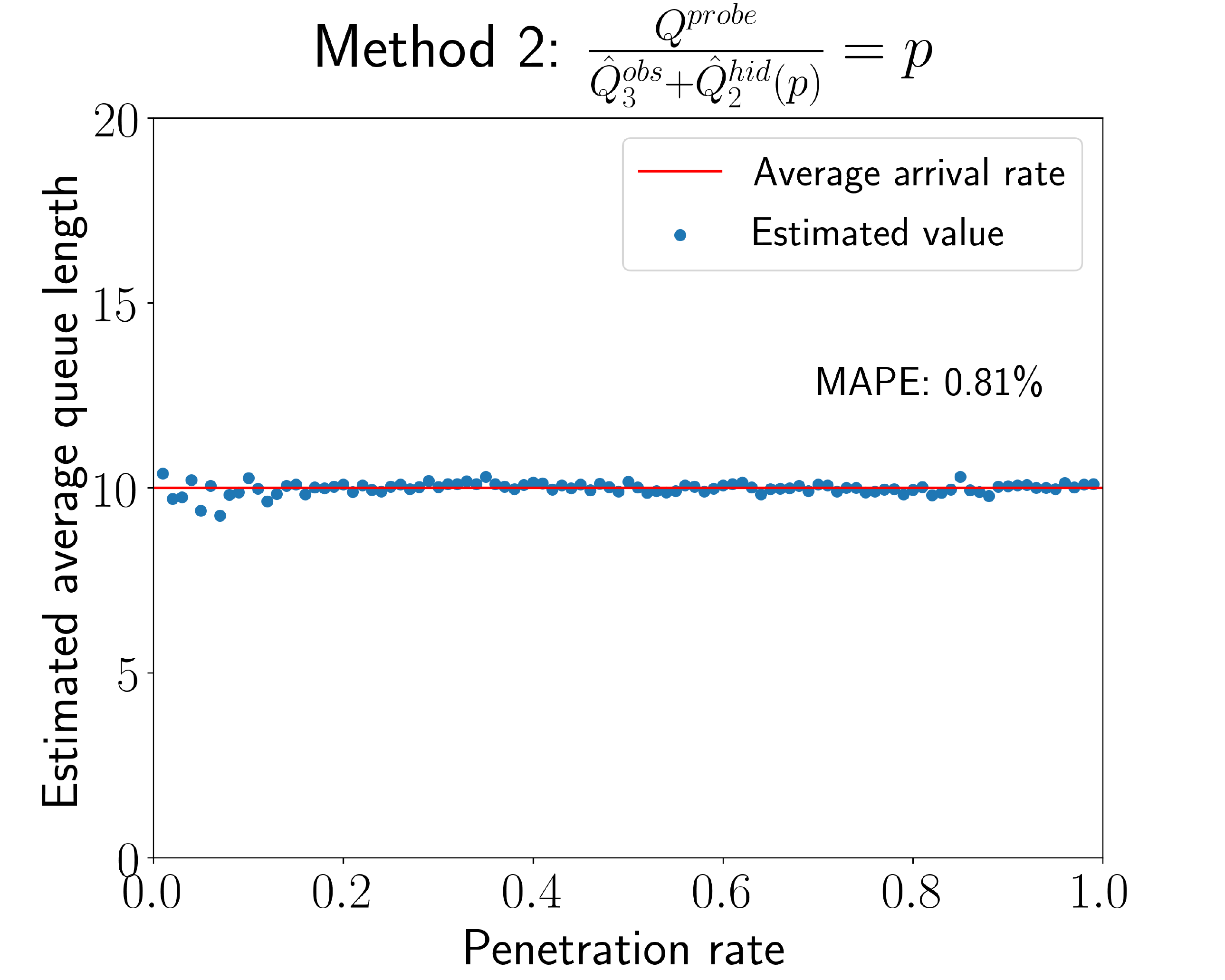}
\par\end{centering}
}
\par\end{centering}
\caption{\label{fig: q_results}The results of queue length estimation using
different methods }
\end{figure}

Figure \ref{fig: pr_results} shows the results of penetration rate
estimation using six different submethods introduced in Section \ref{sec: pr}.
The simulation data are generated by a Poisson process with an average
arrival rate during the red phase $\lambda=10$ for 1,000 cycles.
The horizontal axes represent the ground truth of the penetration
rates. The vertical axes represent the estimated values. The used
measure of the estimation accuracy is the mean absolute percentage
error (MAPE). As Figure \ref{fig: pr_results} shows, the dots in
blue are very close to the diagonals, which implies that the methods
can estimate the penetration rate very accurately. Figure \ref{fig: q_results}
shows the results of queue length estimation using the different submethods.
The horizontal axes represent the penetration rates, and the vertical
axes represent the estimated average queue lengths. The results show
that the higher the penetration rate is, the better the estimation
results tend to be. It is intuitive because when the penetration rate
is very low, only a tiny portion of vehicles can be observed. By contrast,
if the penetration rate is very close to 100\%, there will be little
missing information and the estimation results would be more accurate.

In general, method 2 outperforms method 1. To better understand the
mechanism behind method 2, define an inverse proportional function
\[
f(x)=\frac{M}{x},
\]
where $M$ is a positive constant. When $x\gg\sqrt{M}$, the absolute
value of the derivative is $\left|f^{\prime}(x)\right|=\frac{M}{x^{2}}\ll1$.
In method 2, as equation (\ref{eq: est_p_4}) shows, the denominator
of the left-hand side is $\hat{Q}_{i}^{obs}(p)+\hat{Q}_{j}^{hid}(p)$,
which is much larger than $\sqrt{Q^{probe}}$. Therefore, due to the
property of the inverse proportional function, the error in $\hat{Q}_{i}^{obs}(p)+\hat{Q}_{j}^{hid}(p)$
only results in an error of $p$ which is orders of magnitude smaller.
That is why method 2 generally outperforms method 1. 

Among the estimators of $Q^{obs}$, $\hat{Q}_{1}^{obs}$ scales up
the stopping positions of the first probe vehicle by a relatively
large scaling factor $(n_{i}+1)$ in each cycle, and thus usually
results in large variances when estimating $Q^{obs}$. $\hat{Q}_{2}^{obs}$
scales up the stopping positions of the last probe vehicle with a
relatively smaller scaling factor $\frac{n_{i}+1}{n_{i}}$, which
results in smaller variances than $\hat{Q}_{1}^{obs}$. $\hat{Q}_{3}^{obs}$
estimates $Q^{obs}$ by summing up the stopping positions of the first
and the last probe vehicles in each cycle. Since there is no scaling
up factor, the estimation accuracy is even better. $\hat{Q}_{4}^{obs}$
is a function of the penetration rate $p$. The queue length distribution
required in the calculation is approximated by aggregating the stopping
positions of all the probe vehicles. The performance of $\hat{Q}_{4}^{obs}$
is similar to $\hat{Q}_{3}^{obs}$. As for the estimators of $Q^{hid}$,
$\hat{Q}_{2}^{hid}$ generally has an edge over $\hat{Q}_{1}^{hid}$,
as it usually gives better results than $\hat{Q}_{1}^{hid}$. In addition,
$\hat{Q}_{1}^{hid}$ requires the signal timing information such as
the number of cycles which is not necessarily needed by $\hat{Q}_{2}^{hid}$.

\subsubsection{The effect of sample size}

\begin{figure}[H]
\begin{centering}
\subfloat[]{\begin{centering}
\includegraphics[width=0.4\textwidth]{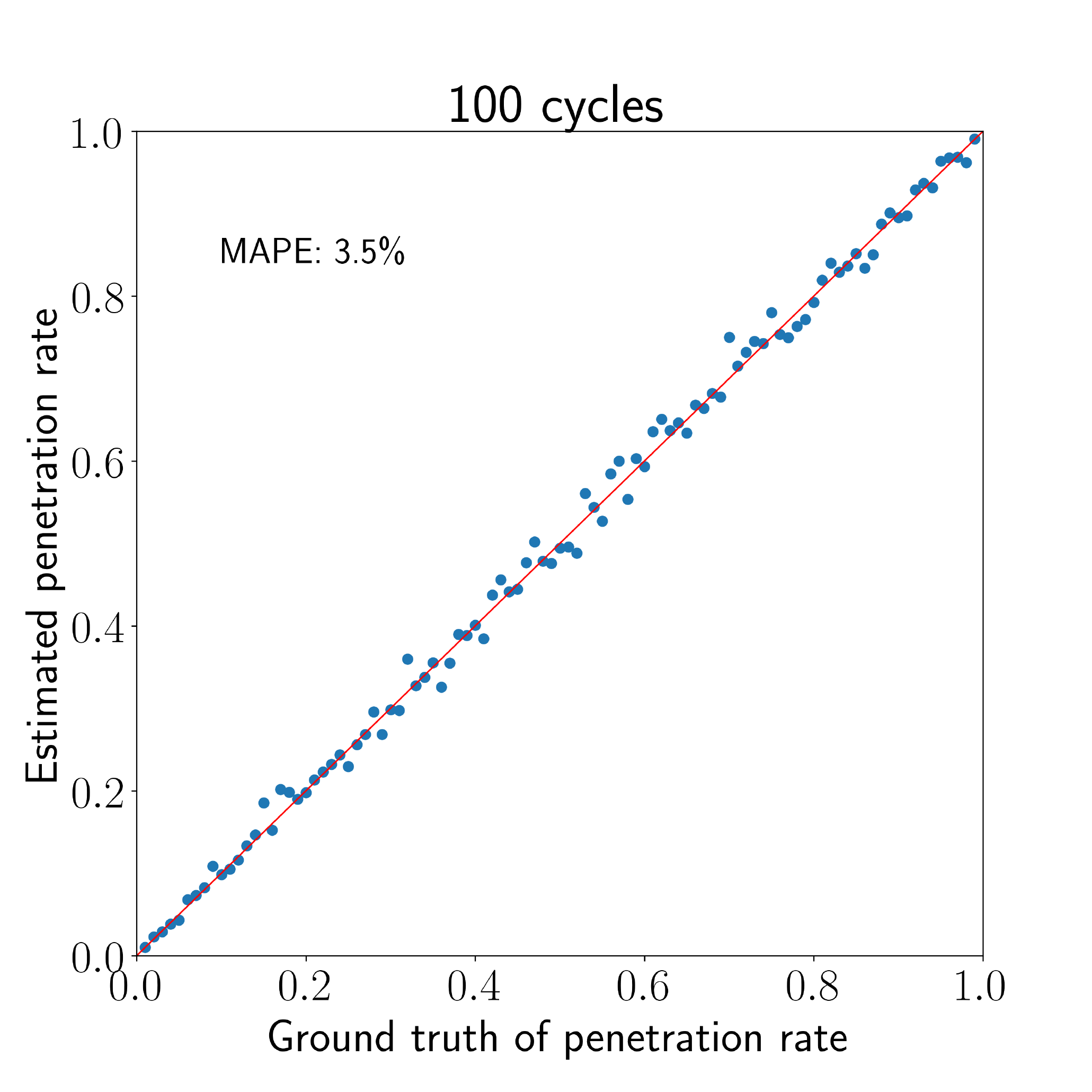}
\par\end{centering}
}\subfloat[]{\begin{centering}
\includegraphics[width=0.4\textwidth]{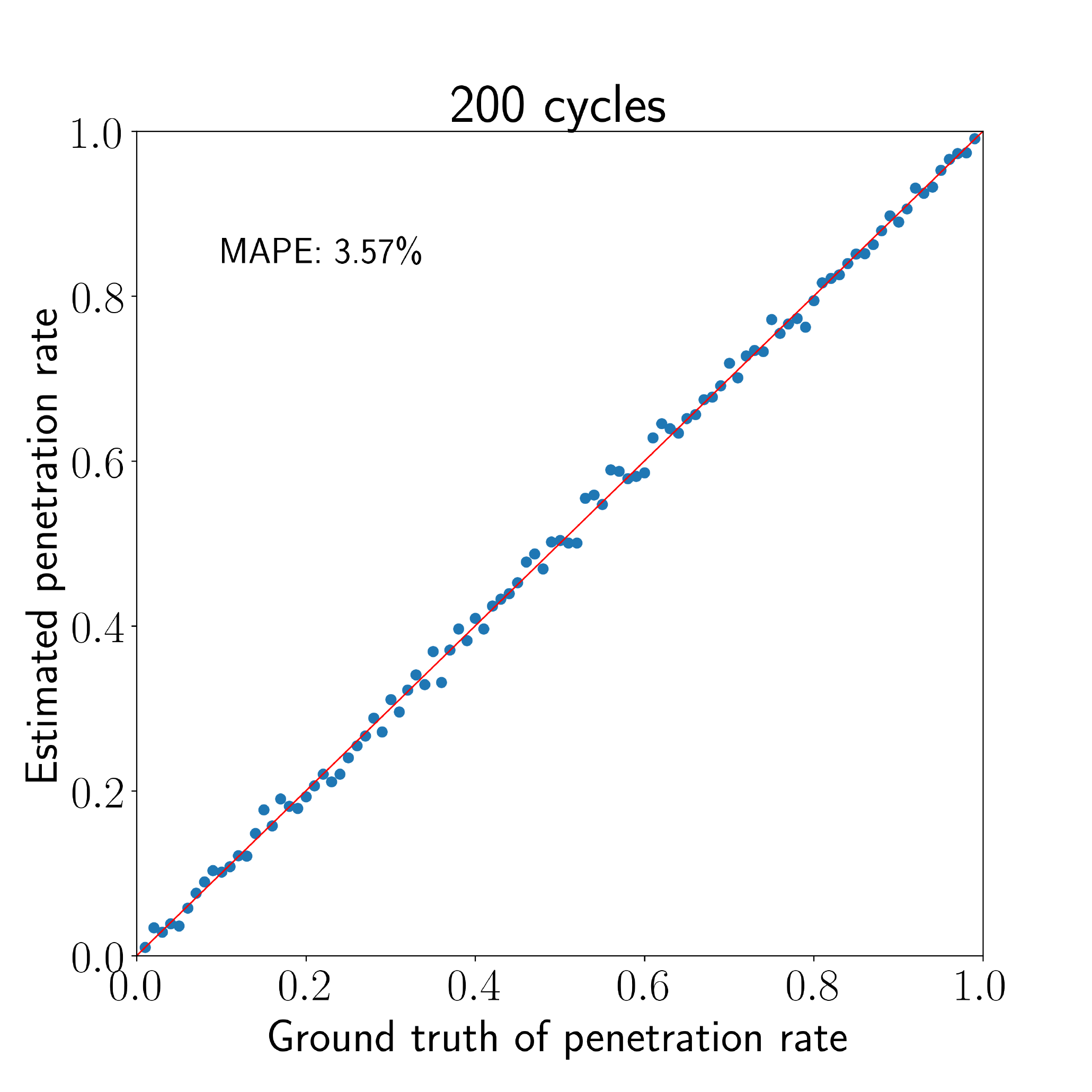}
\par\end{centering}
}
\par\end{centering}
\begin{centering}
\subfloat[]{\begin{centering}
\includegraphics[width=0.4\textwidth]{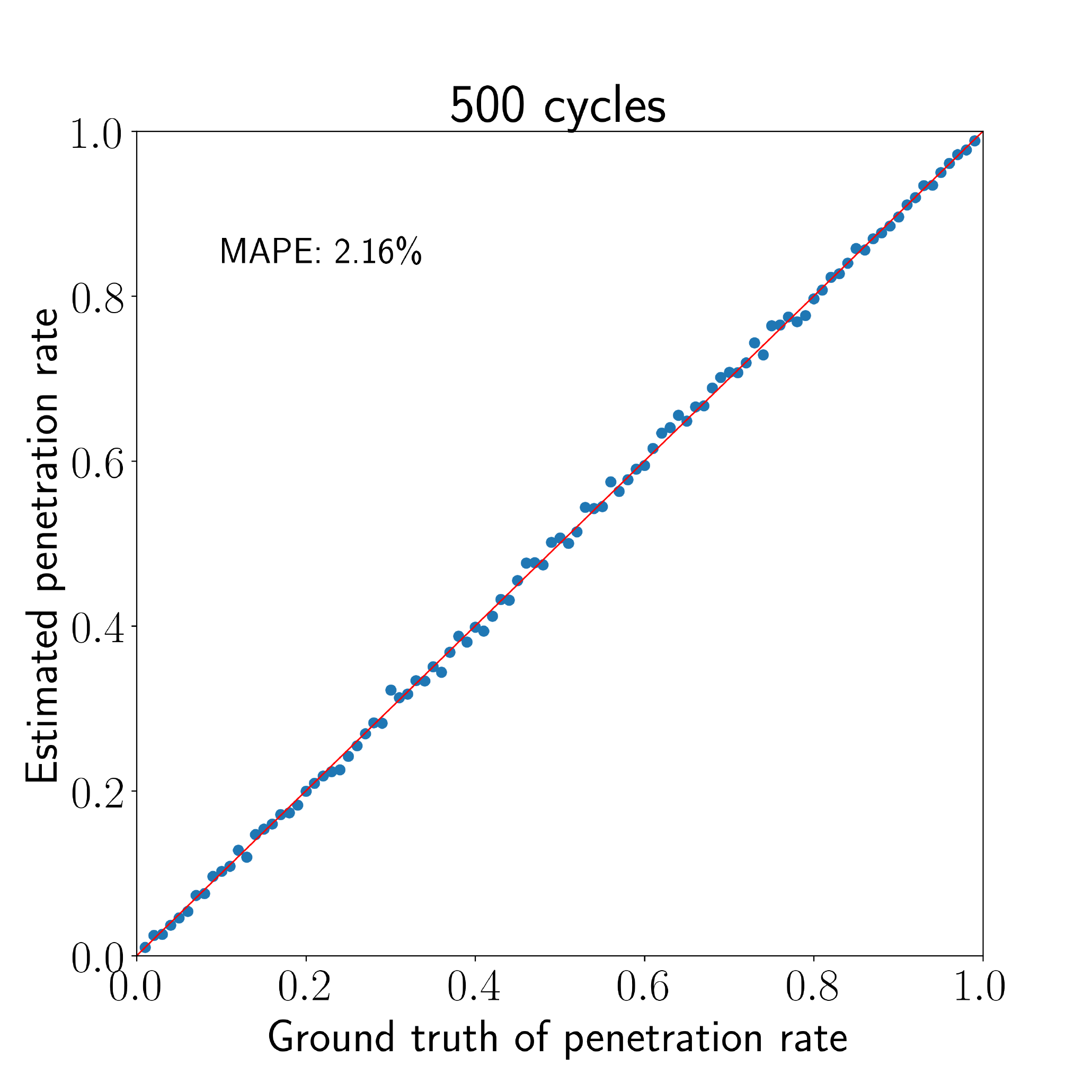}
\par\end{centering}
}\subfloat[]{\begin{centering}
\includegraphics[width=0.4\textwidth]{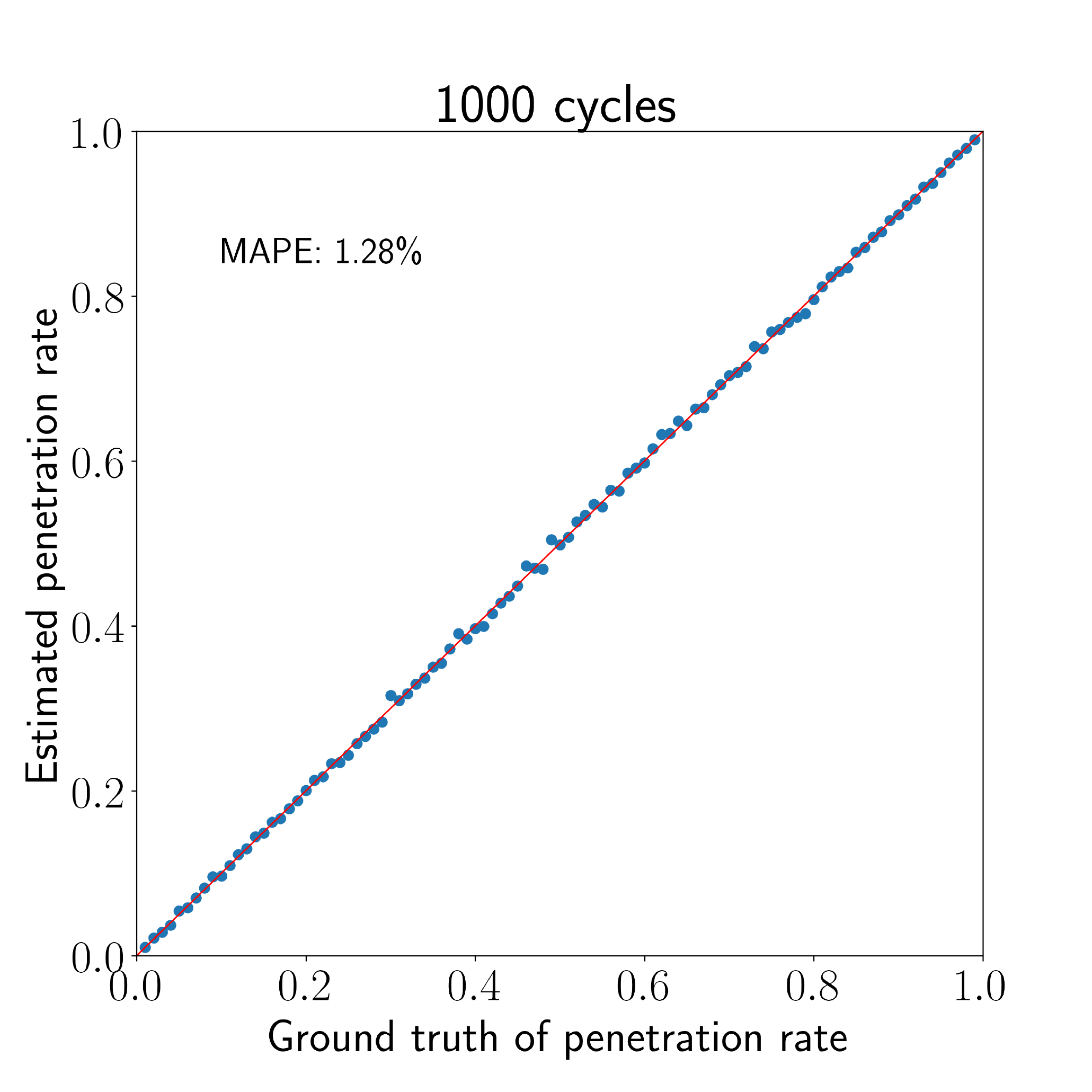}
\par\end{centering}
}
\par\end{centering}
\caption{\label{fig: pr_results_sample_size}The results of penetration rate
estimation with different sample sizes }
\end{figure}

\begin{figure}[H]
\begin{centering}
\subfloat[]{\begin{centering}
\includegraphics[width=0.4\textwidth]{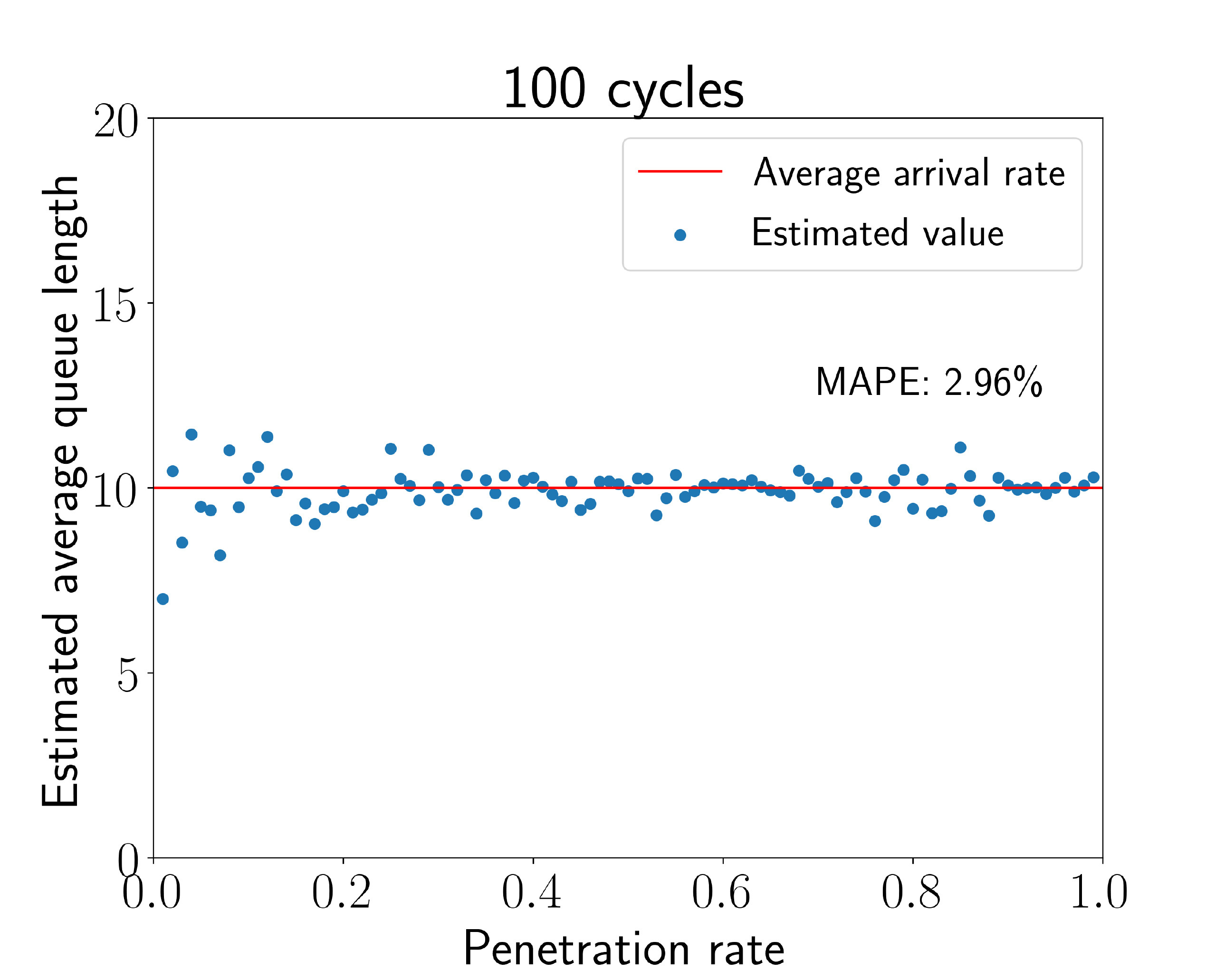}
\par\end{centering}
}\subfloat[]{\begin{centering}
\includegraphics[width=0.4\textwidth]{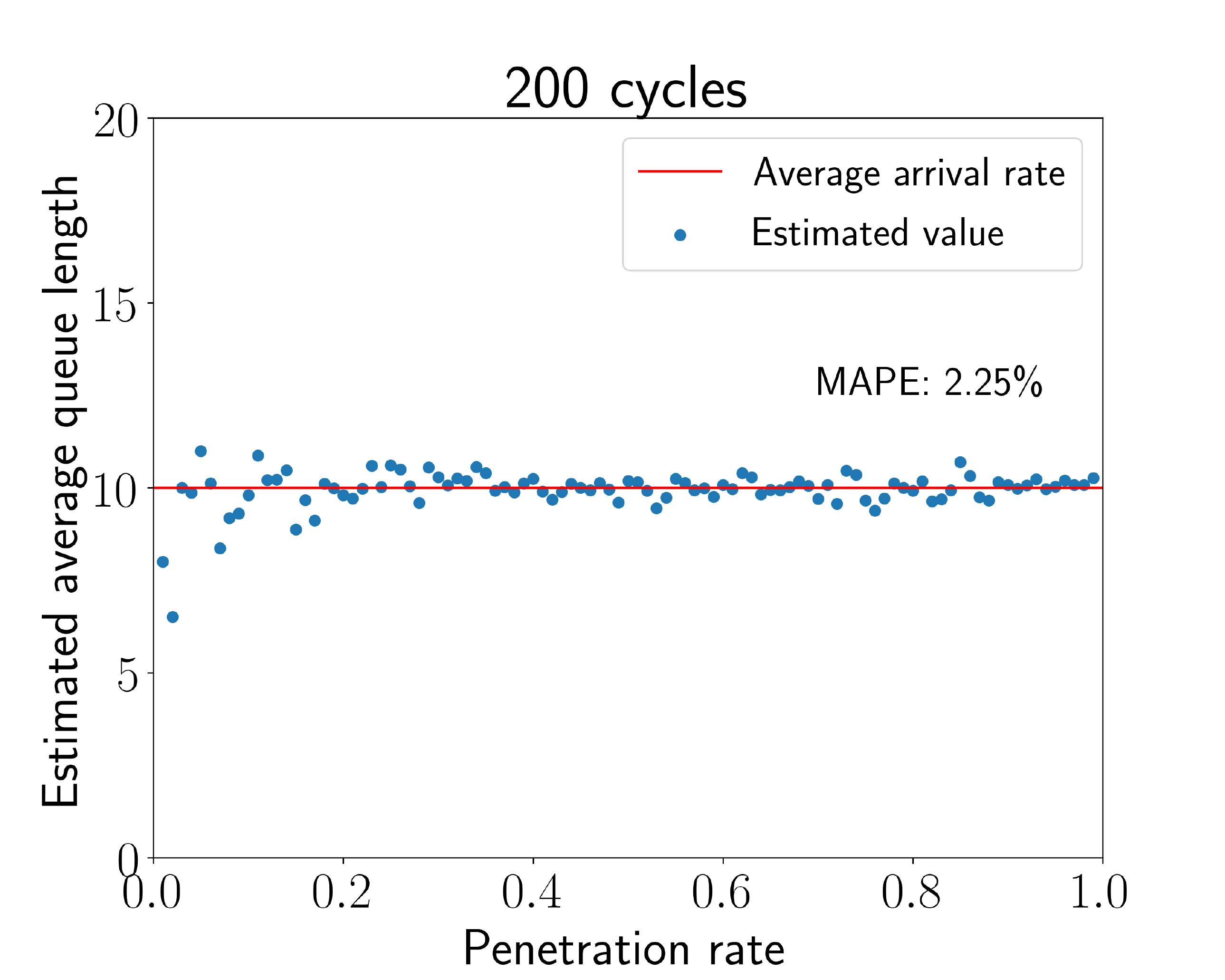}
\par\end{centering}
}
\par\end{centering}
\begin{centering}
\subfloat[]{\begin{centering}
\includegraphics[width=0.4\textwidth]{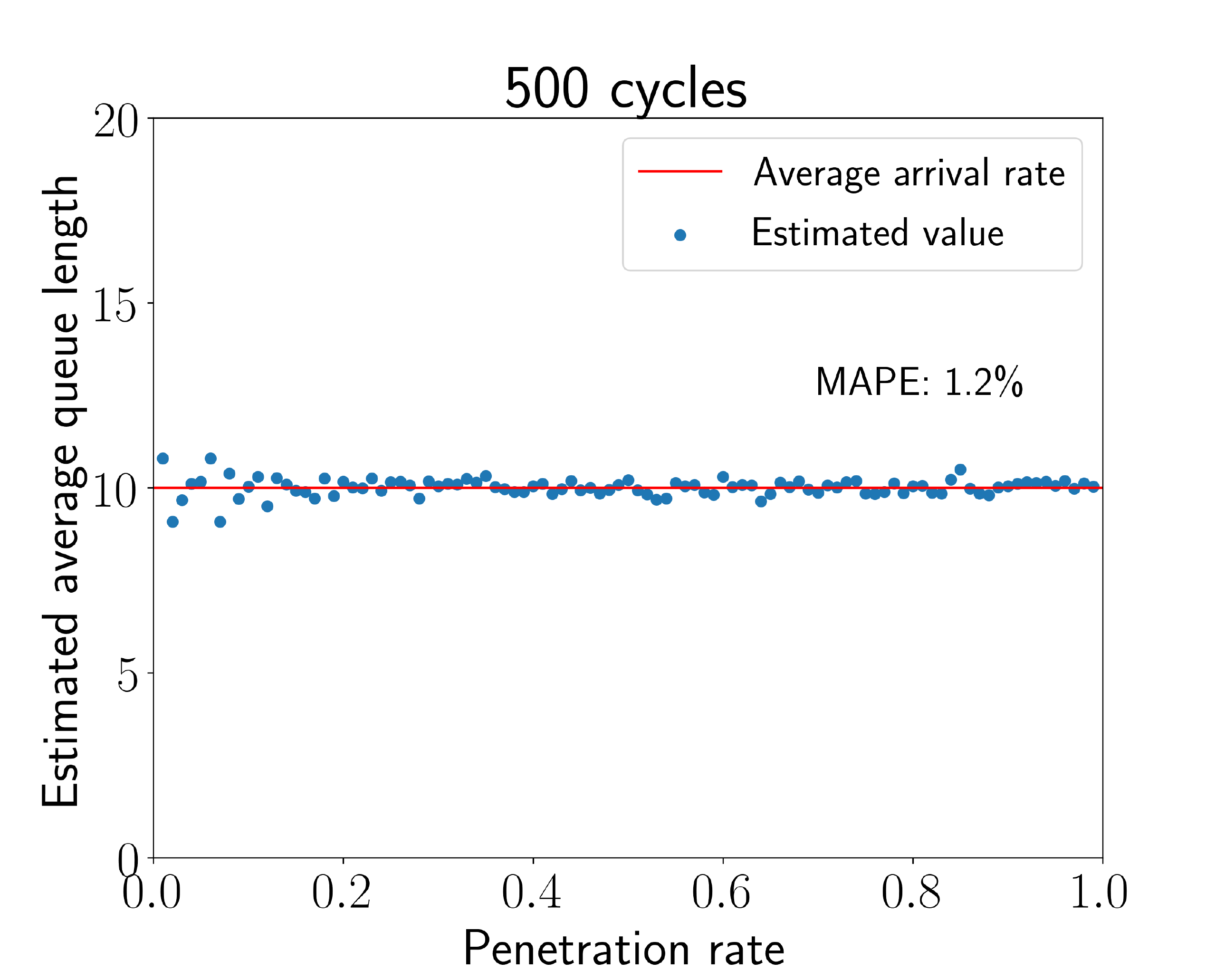}
\par\end{centering}
}\subfloat[]{\begin{centering}
\includegraphics[width=0.4\textwidth]{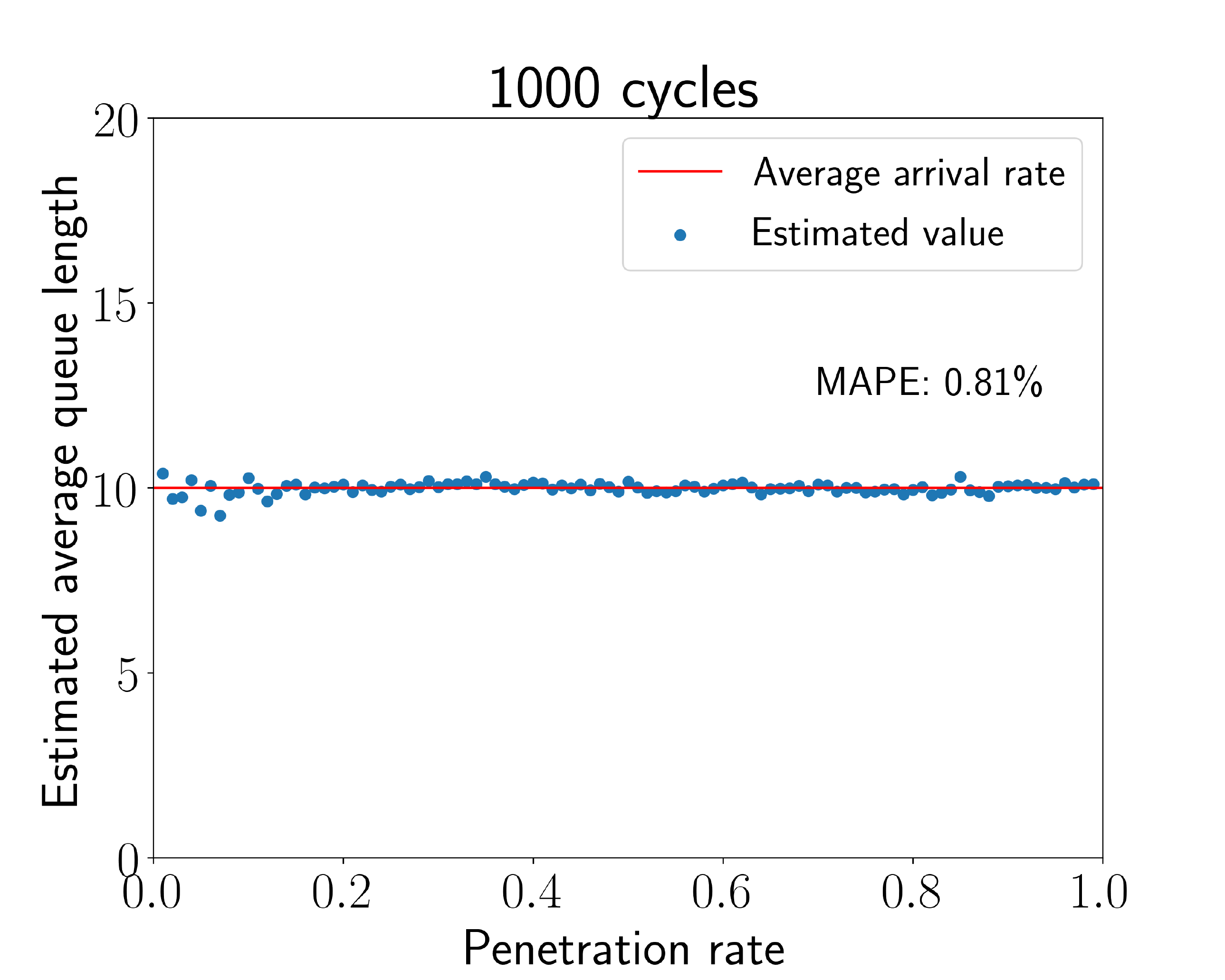}
\par\end{centering}
}
\par\end{centering}
\caption{\label{fig: q_results_sample_size}The results of queue length estimation
with different sample sizes }
\end{figure}

In order to demonstrate the impact of sample size on the estimation
accuracy, the data of 100 cycles, 200 cycles, 500 cycles, and 1,000
cycles are used in four rounds of tests, respectively. The submethod
$\frac{Q^{probe}}{\hat{Q}_{3}^{obs}(p)+\hat{Q}_{2}^{hid}(p)}=p$ is
applied. The results in Figure \ref{fig: pr_results_sample_size}
and Figure \ref{fig: q_results_sample_size} show that better results
can be obtained when the sample size is larger.

\subsubsection{The effect of the arrival rate during the red phase}

\begin{figure}[H]
\begin{centering}
\subfloat[]{\begin{centering}
\includegraphics[width=0.4\textwidth]{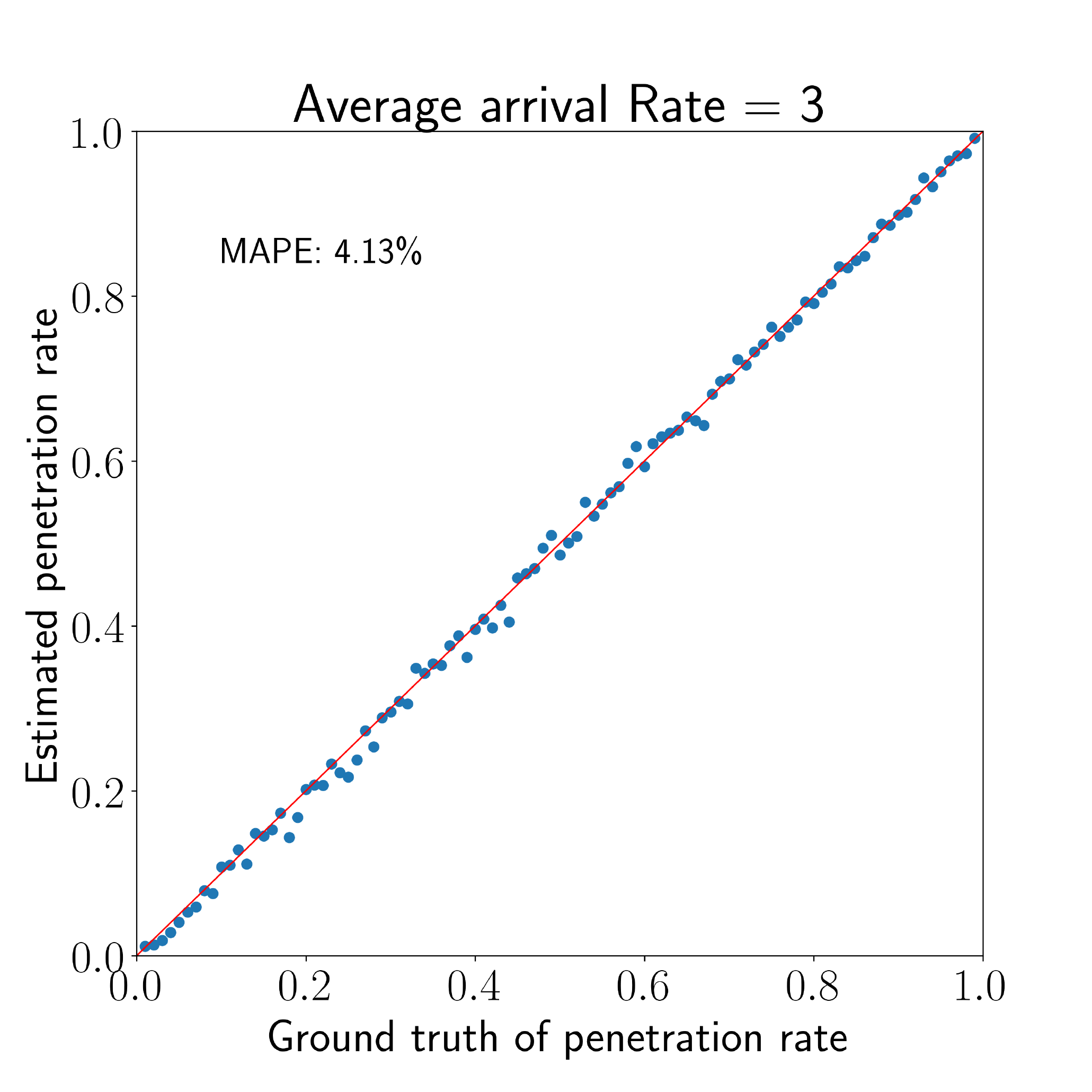}
\par\end{centering}
}\subfloat[]{\begin{centering}
\includegraphics[width=0.4\textwidth]{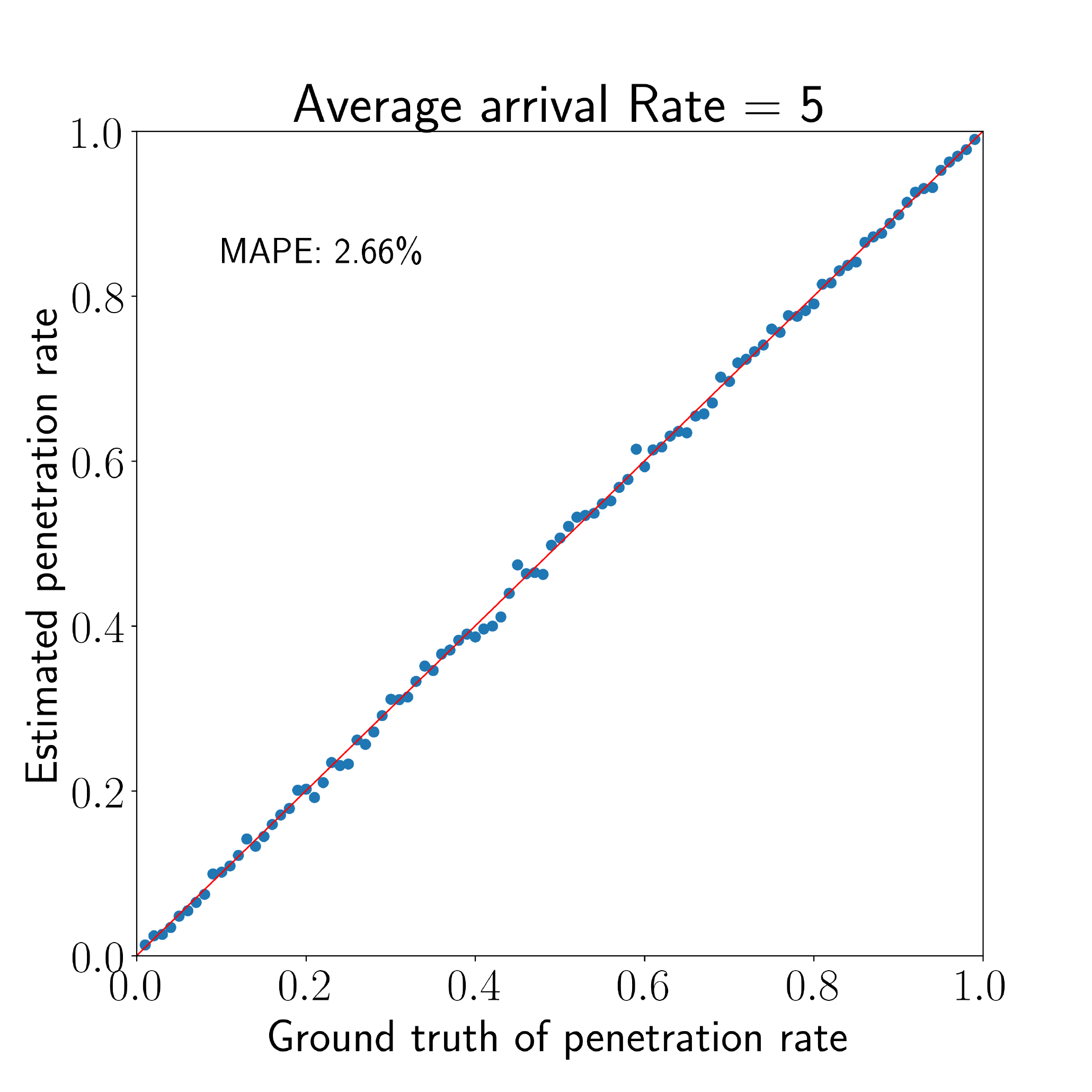}
\par\end{centering}
}
\par\end{centering}
\begin{centering}
\subfloat[]{\begin{centering}
\includegraphics[width=0.4\textwidth]{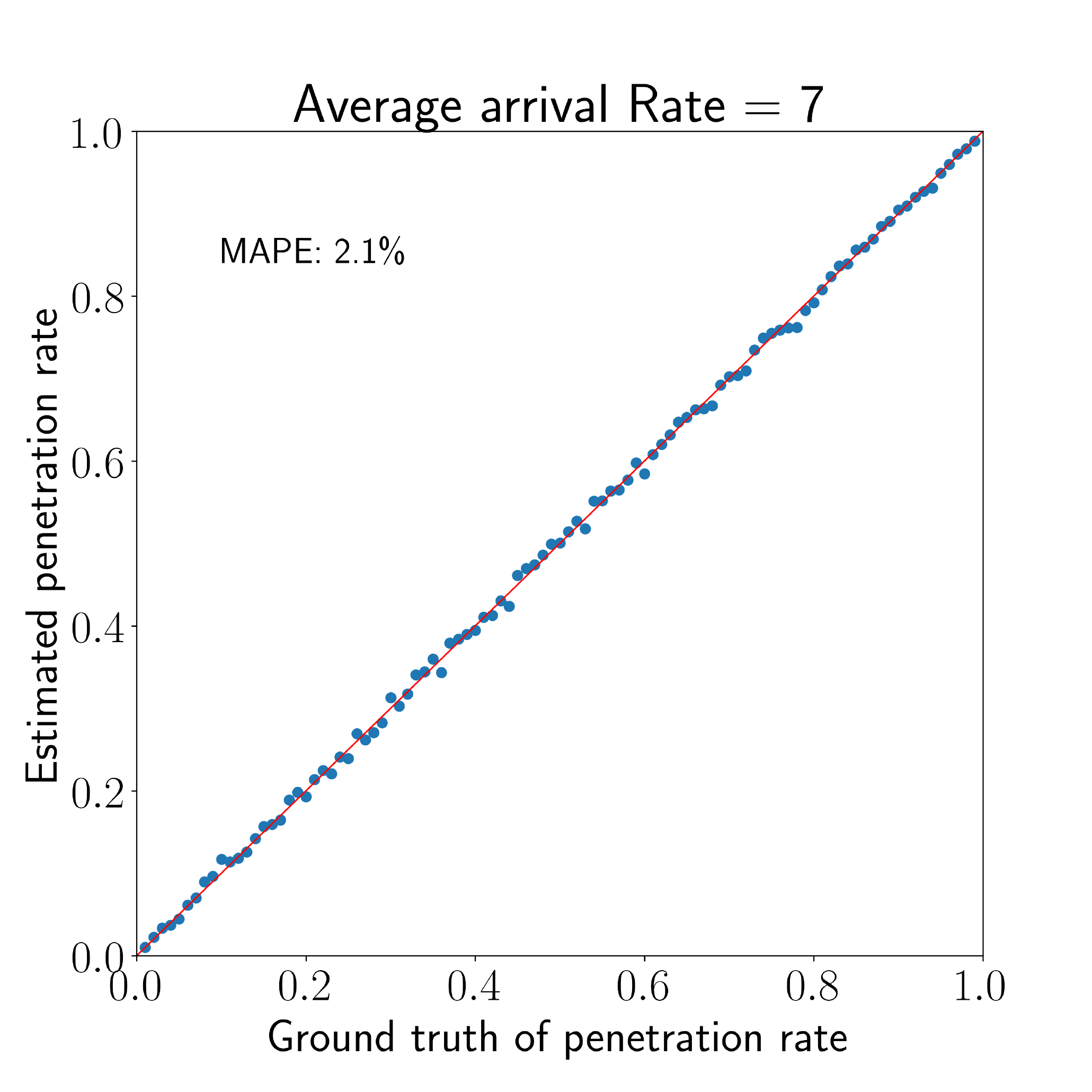}
\par\end{centering}
}\subfloat[]{\begin{centering}
\includegraphics[width=0.4\textwidth]{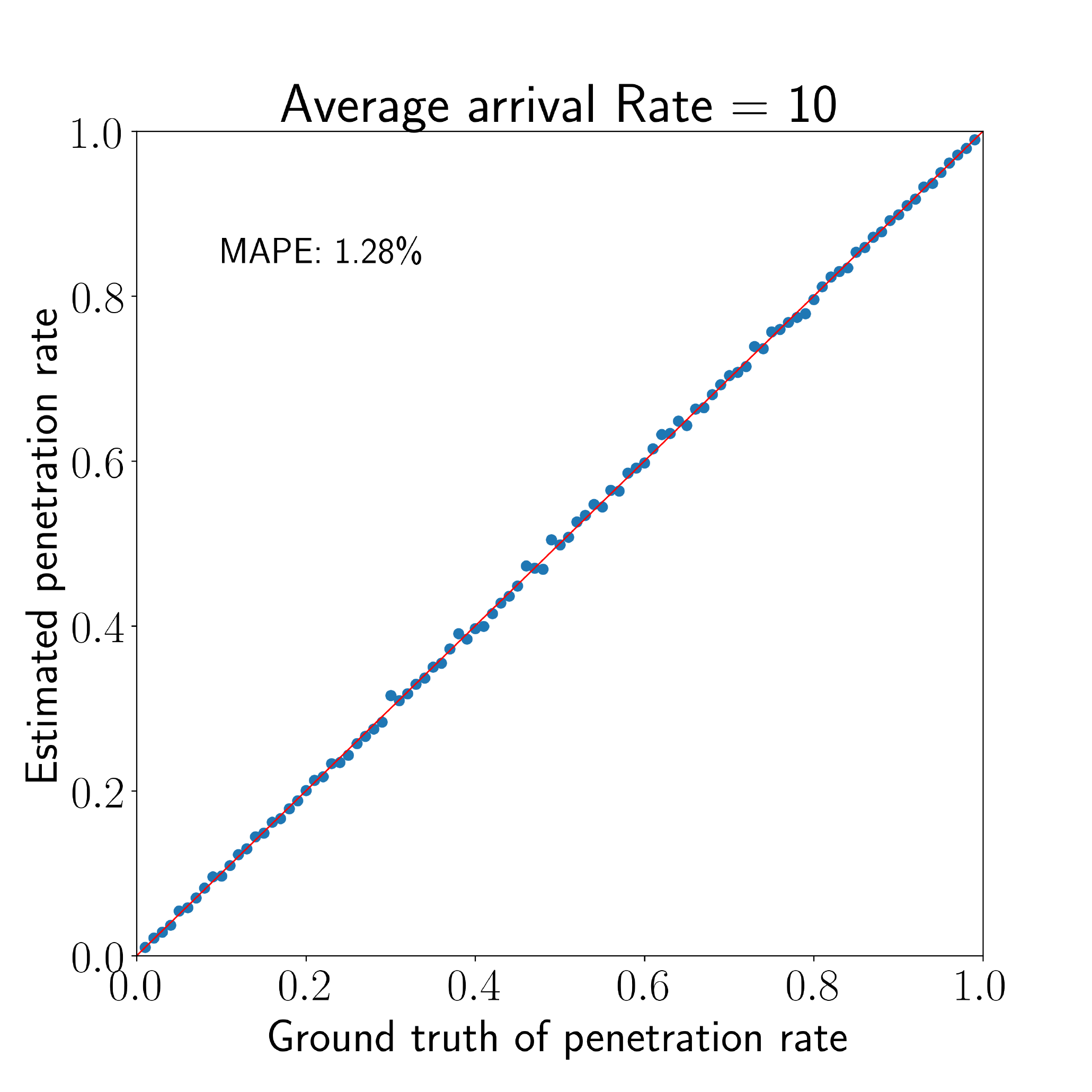}
\par\end{centering}
}
\par\end{centering}
\caption{\label{fig: pr_results_arrival_rate}The results of penetration rate
estimation with different arrival rates}
\end{figure}

\begin{figure}[H]
\begin{centering}
\subfloat[]{\begin{centering}
\includegraphics[width=0.4\textwidth]{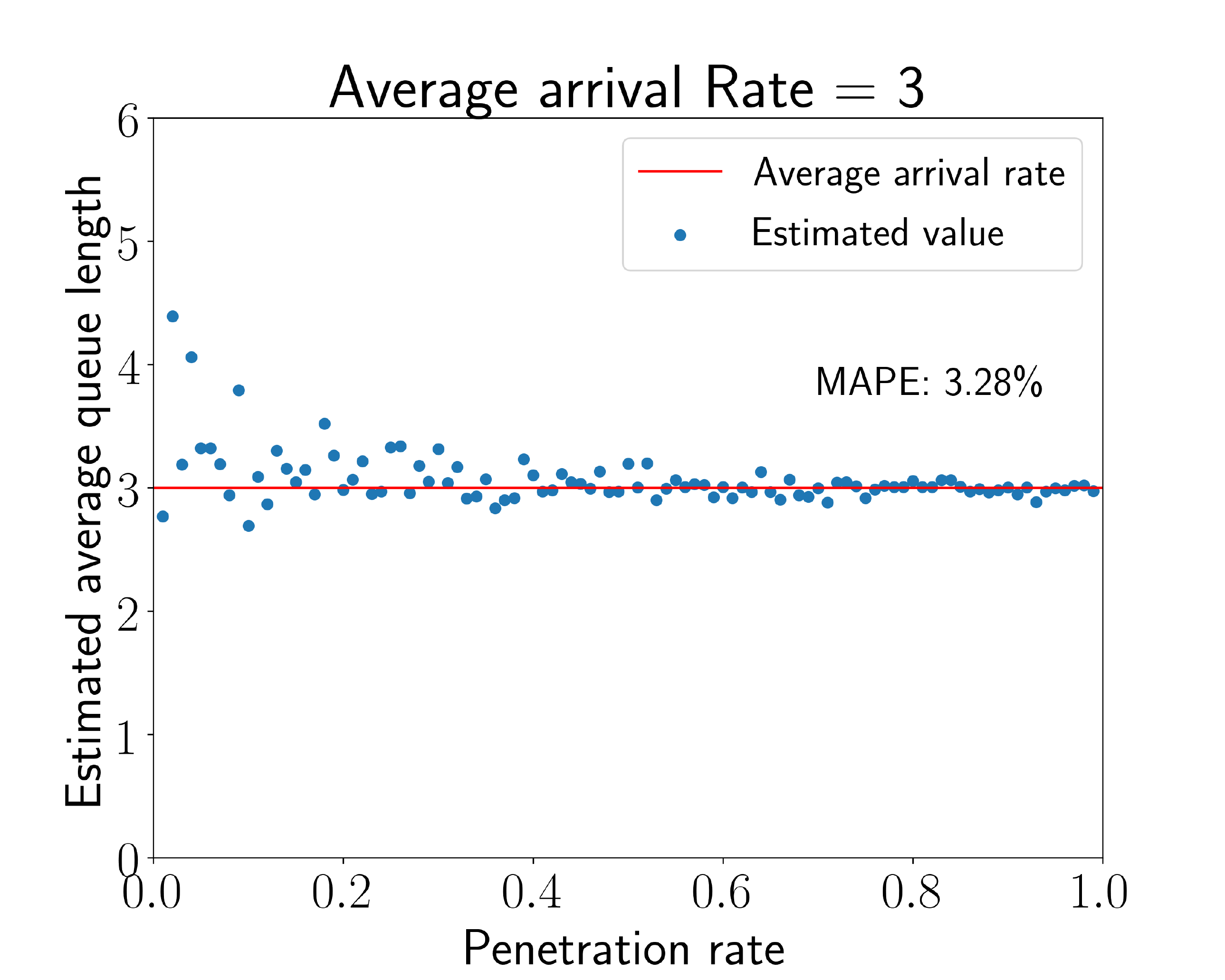}
\par\end{centering}
}\subfloat[]{\begin{centering}
\includegraphics[width=0.4\textwidth]{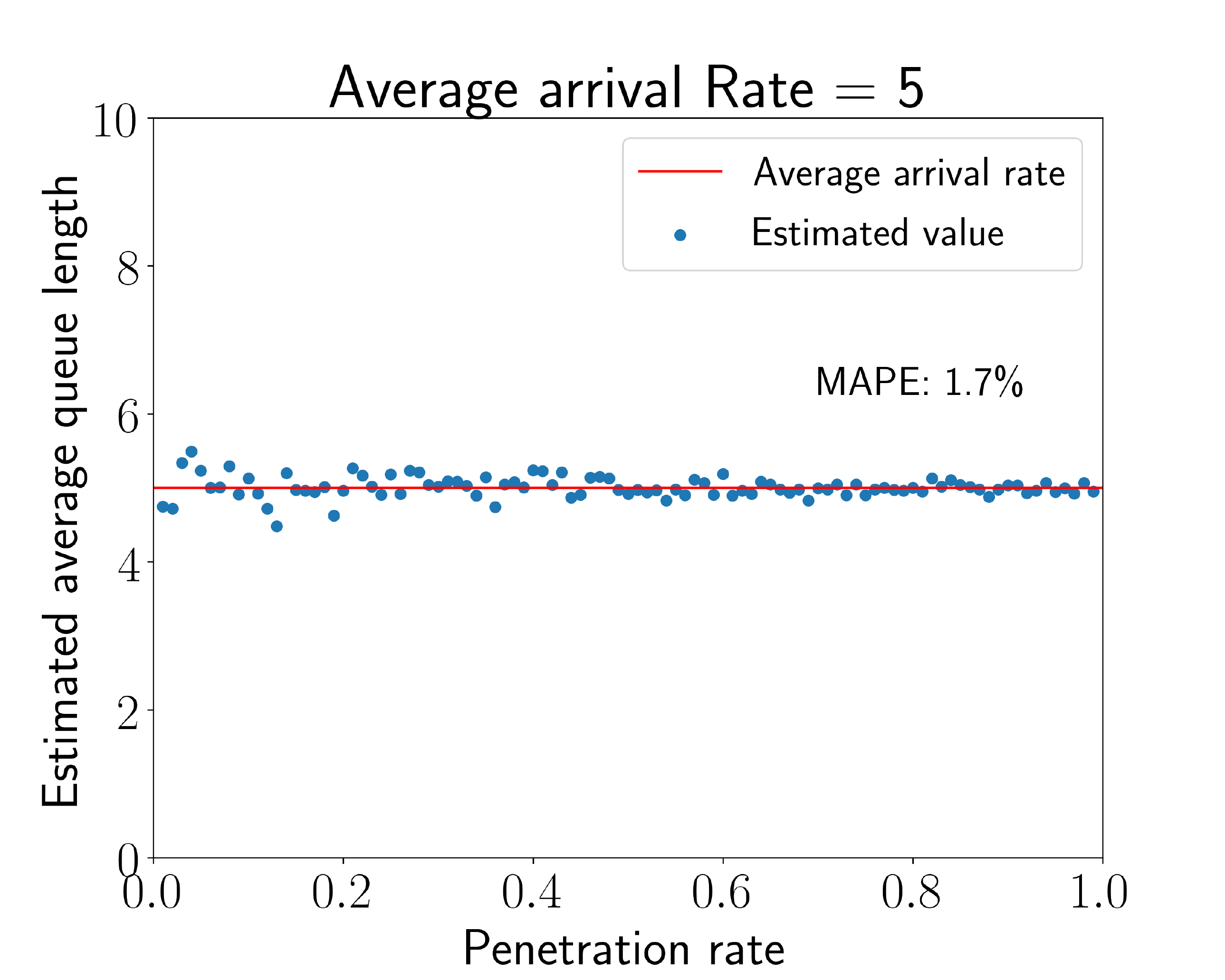}
\par\end{centering}
}
\par\end{centering}
\begin{centering}
\subfloat[]{\begin{centering}
\includegraphics[width=0.4\textwidth]{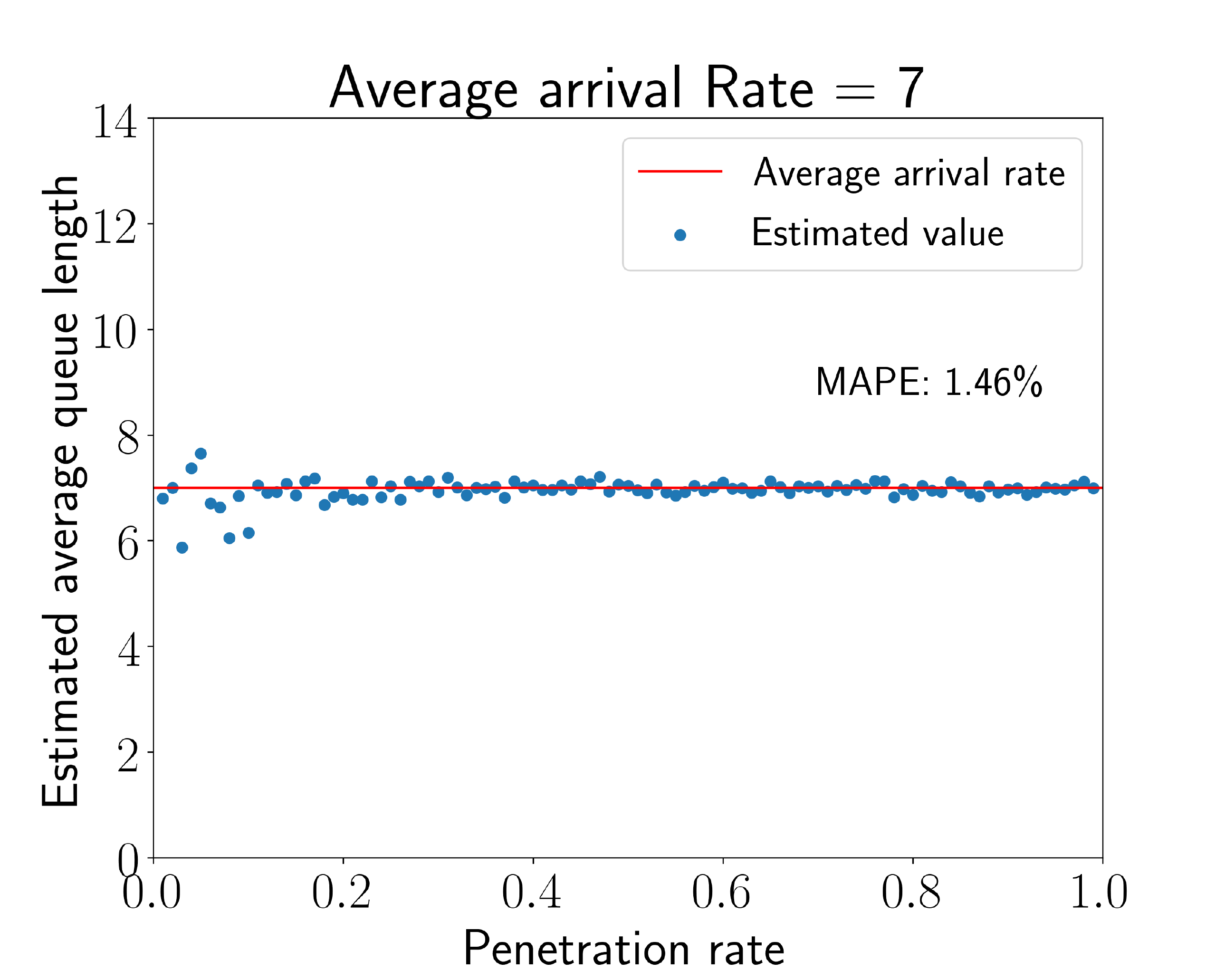}
\par\end{centering}
}\subfloat[]{\begin{centering}
\includegraphics[width=0.4\textwidth]{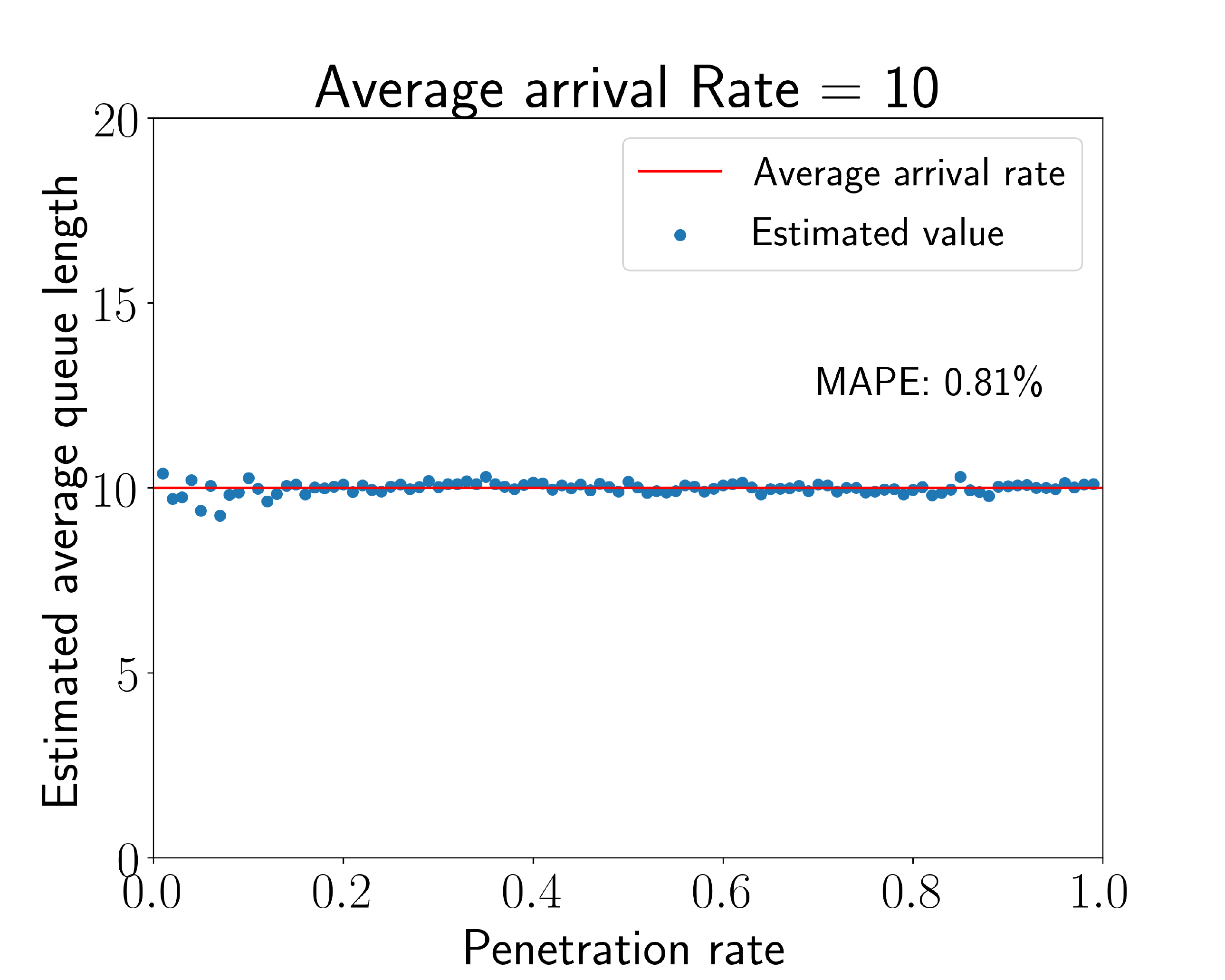}
\par\end{centering}
}
\par\end{centering}
\caption{\label{fig: q_results_arrival_rate}The results of queue length estimation
with different arrival rates}
\end{figure}

To study the impact of the arrival rate on the estimation accuracy,
the same submethod is applied to four different Poisson processes
of which the average arrival rates are 3, 5, 10, and 15, respectively.
In each test, 1,000 cycles of data are used. The results in Figure
\ref{fig: pr_results_arrival_rate} and Figure \ref{fig: q_results_arrival_rate}
show that the larger the arrival rate is, the more accurate the estimation
tends to be. The reason is that a higher arrival rate implies more
observations of the probe vehicles, which could generally improve
the estimation accuracy.

\subsubsection{The effect of overflow queues}

\begin{figure}[H]
\begin{centering}
\subfloat[]{\begin{centering}
\includegraphics[width=0.4\textwidth]{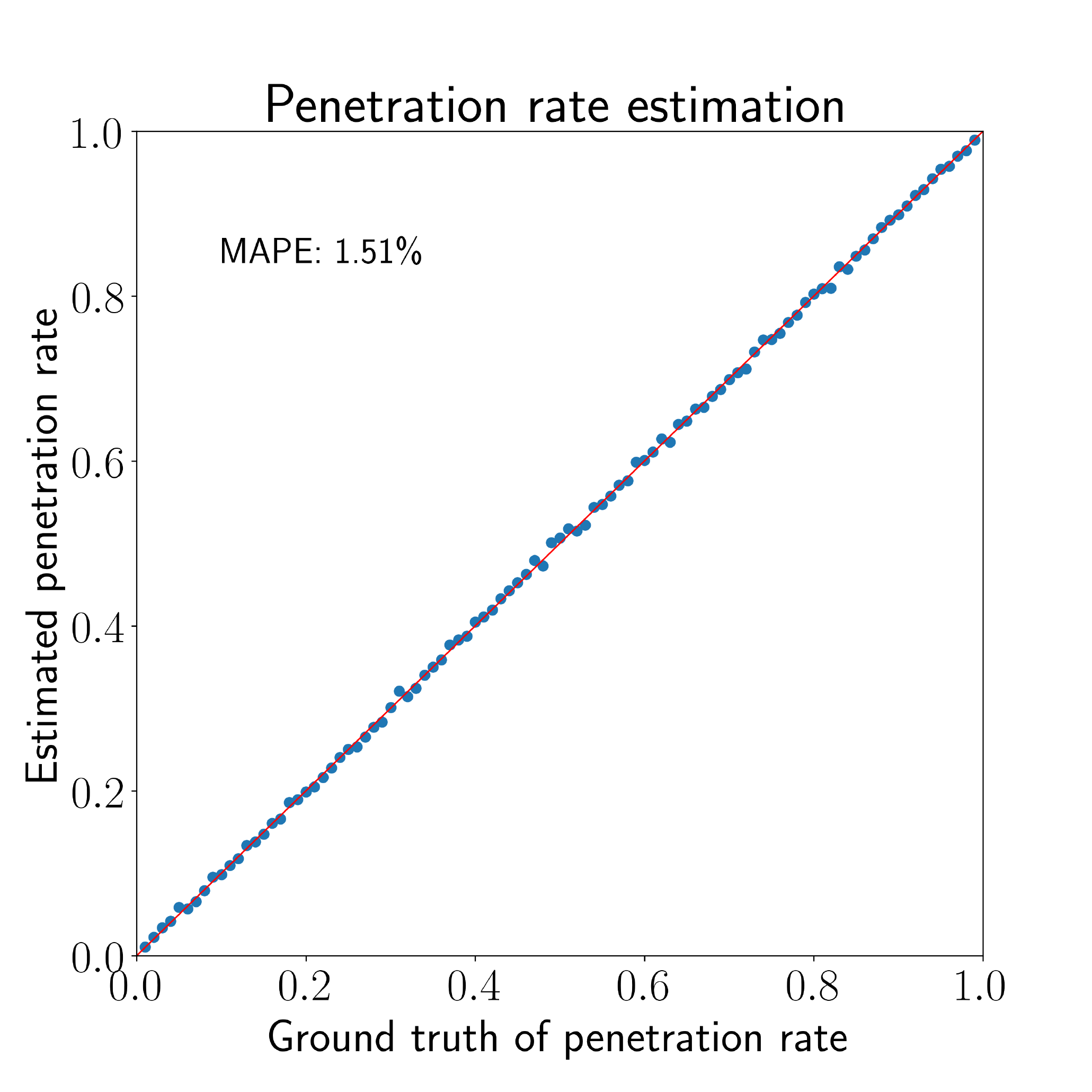}
\par\end{centering}
}\subfloat[]{\begin{centering}
\includegraphics[width=0.4\textwidth]{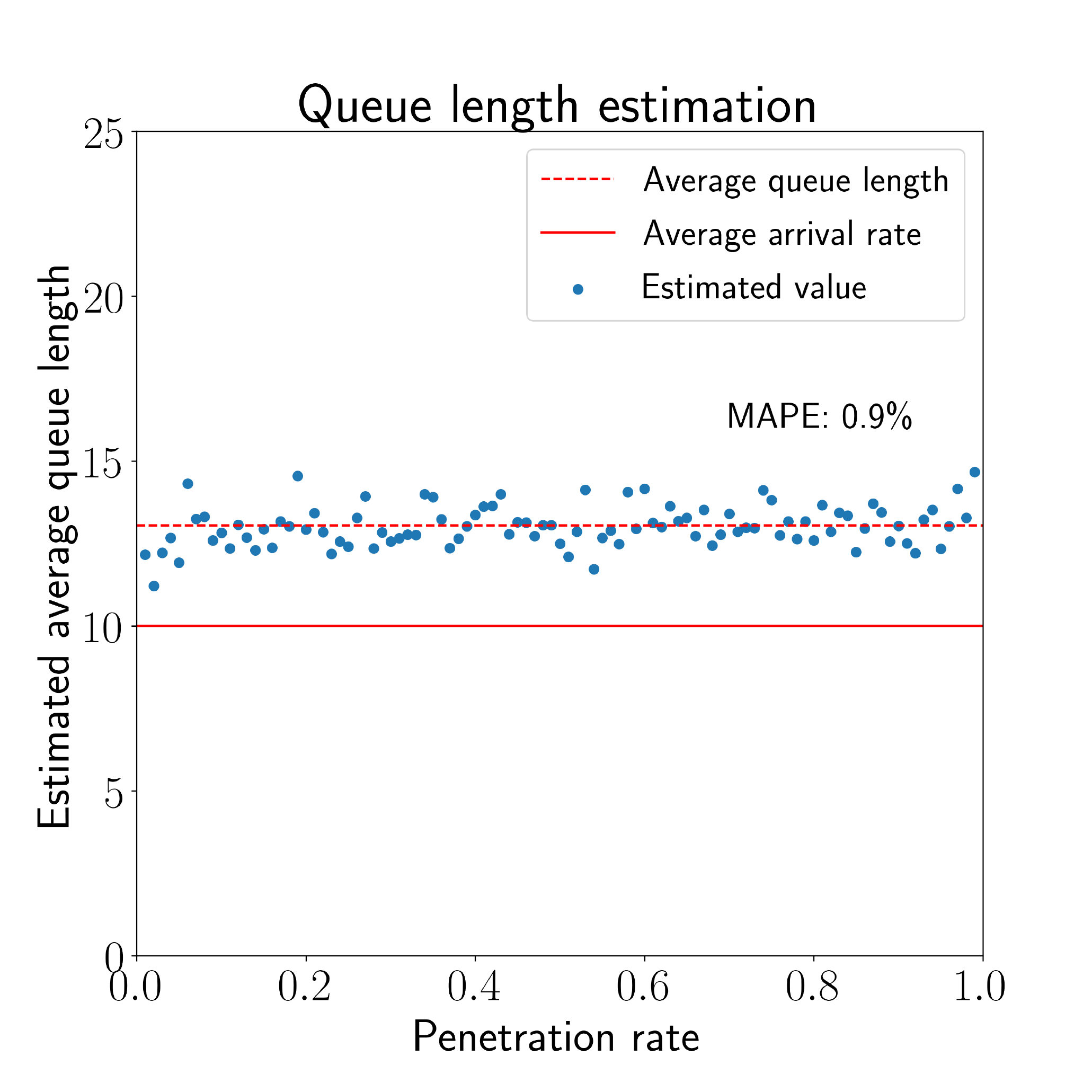}
\par\end{centering}
}
\par\end{centering}
\caption{\label{fig: results_overflow}The estimation results of the cases
with overflow queues: (a) penetration rate estimation, (b) queue length
estimation}
\end{figure}

In previous subsections, it is assumed that the queue in each cycle
can be entirely discharged, that is, there are no overflow queues.
In the real world, the number of vehicles arriving at an intersection
in a cycle might exceed the number of vehicles the traffic signal
can serve. To investigate the impact of the overflow queues on the
estimation accuracy, the cases with overflow queues are also simulated.
The simulation set-up of the overflow queues is similar to \citet{comert2009queue}.
The average arrival rates in the green phase and in the red phase
are set to 10. The maximum number of vehicles that can be served in
each cycle is set to 22. The estimation results for penetration rates
and queue lengths are shown in Figure \ref{fig: results_overflow}.
Since the simulation captures the effect of overflow queues, the average
queue length is different from the average arrival rate in the red
phase. 

\subsection{Real-world data}

\begin{figure}[H]
\begin{centering}
\includegraphics[width=6in]{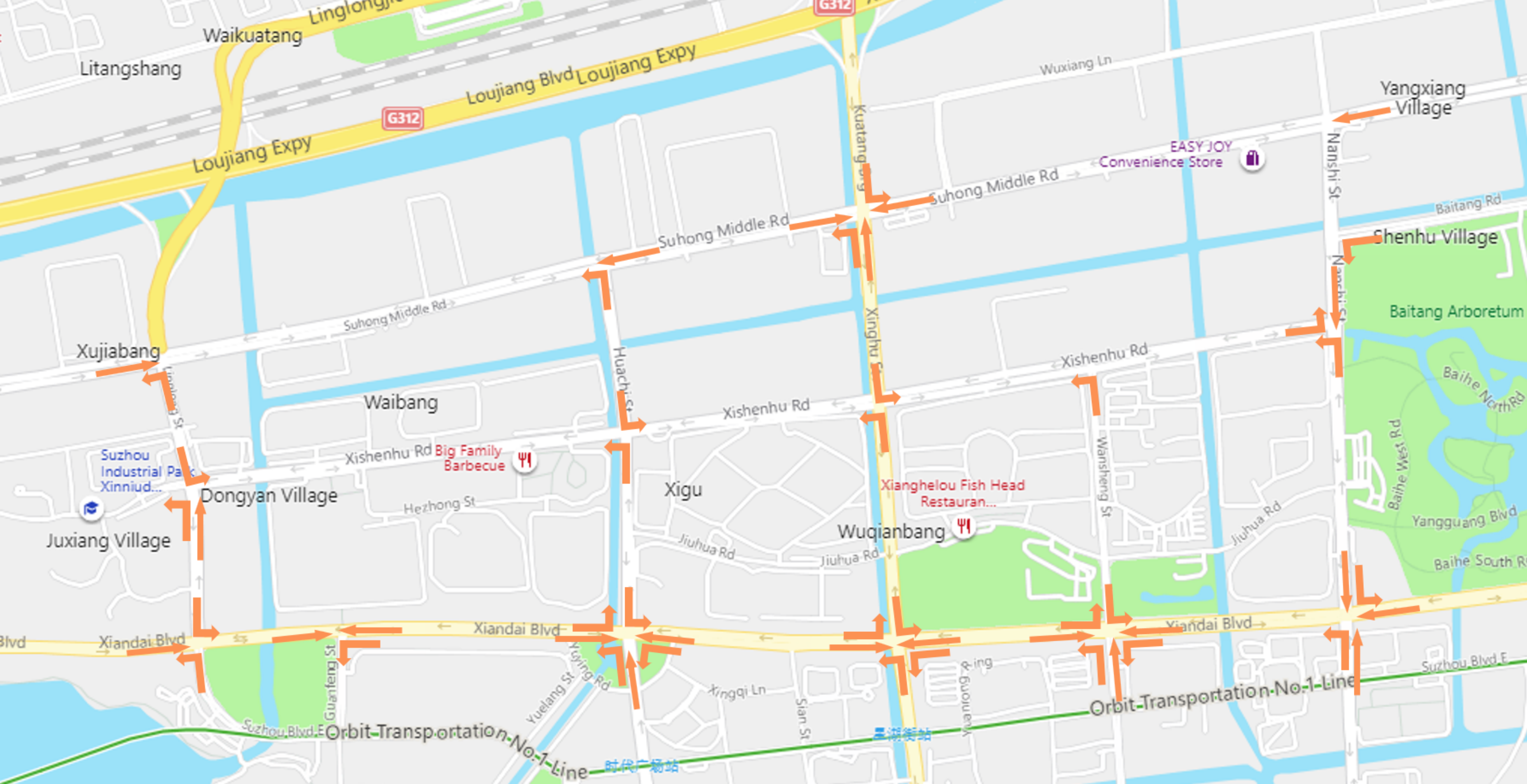}
\par\end{centering}
\noindent \centering{}\caption{\label{fig: The-studied-movements}The studied movements in Suzhou}
\end{figure}

The proposed methods are also tested using real-world data. The focus
of this test is on traffic volume estimation. Queue length estimation
is not validated using real-world data because the ground truth of
queue lengths is not available. The trajectory data are collected
by Didi Chuxing from the vehicles offering its ride-hailing services
in an area in Suzhou, Jiangsu Province, China, shown in Figure \ref{fig: The-studied-movements}.
The data of the 15 workdays from May 8, 2018, to May 28, 2018, are
used for validation. The GPS trajectories of the Didi vehicles in
the selected area are mapped onto the transportation network by a
map matching algorithm \citep{newson2009hidden}. For each movement
and each one-hour time slot, the ``snapshots'' of the trajectory
data are taken to extract the observed partial queues. Due to the
accuracy of the trajectory data, the average space headway for the
queueing vehicles could not be easily estimated. Therefore, its value
is empirically set to 7.5 m/veh for the peak hours and 8.0 m/veh for
the off-peak hours. For the movements with multiple lanes, since the
accuracy of the trajectory data cannot reach the lane level, the stopping
vehicles are randomly assigned to the different lanes. The random
assignment process is repeated for 50 times to get an average estimate.
Signal timing information from other data sources is not necessarily
needed, as\textcolor{black}{{} the trajectory data of the probe vehicles
already contain some signal timing information. For instance, if the
observed partial queue changes from $(0,0,1,0,0,0,1)$ to $(0,0,0,0,0,0,1)$,
then it can be inferred that the latter queue is in the green phase
because the first probe vehicle in the former queue has moved away}.
In general, the cycle length of the traffic signal ranges from 2 min
to 3 min in the selected area. Therefore, for each movement and each
time slot, the number of signal cycles of the evaluation period is
in the range of 300 to 450.

The methodology in this paper does not apply to the right-turn movements,
as there might not be queues. Also, due to the accuracy of the data,
it is almost impossible to deal with the lanes with mixed movements.
The studied movements are represented by the arrows in Figure \ref{fig: The-studied-movements}.
In total, 22 through movements and 31 left-turn movements are studied. 

\begin{figure}[H]
\begin{centering}
\subfloat[]{\begin{centering}
\includegraphics[scale=0.45]{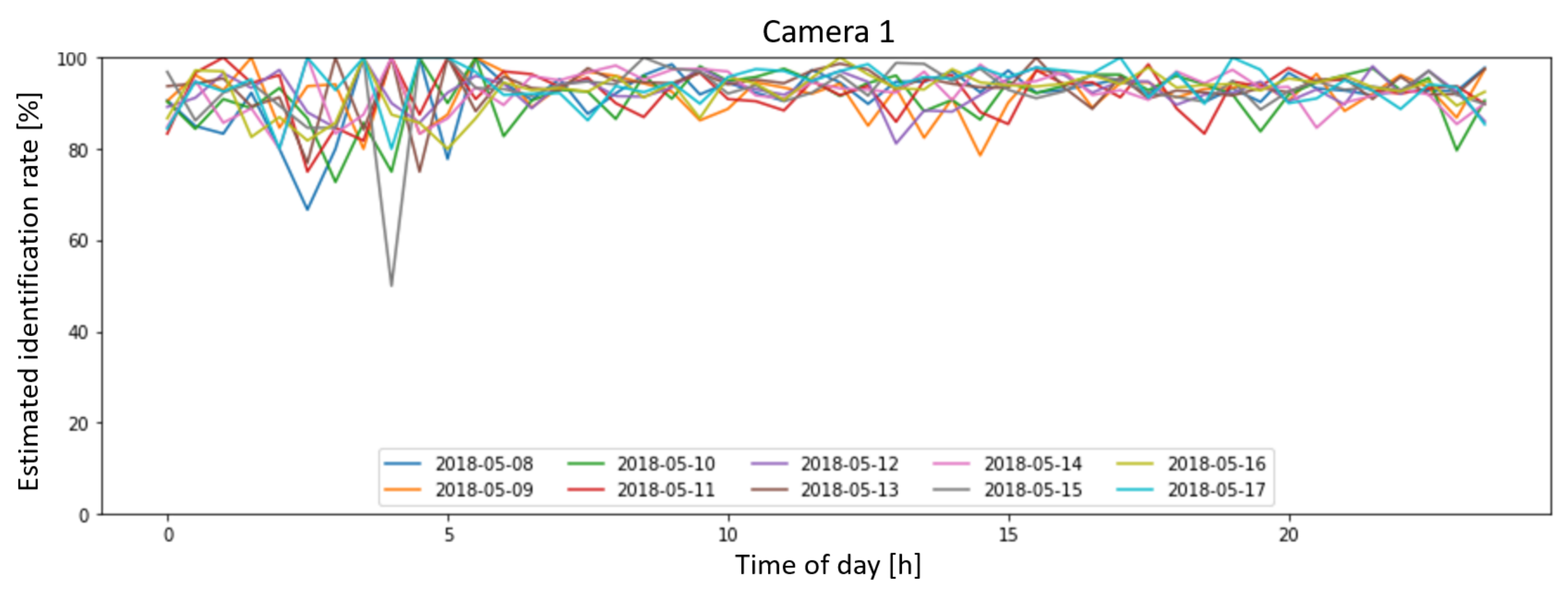}
\par\end{centering}
}
\par\end{centering}
\begin{centering}
\subfloat[]{\begin{centering}
\includegraphics[scale=0.45]{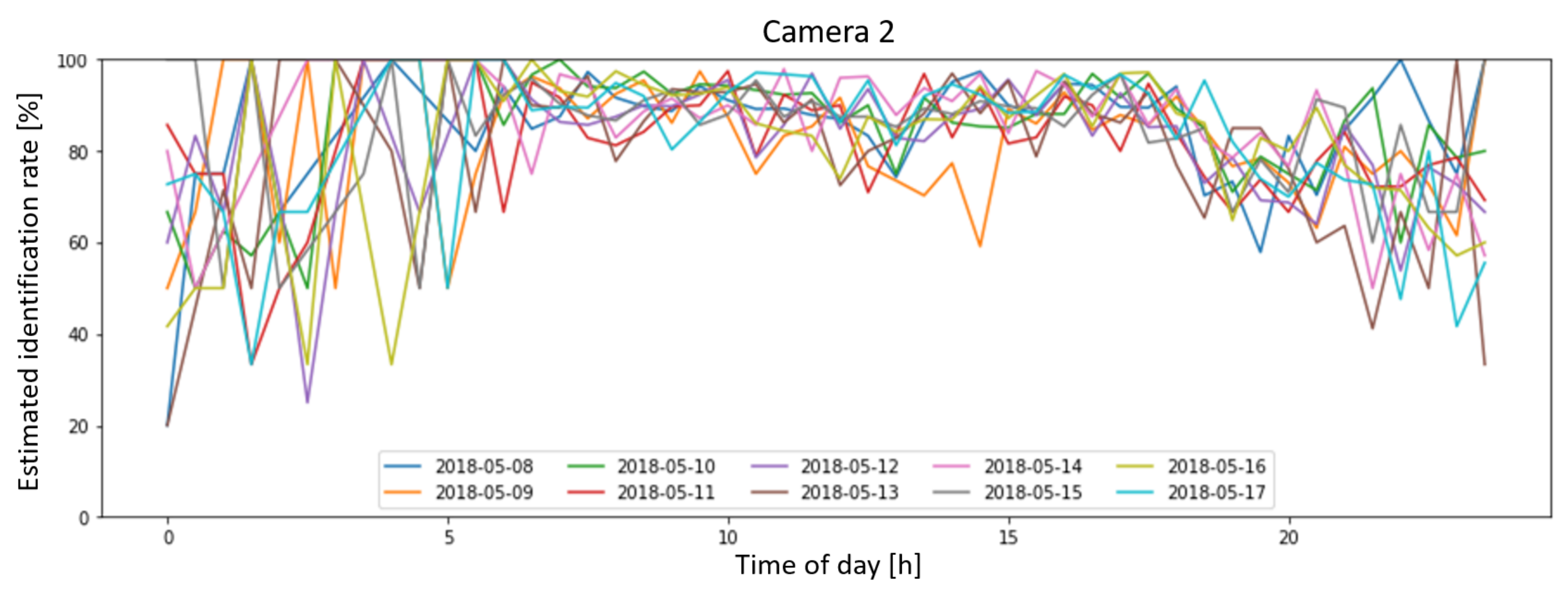}
\par\end{centering}
}
\par\end{centering}
\begin{centering}
\subfloat[]{\begin{centering}
\includegraphics[scale=0.45]{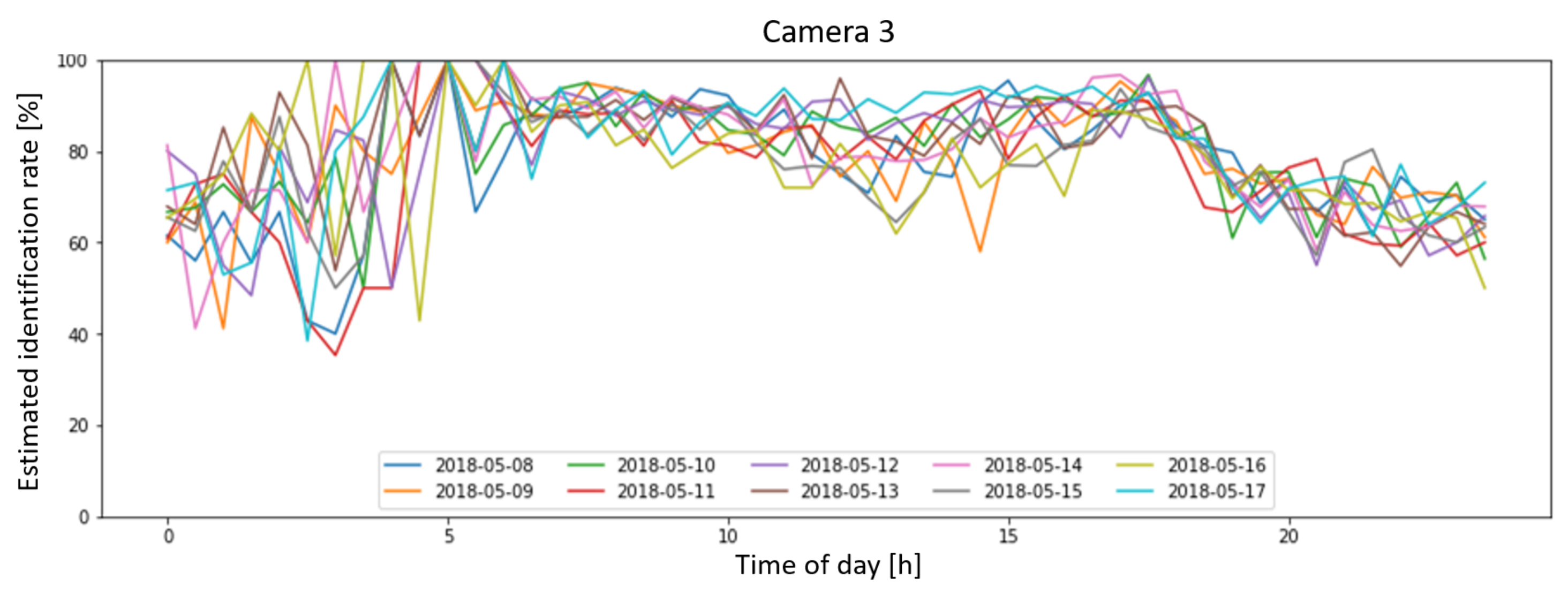}
\par\end{centering}
}
\par\end{centering}
\caption{\label{fig: camera_acura}Accuracy of three typical cameras: (a) camera
1, (b) camera 2, (c) camera 3}
\end{figure}

Most of the signalized intersections in the selected area are covered
by the camera-based automatic vehicle identification systems (AVIS)
that can record the timestamps when vehicles go through the intersections.
Nevertheless, not all the vehicles could be successfully identified
by the cameras, and thus the vehicle counts given by the cameras are
always smaller than the actual traffic volumes. Therefore, for each
camera, its identification rate is estimated by the ratio of the number
of identified Didi Vehicles and the total number of Didi vehicles
passing the camera. Then, the real ``ground truth'' of the traffic
volumes are projected by dividing the vehicle counts by the estimated
identification rates. The estimated identification rates of three
representative cameras during May 8, 2018, to May 15, 2018, are shown
in Figure \ref{fig: camera_acura}. The identification rates mostly
vary from 80\% to 100\% during the day time, whereas the performance
of the cameras becomes very unstable during the night time. Camera
1 outperforms camera 2 and camera 3 at night likely due to better
lighting conditions. Considering the unstable accuracy during the
night time, we only used the data collected from 8:00 to 19:00 for
the validation. 

\subsubsection{Results}

\begin{figure}[H]
\begin{centering}
\subfloat[]{\begin{centering}
\includegraphics[width=0.5\textwidth]{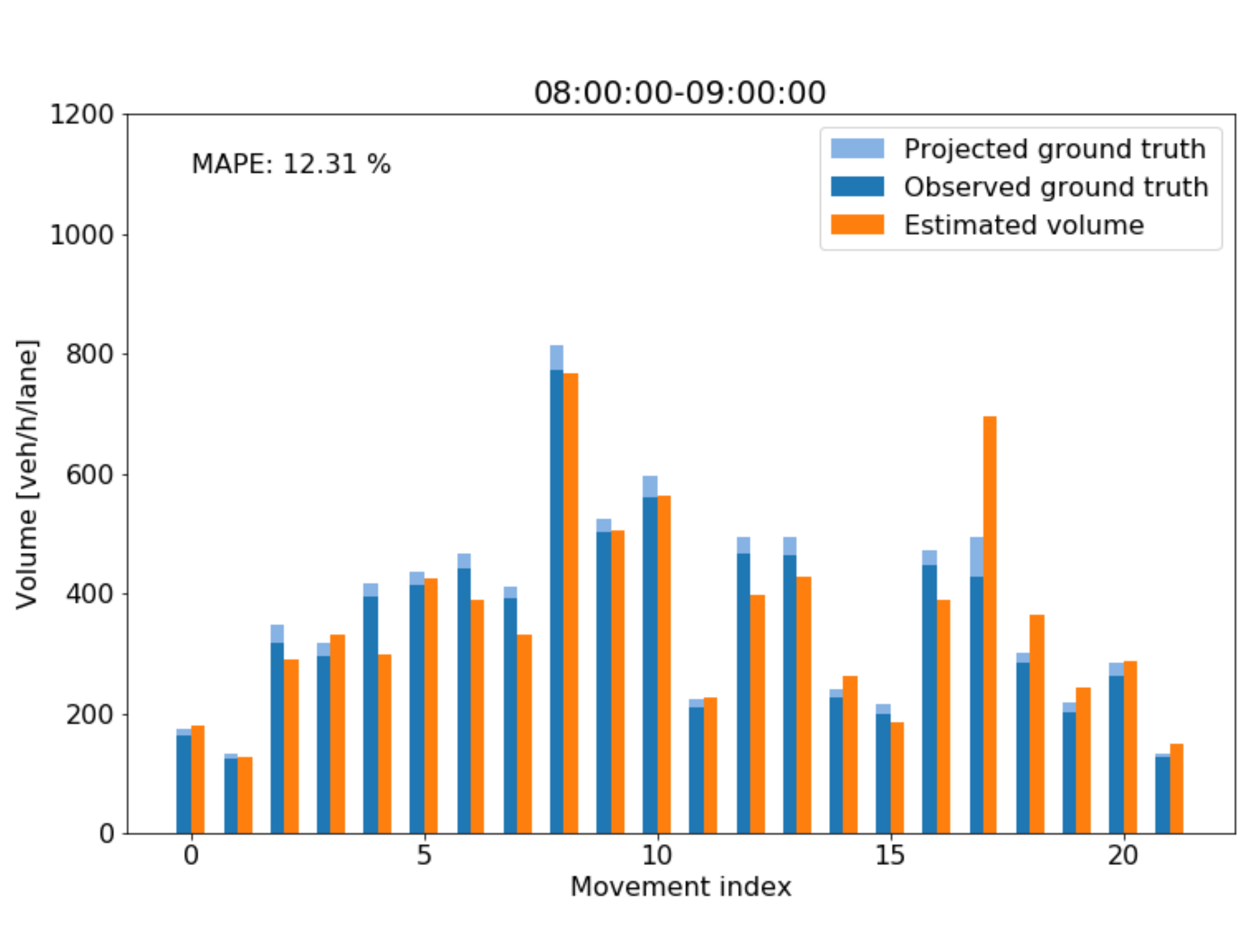}
\par\end{centering}
}\subfloat[]{\begin{centering}
\includegraphics[width=0.5\textwidth]{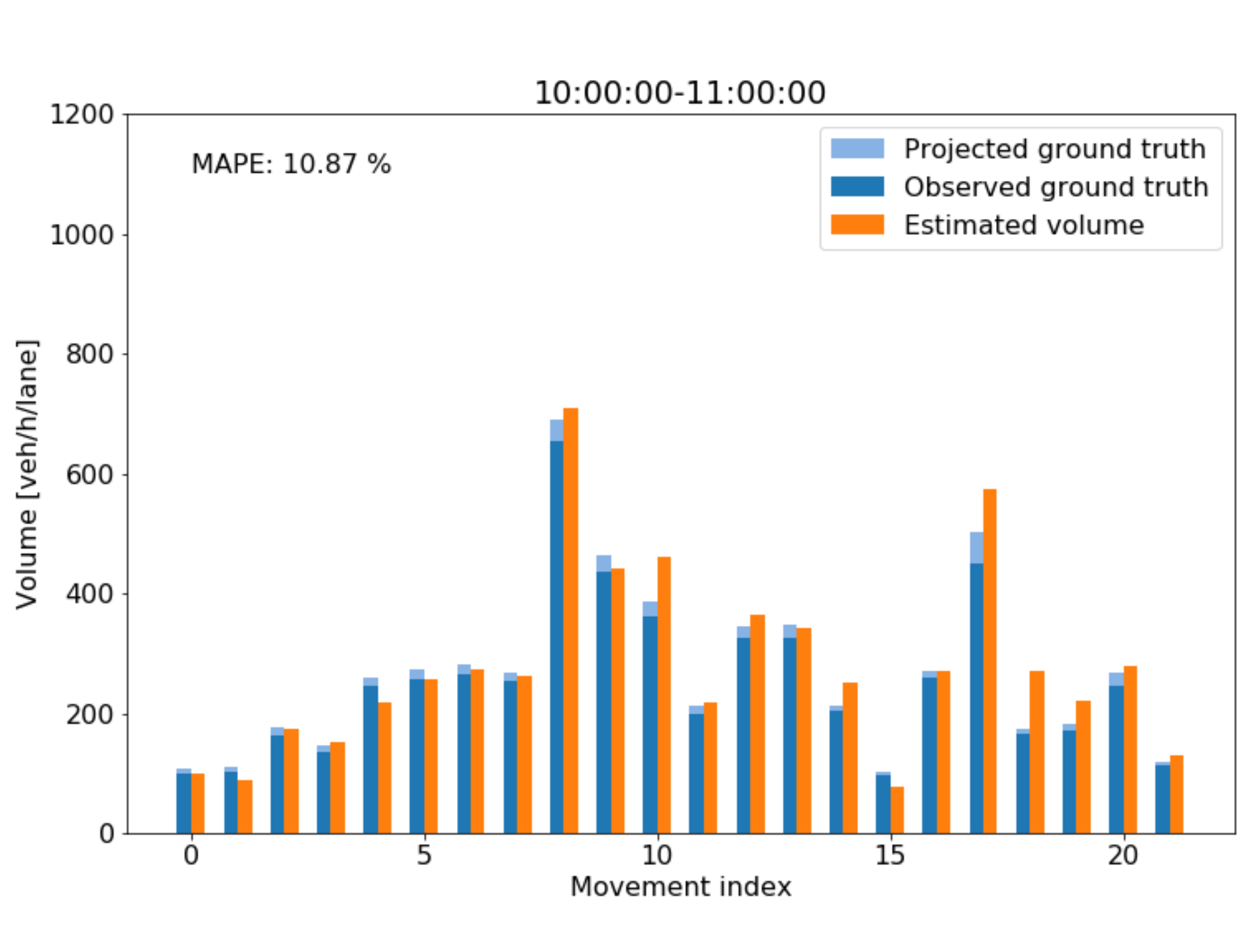}
\par\end{centering}
}
\par\end{centering}
\begin{centering}
\subfloat[]{\begin{centering}
\includegraphics[width=0.5\textwidth]{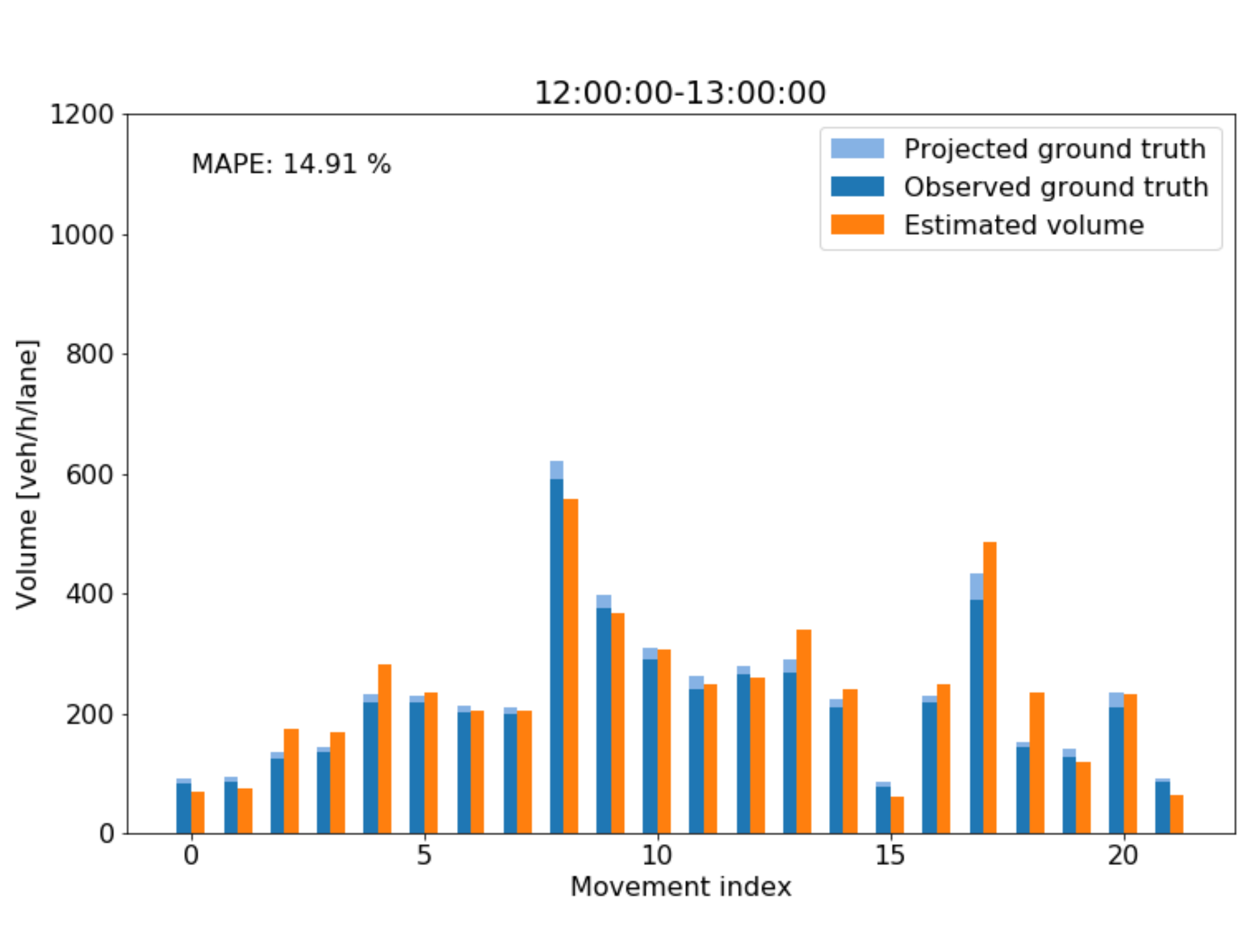}
\par\end{centering}
}\subfloat[]{\begin{centering}
\includegraphics[width=0.5\textwidth]{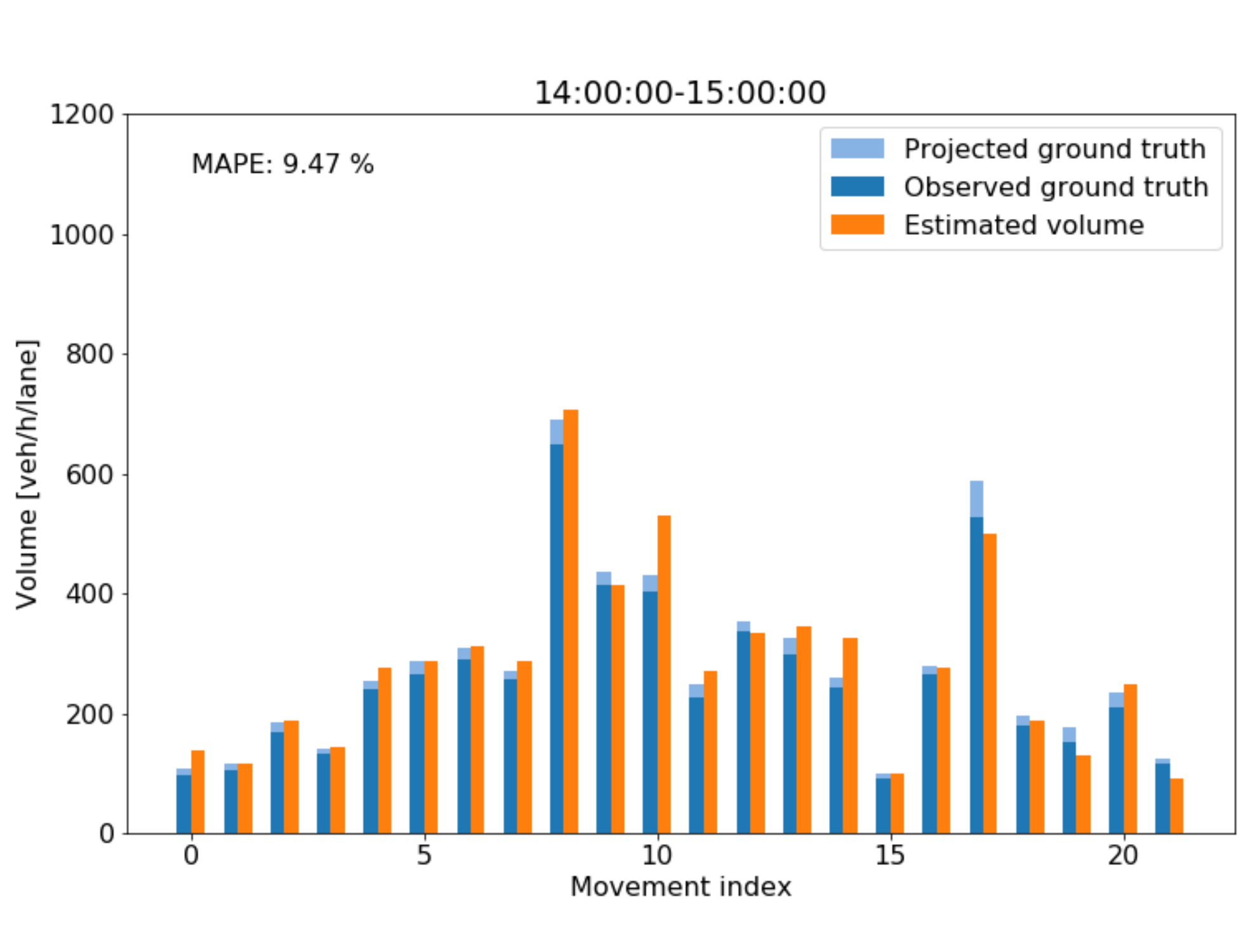}
\par\end{centering}
}
\par\end{centering}
\begin{centering}
\subfloat[]{\begin{centering}
\includegraphics[width=0.5\textwidth]{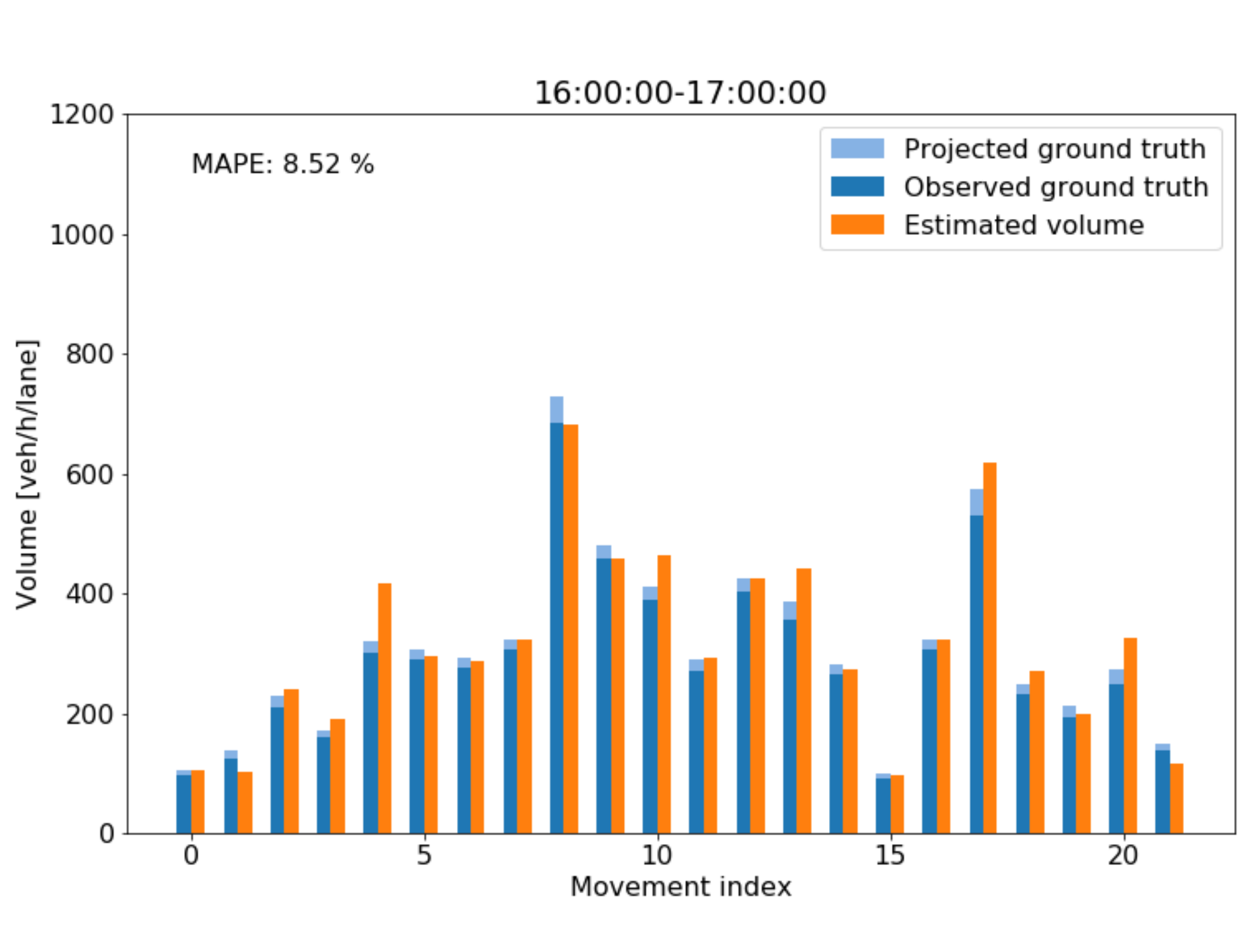}
\par\end{centering}
}\subfloat[]{\begin{centering}
\includegraphics[width=0.5\textwidth]{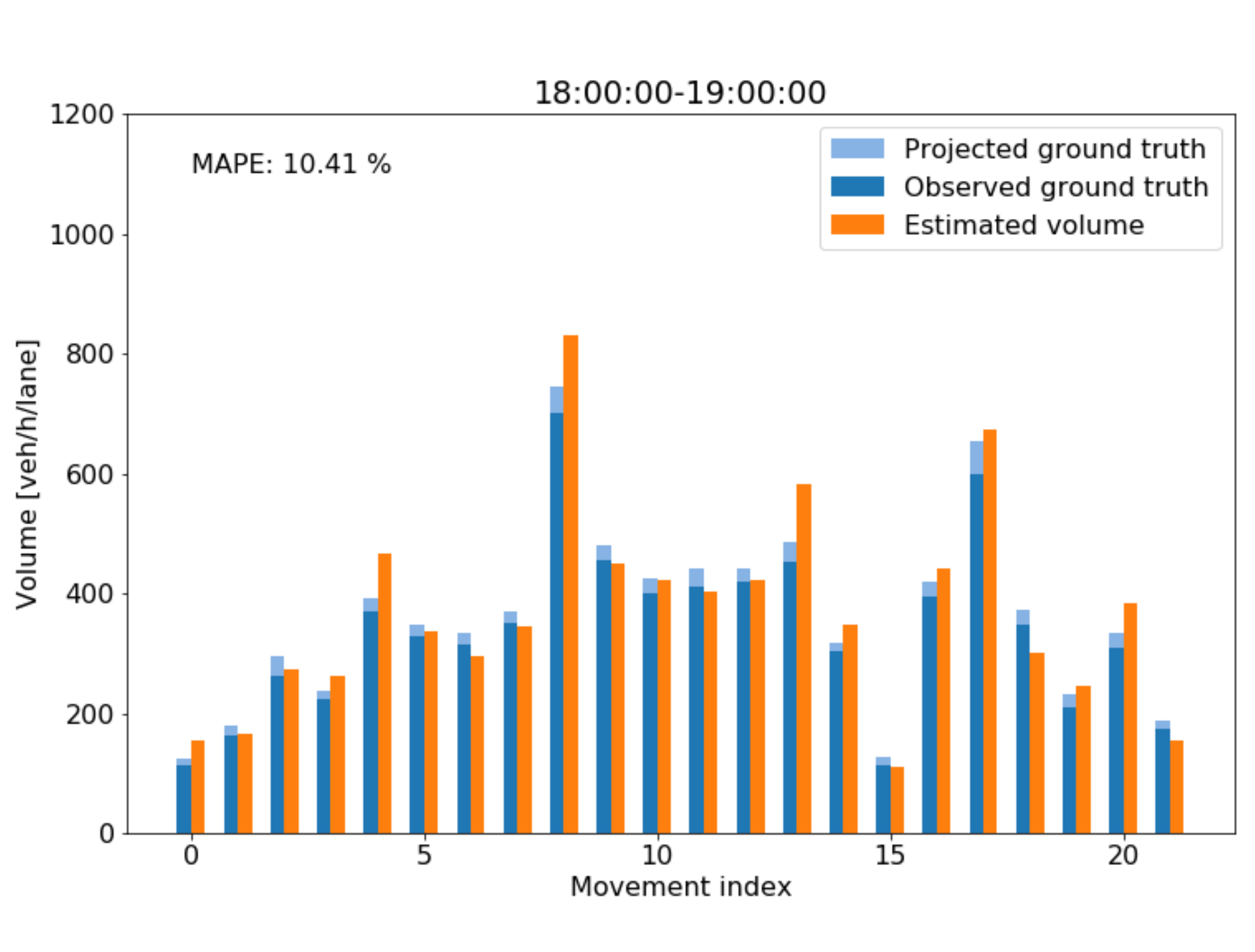}
\par\end{centering}
}
\par\end{centering}
\caption{\label{fig: Results-for-through}Traffic volume estimation results
for the through movements in different TODs: (a) 08:00-09:00, (b)
10:00-11:00, (c) 12:00-13:00, (d) 14:00-15:00, (e) 16:00-17:00, (f)
18:00-19:00}
\end{figure}

\begin{figure}[H]
\noindent \begin{centering}
\subfloat[]{\noindent \begin{centering}
\includegraphics[width=0.5\textwidth]{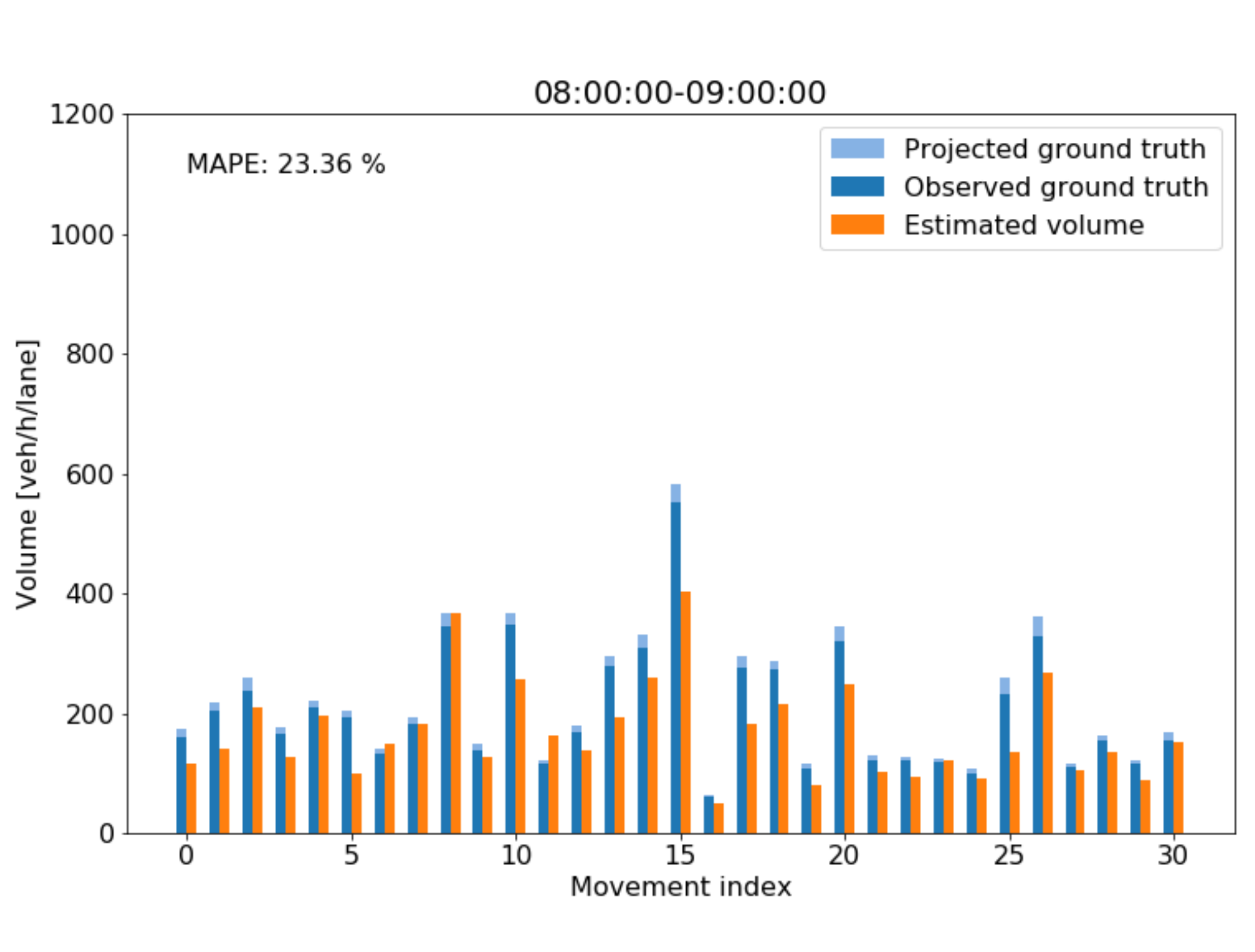}
\par\end{centering}
}\subfloat[]{\noindent \begin{centering}
\includegraphics[width=0.5\textwidth]{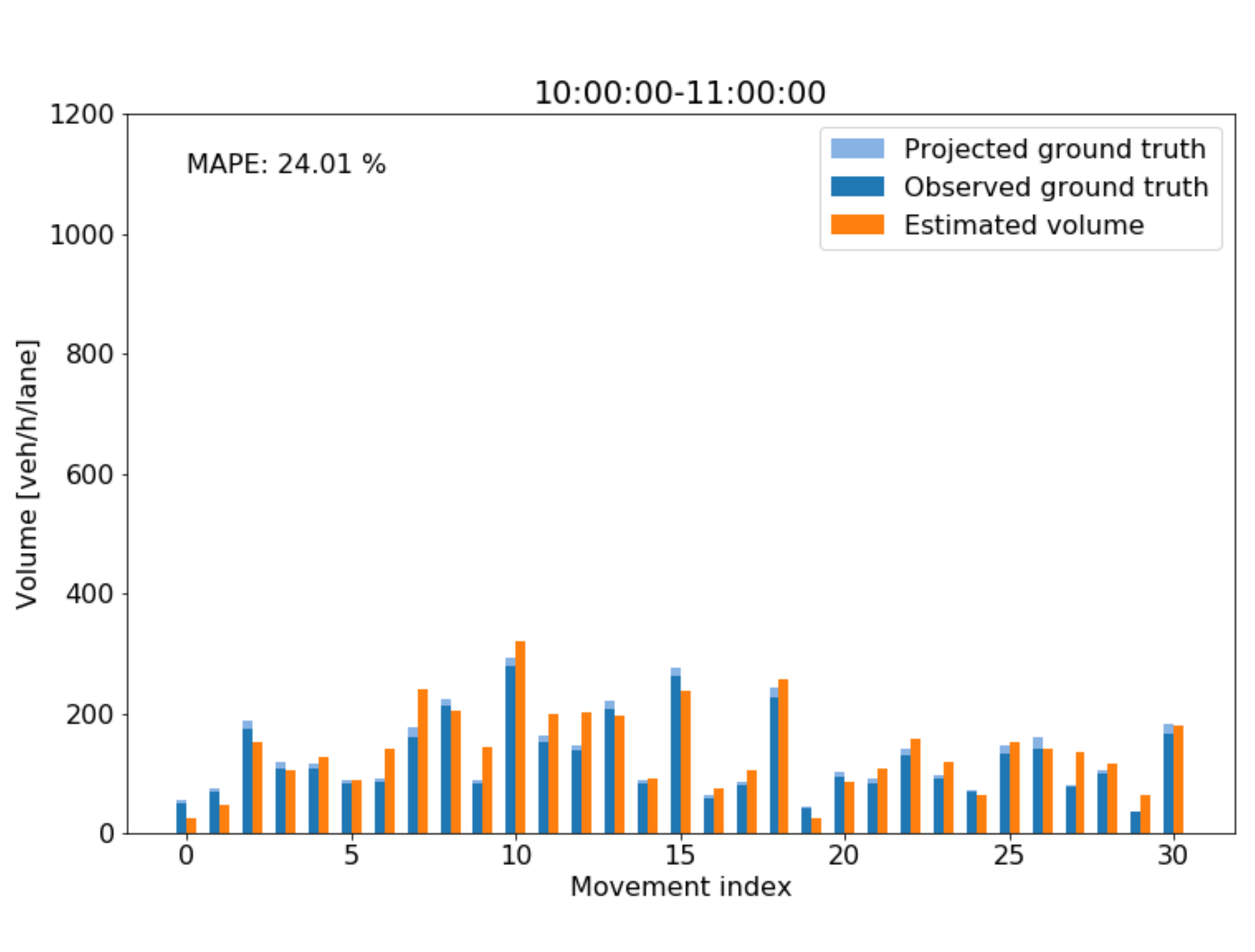}
\par\end{centering}
}
\par\end{centering}
\begin{centering}
\subfloat[]{\begin{centering}
\includegraphics[width=0.5\textwidth]{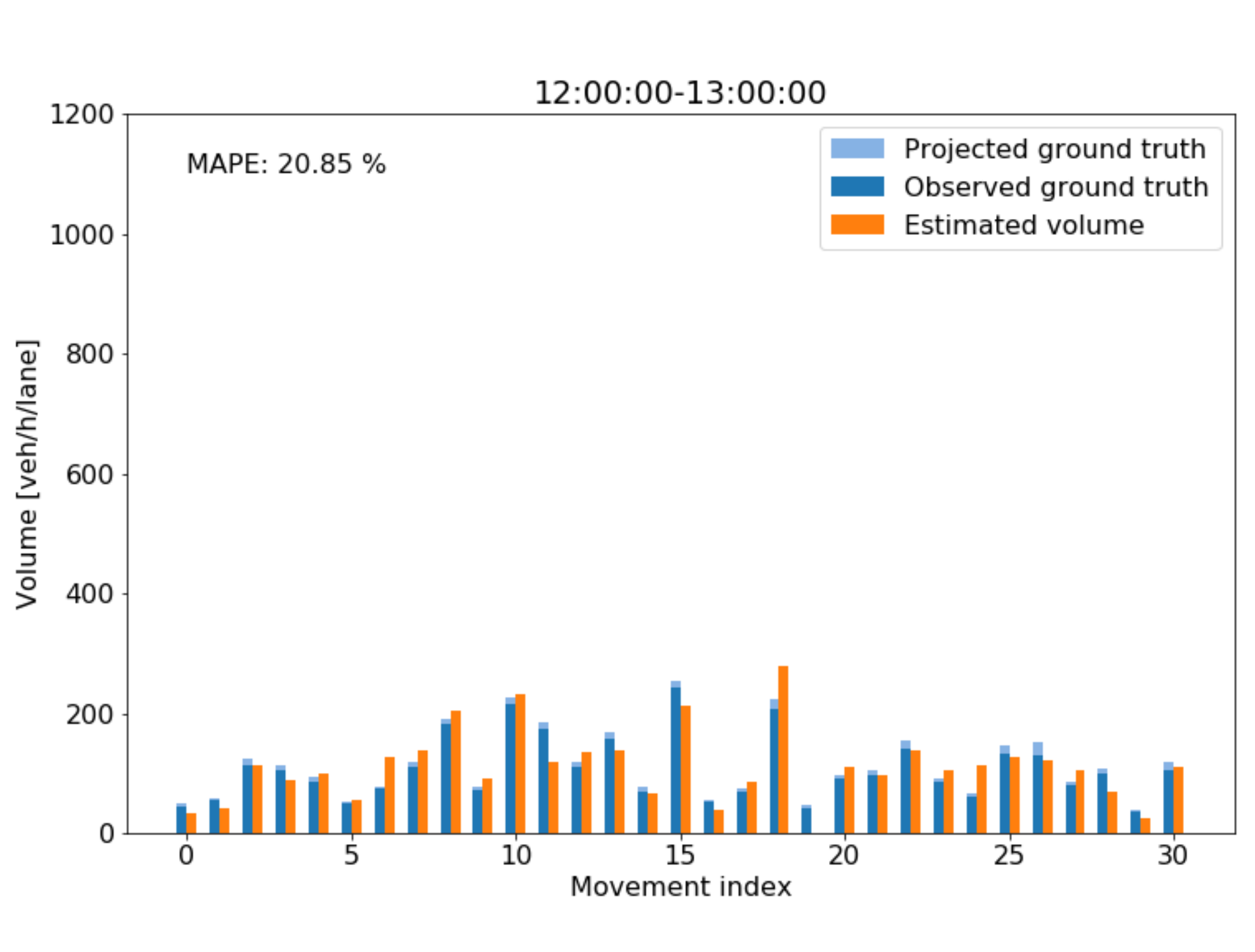}
\par\end{centering}
}\subfloat[]{\begin{centering}
\includegraphics[width=0.5\textwidth]{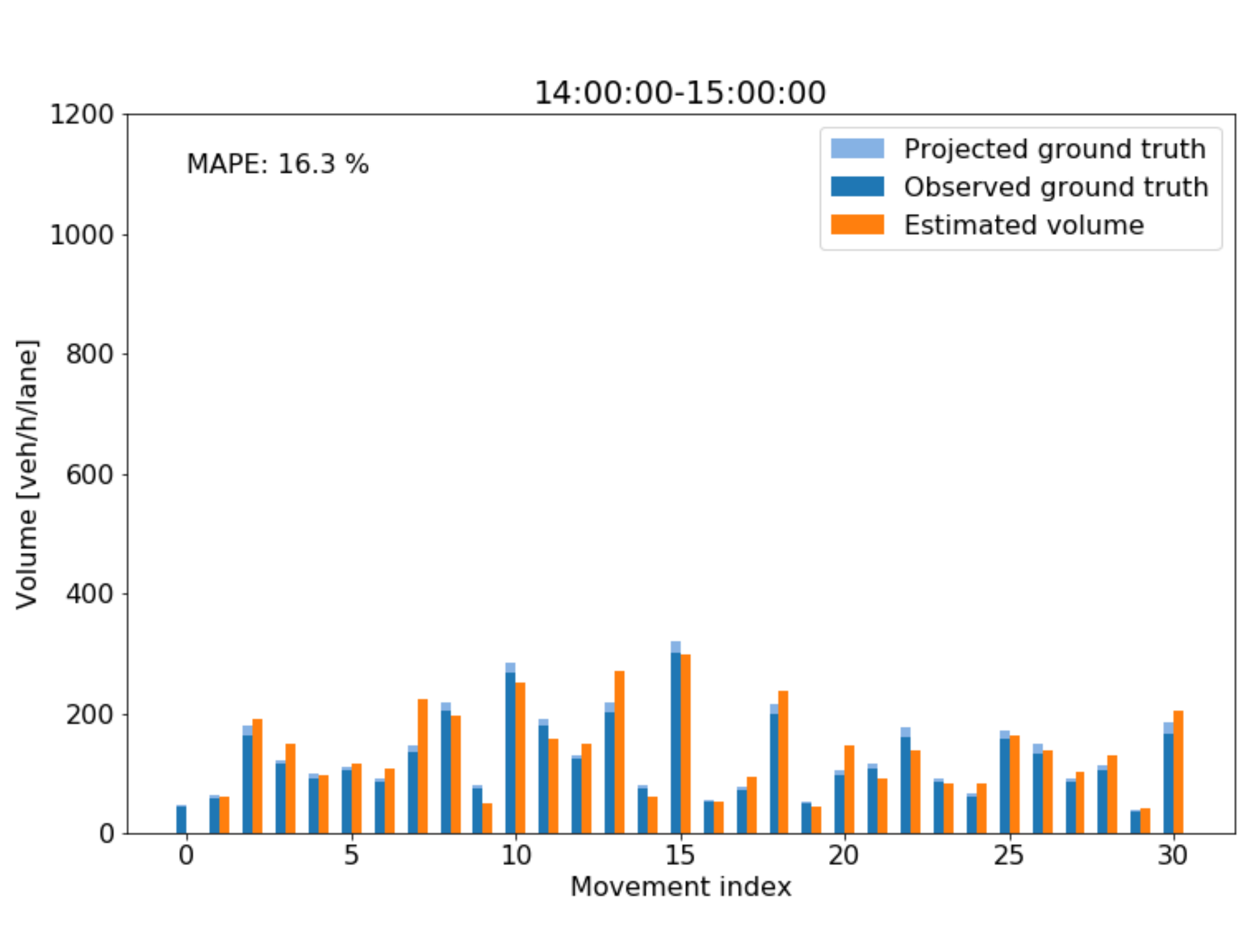}
\par\end{centering}
}
\par\end{centering}
\begin{centering}
\subfloat[]{\begin{centering}
\includegraphics[width=0.5\textwidth]{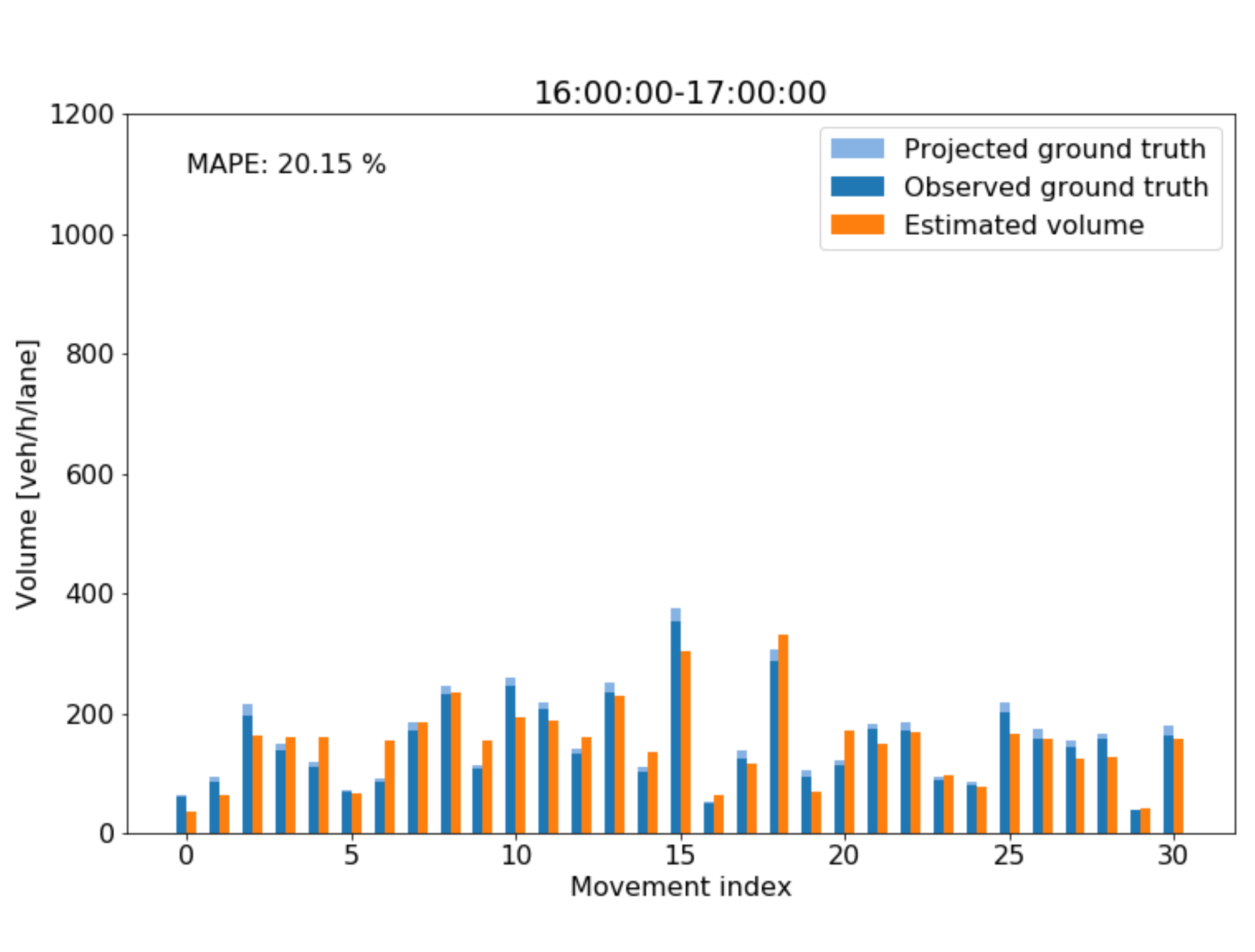}
\par\end{centering}
}\subfloat[]{\begin{centering}
\includegraphics[width=0.5\textwidth]{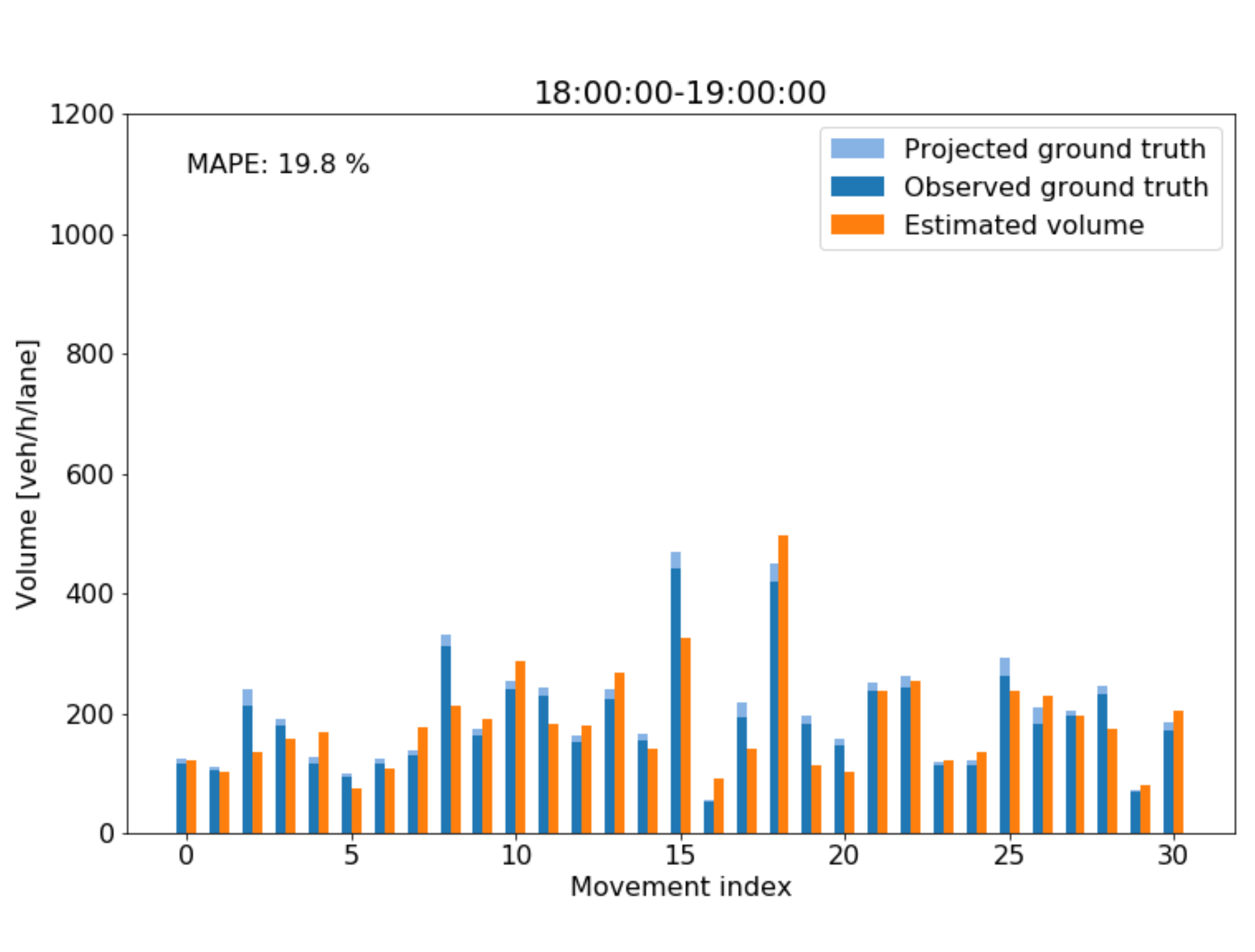}
\par\end{centering}
}
\par\end{centering}
\caption{\label{fig: Results-for-left}Traffic volume estimation results for
the left-turn movements in different TODs: (a) 08:00-09:00, (b) 10:00-11:00,
(c) 12:00-13:00, (d) 14:00-15:00, (e) 16:00-17:00, (f) 18:00-19:00}
\end{figure}

Figure \ref{fig: Results-for-through} shows the results of traffic
volume estimation for the studied through movements in six different
time slots. The estimation results show that the applied method $\frac{Q^{probe}}{\hat{Q}_{3}^{obs}+\hat{Q}_{2}^{hid}(p)}=p$
can estimate traffic volume very accurately, which would be sufficient
for most applications of mid-term or long-term signal control and
performance measures. Figure \ref{fig: Results-for-left} shows the
results for the left-turn movements. The undermined performance further
verifies the effect of the arrival rate on the estimation accuracy
studied using the simulation data, since the traffic volumes of the
left-turn movements are much smaller compared to the through movements.

Compared to the results of the simulation data, the estimation accuracy
is undermined when the method is applied to the real-world data, due
to the following reasons. First, although the map matching algorithm
\citep{newson2009hidden} can mitigate the effect of GPS errors at
the data preprocessing stage, the errors in the real-world trajectory
data could still influence the estimation accuracy. Second, in the
real world, for each movement and each one-hour time slot, the penetration
rate and the queueing pattern might slightly vary during the studied
15 workdays. Third, the average space headway for the queueing vehicles
is set empirically, which might introduce some biases into the results.
If the data with better accuracy are available, the value of the average
space headway should be estimated independently for each movement
and each time slot.

\section{Conclusions\label{sec: discussion}}

This paper proposes a general framework and a series of methods for
the trajectory-based queue length and traffic volume estimation. For
each specific movement and each specific time slot, the penetration
rate of the probe vehicles is estimated by using the aggregated historical
trajectory data of the probe vehicles. Once the penetration rate is
estimated, it can be used to project the queue length and the traffic
volume. 

The proposed methods do not assume the type of vehicle arrival process
or the queueing process. Therefore, the proposed methods are adaptable
to both under-saturation and over-saturation cases. The proposed methods
do not require high penetration rates and would be feasible for use
in reality nowadays. The tests by both the simulation and the real-world
data show good estimation accuracy, indicating that the proposed methods
could be used for traffic signal control and performance measures
at signalized intersections.

There are certain limitations in the current work that should be addressed
in the future. For instance, the proposed methods in this paper take
the stopping positions of the probe vehicles as the features to infer
the penetration rate of the probe vehicles. However, there might not
be queues forming at the non-signalized intersections or in the right-turn
movements. Also, the queueing patterns in the shared left-through
(right-through) lanes could be different from other left-turn (right-turn)
lanes or through lanes. Therefore, when applying the proposed methods
to these cases, additional care is required. 

\pagebreak{}

\bibliographystyle{elsarticle-harv}
\bibliography{reference}

\pagebreak{}

\appendix

\section*{Appendix A \label{sec: Appendix-A}}

\setcounter{equation}{0} 
\renewcommand\theequation{A.\arabic{equation}} 

\subsection*{Definitions}

For $k,n\in\mathbb{N}$ and $n\ge k$, 
\begin{equation}
C_{n}^{k}=\frac{n!}{k!(n-k)!},
\end{equation}
\begin{equation}
A_{n}^{k}=\frac{n!}{(n-k)!}.
\end{equation}

\subsection*{Theorem 1}

For conciseness, $Q_{i},N_{i},S_{i},T_{i},n_{i},s_{i},t_{i}$ are
represented by $Q,N,S,T,n,s,t$, respectively.

\begin{align}
\mathbb{E}(S\mid N=n,Q=l) & =\frac{l+1}{n+1},
\end{align}

\begin{equation}
\mathbb{E}(Q\mid N=n)=\mathbb{E}(S\mid N=n)(n+1)-1,
\end{equation}

where $n\ge1$.

Proof: 
\begin{align}
\mathbb{E}(S\mid N=n,Q=l) & =\sum_{j=1}^{l-n+1}P(S=j\mid N=n,Q=l)j\\
 & =\sum_{j=1}^{l-n+1}\frac{nC_{l-n}^{j-1}A_{j-1}^{j-1}A_{l-j}^{l-j}}{A_{l}^{l}}j\\
 & =\sum_{j=1}^{l-n+1}\frac{nA_{l-j}^{n-1}}{A_{l}^{n}}i\\
 & =\frac{n}{A_{l}^{n}}\sum_{j=1}^{l-n+1}A_{l-j}^{n-1}j\\
 & =\frac{n}{A_{l}^{n}}\sum_{k=0}^{l-n}A_{n+k-1}^{n-1}(l-n+1-k)\\
 & =\frac{n}{A_{l}^{n}}\sum_{k=0}^{l-n}A_{n+k-1}^{n-1}(l+1)-\frac{n}{A_{l}^{n}}\sum_{k=0}^{l-n}A_{n+k-1}^{n-1}(n+k)\\
 & =(l+1)\sum_{k=0}^{l-n}\frac{(n+k-1)!(l-n)!n!}{k!l!(n-1)!}-\frac{n}{A_{l}^{n}}\sum_{k=0}^{l-n}A_{n+k}^{n}\\
 & =\frac{l+1}{C_{l}^{n}}\sum_{k=0}^{l-n}C_{n+k-1}^{n-1}-\frac{n}{C_{l}^{n}}\sum_{k=0}^{l-n}C_{n+k}^{n}\label{eq: thm1_1}\\
 & =(l+1)\frac{C_{l}^{n}}{C_{l}^{n}}-n\frac{C_{l+1}^{n+1}}{C_{l}^{n}}\label{eq: thm1_2}\\
 & =(l+1)-n\frac{l+1}{n+1}\\
 & =\frac{l+1}{n+1}
\end{align}

Chu's theorem \citep{merris2003combinatorics} is applied when converting
equation (\ref{eq: thm1_1}) to equation (\ref{eq: thm1_2}).

Then, based on the results above, 
\begin{align}
\mathbb{E}(S\mid N=n) & =\sum_{j=1}^{L_{max}}P(S=j\mid N=n)j\\
 & =\sum_{j=1}^{L_{max}}\sum_{l=j+n-1}^{L_{max}}P(S=j\mid N=n,Q=l)P(Q=l\mid N=n)j\\
 & =\sum_{l=n}^{L_{max}}\sum_{j=1}^{l-n+1}P(S=j\mid N=n,Q=l)P(Q=l\mid N=n)j\\
 & =\sum_{l=n}^{L_{max}}P(Q=l\mid N=n)\sum_{j=1}^{l-n+1}P(S=j\mid N=n,Q=l)j\\
 & =\sum_{l=n}^{L_{max}}P(Q=l\mid N=n)\mathbb{E}(S\mid N=n,Q=l)\\
 & =\sum_{l=n}^{L_{max}}P(Q=l\mid N=n)\frac{l+1}{n+1}\\
 & =\frac{1}{n+1}\sum_{l=n}^{L_{max}}P(Q=l\mid N=n)(l+1)\\
 & =\frac{1}{n+1}\left(\mathbb{E}(Q\mid N=n)+1\right).
\end{align}

This is equivalent to

\begin{equation}
\mathbb{E}(Q\mid N=n)=\mathbb{E}(S\mid N=n)(n+1)-1.
\end{equation}

\subsection*{Theorem 2}

For conciseness, $Q_{i},N_{i},S_{i},T_{i},n_{i},s_{i},t_{i}$ are
represented by $Q,N,S,T,n,s,t$, respectively.

\begin{align}
\mathbb{E}(T\mid N=n,Q=l) & =n\frac{l+1}{n+1},
\end{align}

\begin{equation}
\mathbb{E}(Q\mid N=n)=\mathbb{E}(T\mid N=n)\frac{n+1}{n}-1,
\end{equation}

where $n\ge1$.

Proof:

\begin{align}
\mathbb{E}(T\mid N=n,Q=l) & =\sum_{j=n}^{l}P(T=j\mid N=n,Q=l)j\\
 & =\sum_{j=n}^{l}\frac{nC_{l-n}^{l-j}A_{j-1}^{j-1}A_{l-j}^{l-j}}{A_{l}^{l}}j\\
 & =\sum_{j=n}^{l}\frac{nA_{j-1}^{n-1}}{A_{l}^{n}}j\\
 & =n\sum_{j=n}^{l}\frac{A_{j}^{n}}{A_{l}^{n}}\\
 & =n\sum_{j=n}^{l}\frac{C_{j}^{n}}{C_{l}^{n}}\\
 & =\frac{n}{C_{l}^{n}}\sum_{k=0}^{l-n}C_{n+k}^{n}\\
 & =\frac{nC_{l+1}^{n+1}}{C_{l}^{n}}\\
 & =n\frac{l+1}{n+1}
\end{align}

Then, based on the results above, 
\begin{align}
\mathbb{E}(T\mid N=n) & =\sum_{j=n}^{L_{max}}P(T=j\mid N=n)j\\
 & =\sum_{j=n}^{L_{max}}\sum_{l=j}^{L_{max}}P(T=j\mid N=n,Q=l)P(Q=l\mid N=n)j\\
 & =\sum_{l=n}^{L_{max}}\sum_{j=n}^{l}P(T=j\mid N=n,Q=l)P(Q=l\mid N=n)j\\
 & =\sum_{l=n}^{L_{max}}P(Q=l\mid N=n)\sum_{j=n}^{l}P(T=j\mid N=n,Q=l)j\\
 & =\sum_{l=n}^{L_{max}}P(Q=l\mid N=n)\mathbb{E}(T\mid N=n,Q=l)\\
 & =\sum_{l=n}^{L_{max}}P(Q=l\mid N=n)n\frac{l+1}{n+1}\\
 & =\frac{n}{n+1}\sum_{l=n}^{L_{max}}P(Q=l\mid N=n)(l+1)\\
 & =\frac{n}{n+1}\left(\mathbb{E}(Q\mid N=n)+1\right).
\end{align}

This is equivalent to 
\begin{equation}
\mathbb{E}(Q\mid N=n)=\mathbb{E}(T\mid N=n)\frac{n+1}{n}-1.
\end{equation}

\subsection*{Theorem 3}

For conciseness, $Q_{i},N_{i},S_{i},T_{i},n_{i},s_{i},t_{i}$ are
represented by $Q,N,S,T,n,s,t$, respectively.

\begin{equation}
\mathbb{E}(Q\mid N\ge1)=\mathbb{E}(S\mid N\ge1)+\mathbb{E}(T\mid N\ge1)-1
\end{equation}

Proof:

First of all, 
\begin{equation}
P(S=j\mid N\ge1,Q=l)=p(1-p)^{j-1},\text{ if }1\le j\le l.
\end{equation}

\begin{align}
P(T=l-j+1\mid N\ge1,Q=l) & =p(1-p)^{l-(l-j+1)}\\
 & =p(1-p)^{j-1}\\
 & =P(S=j\mid N\ge1,Q=l),\text{ if }1\le j\le l.
\end{align}

Then, 
\begin{align}
\mathbb{E}(S\mid N\ge1) & =\sum_{j=1}^{L_{max}}P(S=j\mid N\ge1)j\\
 & =\sum_{j=1}^{L_{max}}\sum_{l=j}^{L_{max}}P(S=j\mid N\ge1,Q=l)P(Q=l\mid N\ge1)j\\
 & =\sum_{l=1}^{L_{max}}\sum_{j=1}^{l}P(S=j\mid N\ge1,Q=l)P(Q=l\mid N\ge1)j\\
 & =\sum_{l=1}^{L_{max}}\sum_{j=1}^{l}P(T=l-j+1\mid N\ge1,Q=l)P(Q=l\mid N\ge1)j\\
 & =\sum_{l=1}^{L_{max}}\sum_{k=1}^{l}P(T=k\mid N\ge1,Q=l)P(Q=l\mid N\ge1)(l-k+1)
\end{align}

\begin{align}
\mathbb{E}(T\mid N\ge1) & =\sum_{k=1}^{L_{max}}P(T=k\mid N\ge1)k\\
 & =\sum_{k=1}^{L_{max}}\sum_{l=k}^{L_{max}}P(T=k\mid N\ge1,Q=l)P(Q=l\mid N\ge1)k\\
 & =\sum_{l=1}^{L_{max}}\sum_{k=1}^{l}P(T=k\mid N\ge1,Q=l)P(Q=l\mid N\ge1)k
\end{align}

Therefore, 
\begin{align}
\mathbb{E}(S\mid N\ge1)+\mathbb{E}(T\mid N\ge1)-1 & =\sum_{l=1}^{L_{max}}\sum_{k=1}^{l}P(T=k\mid N\ge1,Q=l)P(Q=l\mid N\ge1)(l-1)+1\\
 & =\sum_{l=1}^{L_{max}}P(Q=l\mid N\ge1)(l-1)+1\\
 & =\sum_{l=1}^{L_{max}}P(Q=l\mid N\ge1)l\\
 & =\mathbb{E}(Q\mid N\ge1)
\end{align}

Alternatively, Theorem 3 can also be proved by combining Theorem 1
and Theorem 2.
\end{document}